\documentclass[a4paper,11pt]{article}
\pdfoutput=1

\usepackage[a4paper,left=2.73cm,right=2.7cm,top=3cm,bottom=3.5cm]{geometry}

\usepackage{hyperref}
\usepackage{amsmath,amssymb,graphicx,multirow,subcaption,tikz}
\usepackage{colortbl}

\newcommand{\be}{\begin{equation}}
\newcommand{\ee}{\end{equation}}

\newcommand{\bal}{\begin{aligned}}
\newcommand{\eal}{\end{aligned}}

\def\cN{{\cal N}}

\numberwithin{equation}{section}

\usepackage{bm} 
\allowdisplaybreaks  

\def\sst#1{{\scriptscriptstyle #1}}
\def\0{{\sst{(0)}}}
\def\1{{\sst{(1)}}}
\def\2{{\sst{(2)}}}
\def\3{{\sst{(3)}}}
\def\4{{\sst{(4)}}}
\def\5{{\sst{(5)}}}
\def\6{{\sst{(6)}}}
\def\7{{\sst{(7)}}}
\def\8{{\sst{(8)}}}

\begin{document}

 \begin{titlepage}

\thispagestyle{empty}

\begin{flushright}
IFT-UAM/CSIC-19-128 \\
HIP-2019-32/TH
\end{flushright}

   \begin{center}
   \baselineskip=20pt

\begin{Large}\textbf{
Flowing to $\mathcal{N}=3$ Chern--Simons-matter theory
}\end{Large}

\vspace{40pt}

{\large  \textbf{Adolfo Guarino}$^{a, b}$ ,\,\, \textbf{Javier Tarr\'io}$^{a,c}$ \,\,and\,\,  \textbf{Oscar Varela}$^{d, e}$}

\vspace{25pt}

$^a${\normalsize  
Departamento de F\'isica, Universidad de Oviedo,\\
Avda. Federico Garc\'ia Lorca 18, 33007 Oviedo, Spain.}
\\[5mm]

$^b${\normalsize  
Instituto Universitario de Ciencias y Tecnolog\'ias Espaciales de Asturias (ICTEA) \\
Calle de la Independencia 13, 33004 Oviedo, Spain.}
\\[5mm]

$^c${\normalsize  Department of Physics and Helsinki Institute of Physics \\ P.O.Box 64
FIN-00014, University of Helsinki, Finland.}
\\[5mm]

$^d${\normalsize  
Department of Physics, Utah State University, Logan, UT 84322, USA.}
\\[5mm]

$^e${\normalsize  
Departamento de F\'\i sica Te\'orica and Instituto de F\'\i sica Te\'orica UAM/CSIC, \\  Universidad Aut\'onoma de Madrid, Cantoblanco, 28049 Madrid, Spain.}
\\[10mm]

\mbox{\texttt{adolfo.guarino@uniovi.es} \, , \, \texttt{javier.tarrio@helsinki.fi} \, , \, \texttt{oscar.varela@usu.edu}}

\vspace{20pt}

\end{center}

\begin{center}
\textbf{Abstract}
\end{center}

\begin{quote}

New renormalisation group flows of three-dimensional Chern--Simons theories with a single gauge group $\textrm{SU}(N)$ and adjoint matter are found holographically. These RG flows have an infrared fixed point given by a CFT with $\mathcal{N}=3$ supersymmetry and $\text{SU}(2)$ flavour symmetry. The ultraviolet fixed point  is again described by a CFT with either $\mathcal{N}=2$ and $\text{SU}(3)$ symmetry or $\mathcal{N}=1$ and $\text{G}_{2}$ symmetry.  The gauge/gravity duals of these RG flows are constructed as domain-wall solutions of a gauged supergravity model in four dimensions that enjoys an embedding into massive IIA supergravity. A concrete RG flow that brings a mass deformation of the  $\mathcal{N}=2$ CFT into the $\mathcal{N}=3$ CFT at low energies is described in detail.

\end{quote}

\vfill

\end{titlepage}

\tableofcontents

\hrulefill
\vspace{10pt}

\section{Motivation and summary of results}
\label{sec.fiedltheory}

Recently, the study of Chern--Simons-matter (CS-matter) theories with a single gauge group $\textrm{SU}(N)$ and matter \cite{Schwarz:2004yj,Gaiotto:2007qi} in the adjoint representation has become holographically accessible at strong coupling and large $N$. This new holographic arena has opened up by the consistent truncation of massive type IIA supergravity \cite{Romans:1985tz} on $S^6$ \cite{Guarino:2015jca,Guarino:2015vca} down to a maximal supergravity in four dimensions with an $\textrm{ISO(7)}$ gauge group \cite{Guarino:2015qaa}. The supergravity gauging is of dyonic type \cite{deWit:2005ub,Dall'Agata:2012bb}, with the magnetic coupling $m$ identified with the Romans mass $F_0$ and with the CS level $k$ as \cite{Guarino:2015jca}
\be
F_0 = m = \frac{k}{2 \pi \ell_s} \ ,
\ee
with $\ell_s$ the string length. The supersymmetric $\textrm{AdS}_{4}$ solutions of the dyonically-gauged $\textrm{ISO(7)}$ supergravity that preserve at least $\textrm{SU(3)}$ or $\textrm{SO(4)}$ residual gauge symmetry have been classified. The supergravity contains four such solutions with $\cN=1$ and $\textrm{G}_2$ gauge symmetry \cite{Borghese:2012qm},  $\cN=2$ and $\textrm{SU(3)}\times\textrm{U(1)}$ gauge symmetry \cite{Guarino:2015jca},  $\cN=3$ and $\textrm{SO(4)}$ gauge symmetry \cite{Gallerati:2014xra},  and  $\cN=1$ and $\textrm{SU(3)}$ gauge symmetry \cite{Guarino:2015qaa} (see Table~\ref{tab.fixedpoints} in Section~\ref{sec:fourchiral}). These give rise to $\textrm{AdS}_4$ solutions of massive IIA \cite{Guarino:2015jca,Varela:2015uca,Pang:2015vna,DeLuca:2018buk} and are respectively dual to three-dimensional superconformal field theories (SCFTs) with $\cN=1$ and $\textrm{G}_2$ flavour symmetry,  $\cN=2$ and $\textrm{SU(3)}$ flavour symmetry,  $\cN=3$ and $\textrm{SU(2)}$ flavour symmetry, and  $\cN=1$ and $\textrm{SU(3)}$ flavour symmetry. The $\cN=2$ and $\cN=3$ gauge/gravity duals have concrete proposals \cite{Guarino:2015jca} in terms of the field theories of \cite{Schwarz:2004yj,Gaiotto:2007qi} with appropriate superpotentials. Two new $\cN=1$ $\textrm{AdS}_4$ solutions with $\textrm{U(1)}$ gauge symmetry are found in appendix~\ref{App:7_chirals}.

\begin{figure}[t]
\begin{center}
\includegraphics[width=0.9\textwidth]{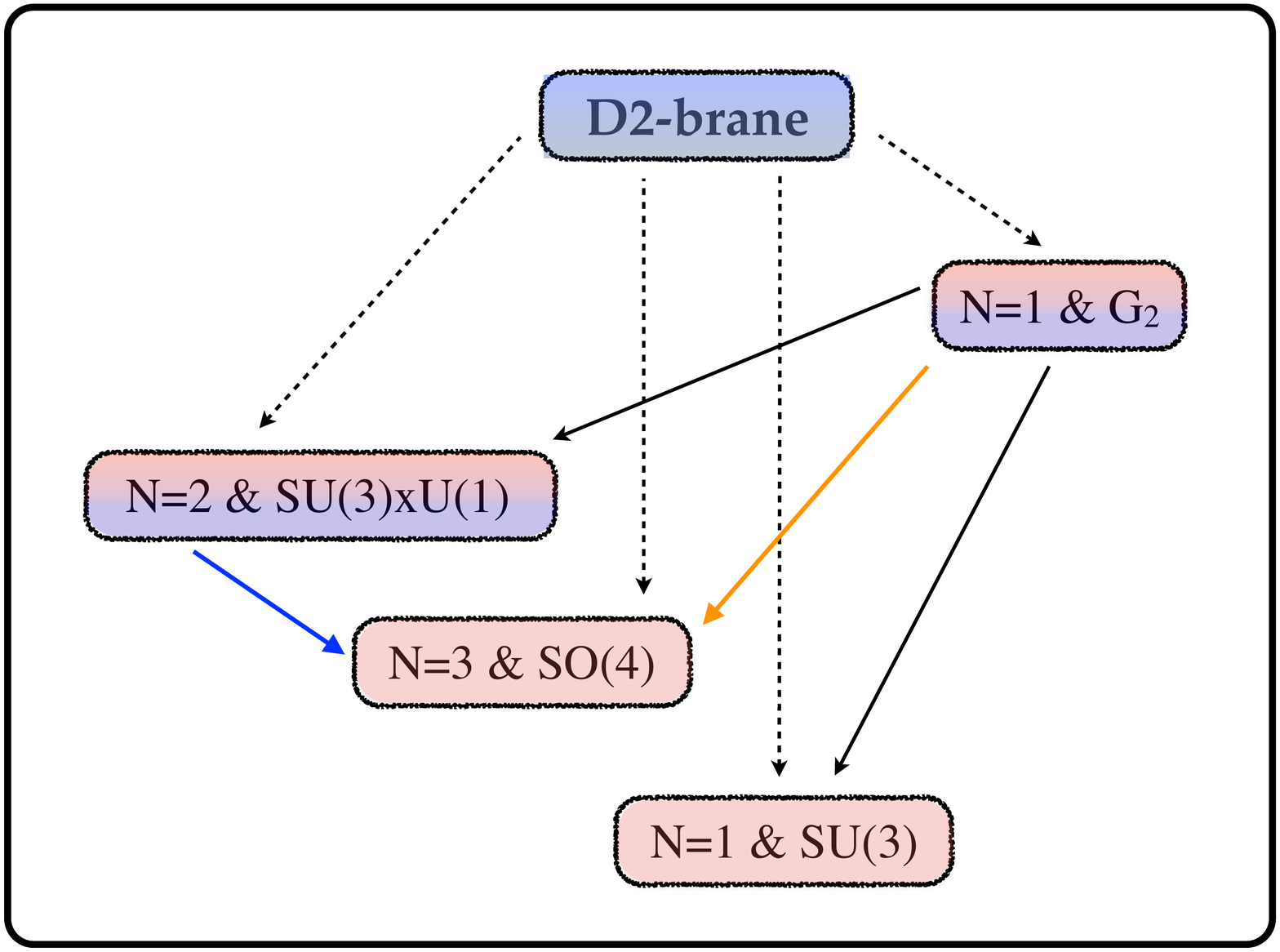}
\caption{Network of domain-walls connecting the D2-brane behaviour (SYM-CS) and the previously known supersymmetric $\textrm{AdS}_{4}$ solutions (CFTs) of the $\textrm{ISO}(7)$ maximal supergravity. The black lines (solid and dashed) correspond to domain-walls previously constructed in the $\textrm{SU}(3)$ invariant sector of the theory \cite{Guarino:2016ynd}. The blue and yellow solid lines are new domain-walls constructed in this paper within the $\textrm{SO}(3)_{\textrm{R}}$ invariant sector of the theory. Actual plots corresponding to these new domain-walls can be found in Figure~\ref{fig.DWflowdomainwall} and Figure~\ref{fig.G2domainwall}.}
\label{Fig.net_DW}
\end{center}
\end{figure}

A network of BPS domain-wall configurations connecting two supersymmetric $\textrm{AdS}_{4}$ solutions with at least $\textrm{SU(3)}$ gauge symmetry was uncovered in \cite{Guarino:2016ynd}. These domain-walls are displayed with black solid lines in Figure~\ref{Fig.net_DW}. The holographic duals of such domain-walls are renormalisation group (RG) flows between two different CFTs with at least $\textrm{SU(3)}$ flavour symmetry. In addition, there are RG flows connecting a non-conformal theory in the ultraviolet (UV) to a CFT in the infrared (IR). The non-conformal theory is identified with the maximally supersymmetric Yang--Mills theory (SYM) in three dimensions deformed with a CS term, and describes the worldvolume of a stack of D2-branes in massive IIA. The domain-walls reaching the D2-branes in the UV are also represented in Figure~\ref{Fig.net_DW} with black dashed lines. Flows involving the $\cN=3$ CFT with $\textrm{SU(2)} \equiv \textrm{SO}(3)_\textrm{R}$ flavour symmetry as a fixed point were excluded from the analysis of \cite{Guarino:2016ynd}. The purpose of this paper is to fill this gap. We will show that this $\cN=3$ CFT can serve as an IR fixed point by constructing new domain-wall solutions that end at this critical point. A natural, simplifying assumption is to request that the flows preserve the $\textrm{SO}(3)_\textrm{R}$ flavour of the IR phase. The $\textrm{SO}(3)_\textrm{R}$-invariant sector of $\cN=8$ $\textrm{ISO}(7)$ supergravity that we recently constructed in \cite{Guarino:2019jef} is thus the natural arena to construct these solutions.\\

\noindent The outcome of our study of supersymmetric four-dimensional domain-walls  is two-fold:

\begin{itemize}

\item On the one hand, we find a one-parameter family of domain-wall solutions corresponding to RG flows that connect the $\cN=1$ CFT with $\textrm{G}_2$ flavour symmetry in the UV to the $\cN=3$ CFT with $\textrm{SU(2)}$ symmetry in the IR. One would expect these RG flows to be triggered by mass deformations of the UV theory. However, due to the lack of a continuous R-symmetry for $\cN=1$ theories in three dimensions, a precise description of these RG flows is not yet available.

\item On the other hand, we find a unique domain-wall solution corresponding to an RG flow that connects the $\cN=2$ CFT with $\textrm{SU(3)}$ flavour symmetry in the UV to the $\cN=3$ CFT with $\textrm{SU(2)}$ symmetry in the IR. This RG flow preserves $\cN=2$ supersymmetry and is created upon deforming the UV CFT with a mass term. This flow is of the type discussed by Gaiotto and Yin (GY) in \cite{Gaiotto:2007qi} and, for this reason, we will refer to it in the following as the GY flow. Interestingly, the GY flow can be ``glued" to another RG flow connecting the $\cN=1$ CFT with $\textrm{G}_2$ symmetry in the UV to the $\cN=2$ CFT with $\textrm{SU(3)}$ symmetry in the IR, whose holographic dual was already constructed in \cite{Guarino:2016ynd}. The combined RG flow then provides a limiting behaviour of the flows referred to above. The full network of available domain-walls is sketched in Figure~\ref{Fig.net_DW}.

\end{itemize}

Being generated by a mass deformation of the $\cN=2$ CFT, the GY flow is similar to the well-known flows of \cite{Freedman:1999gp} and \cite{Benna:2008zy,Ahn:2000aq,Corrado:2001nv}, created by mass deformations of four-dimensional $\cN=4$ SYM \cite{Brink:1976bc} and ABJM \cite{Aharony:2008ug}. This similarity is reflected in the dual type IIA geometry. The latter can be depicted as an $S^2$ bundle over a four-dimensional base, with $S^6$ topology for the total space. In the UV and at intermediate holographic energies along the flow, the $S^2$ fibres are deformed and show only a $\textrm{U(1)}$ symmetry. This is the R-symmetry preserved along the flow. The $S^2$ fibres get squashed inhomogeneously against the base as the flow moves on towards lower energies. Finally, at the IR fixed point the $S^2$ fibres become round and a full $\textrm{SO}(3)$ R-symmetry emerges. All along the flow, the four-dimensional base displays an intact $\mathbb{CP}^1$ acted upon by the $\textrm{SO}(3)_\textrm{R}$ flavour group. Geometrically, the holographic GY flow is thus the analogue to the D3 \cite{Freedman:1999gp} and M2 \cite{Benna:2008zy,Ahn:2000aq,Corrado:2001nv} flows created by similar mass deformations, where the internal $S^5$ and the $S^7$ become squashed along their Hopf fibers \cite{Corrado:2001nv}. The geometries involved in \cite{Benna:2008zy,Ahn:2000aq,Corrado:2001nv}, in \cite{Freedman:1999gp} and in the present case should correspond to configurations of intersecting M-branes and D$\!p$-branes with all allowed (odd or even) values for $p$, respectively. These configurations are not known precisely in any of these cases, and it is beyond the scope of this paper to address this issue. See however \cite{Imamura:2000jj,Bergman:2001qg} for early examples in a related context.

The paper is organised as follows. In Section~\ref{section.minimal} we present a minimal four-dimensional gauged supergravity model that contains the $\textrm{AdS}_{4}$ solutions of interest. This minimal model contains both the gravity dual of the GY flow and a family of domain-walls interpolating between the $\textrm{G}_2$ $\cN=1$ critical point in the UV and $\cN=3$ fixed point in the IR. The former is investigated in Section~\ref{section.GYflow}, and the latter in Section~\ref{section.N=1flows}. Concrete proposals exist for the $\cN=2$ and $\cN=3$ SCFTs linked by the GY flow. We review them in Section~\ref{section.GYflow} and provide field-operator maps relevant to this flow. In Section~\ref{sec.uplift} we discuss the uplift of the domain-wall dual to the GY flow to ten-dimensional massive IIA supergravity. The paper concludes with various appendices with complementary material.

\newpage

\section{Minimal gauged supergravity model}
\label{section.minimal}

Our starting point to holographically investigate RG flows with an $\mathcal{N}=3$ and $\textrm{SU(2)}$ flavour symmetric fixed point in the IR is the half-maximal supergravity coupled to three vector multiplets that we recently constructed in \cite{Guarino:2019jef}. This theory describes the dynamics of the ${\textrm{SU(2)} \sim {\textrm{SO}(3)_{\textrm{R}}}}$ invariant sector of the maximal $\textrm{ISO}(7)$ supergravity \cite{Guarino:2015qaa}. The latter arises upon reduction of massive IIA on $\textrm{S}^6$ \cite{Guarino:2015jca,Guarino:2015vca}. 

\subsection{A four-chiral sector of dyonic ISO(7) supergravity} \label{sec:fourchiral}

Fortunately, the full $\textrm{SO}(3)_{\textrm{R}}$-invariant model of \cite{Guarino:2019jef} is not needed in order to construct the solutions of interest here. The minimal setup that accommodates such solutions consists of a subsector thereof containing the metric field $g_{\mu\nu}$ and four complex scalars $z_{I}$ with $I=1,\dots,4$. The dictionary between the complex scalars $z_{I}$  and the real fields of \cite{Guarino:2019jef} is given by
\begin{equation}
\label{chiralsRealNotation_1}
\bal
z_1 = \frac{b_{11}}{\sqrt{2}} +  i \, e^{-\phi_1/\sqrt{2}}
\hspace{5mm} , \hspace{5mm}
z_2 = - \frac{b_{22}}{\sqrt{2}} +  i \, e^{-\phi_2/\sqrt{2}}
\hspace{5mm} , \hspace{5mm}
z_3 = -\frac{b_{33}}{\sqrt{2}} +  i \, e^{-\phi_3/\sqrt{2}} \ ,
\eal
\end{equation} 
and
\begin{equation}
\label{chiralsRealNotation_2}
z_4 = -\chi +  i \, e^{-\varphi} \ .
\end{equation} 
Here, $\varphi$, $\phi_1$, $\phi_2$, $\phi_3$ are proper scalars and $\chi$, $b_{11}$, $b_{22}$, $b_{33}$ pseudoscalars.  All other fields in the model of \cite{Guarino:2019jef} can be turned off consistently with their equations of motion. In other words, the sector that contains the four complex scalars (\ref{chiralsRealNotation_1}), (\ref{chiralsRealNotation_2}) is a consistent truncation of the $\cN=4$ $\textrm{SO}(3)_{\textrm{R}}$-invariant sector \cite{Guarino:2019jef} of $\cN=8$ dyonic $\textrm{ISO}(7)$ supergravity \cite{Guarino:2015qaa}. See appendix~\ref{App:7_chirals} for an alternative derivation of this minimal eight-scalar model directly from the full $\cN=8$ supergravity. 

This simple model can be recast as a minimal ($\mathcal{N}=1$) supergravity coupled to four chiral fields. The complex scalars $z_{I}$ serve as coordinates on the scalar geometry $\left[\textrm{SL}(2)/\textrm{SO}(2)\right]^4$, equipped with the K\"ahler potential
\begin{equation}
\label{K_4}
K = - 2 \, \displaystyle\sum_{i=1}^3   \log[-i(z_{i}-\bar{z}_{i})]  - \log[-i(z_{4}-\bar{z}_{4})] \ .
\end{equation}
Interactions are codified in a cubic holomorphic superpotential
\begin{equation}
\label{W_SO(3)_R}
W = 2 \, m + 2\, g\, \left[ \, 4 \, z_1 \, z_2 \, z_3 +  \left( z_1^2 + z_2^2 + z_3^2 \right) \, z_{4}  \, \right] \ ,
\end{equation}
where $g$ and $m$ are the coupling constants of the parent $\cN=8$ supergravity \cite{Guarino:2015qaa}. The (bosonic) action is then of Einstein-scalar form
\be
\label{eq.action}
S_{\textrm{bos}} = \int d^{4}x \, \sqrt{g} \,  \left( \, \frac{1}{2} \, R - V - \frac{1}{2} \, K_{I \bar{J}} \, \partial_\mu z^I  \, \partial^\mu \bar{z}^{\bar{J}} \, \right)  \ ,
\ee
where $K_{I \bar{J}}=\partial_{z^I}\partial_{\bar{z}^{\bar{J}}}K$ is the K\"ahler metric on the scalar geometry
\be
K_{I \bar{J}} \, d z^I \,d \bar{z}^{\bar{J}} =  - \sum_{i=1}^3 \frac{2}{\left( z_i - \bar{z}_i \right)^2} dz_i \, d\bar{z}_i   - \frac{1}{\left( z_4 - \bar{z}_4 \right)^2} dz_4 \, d\bar{z}_4 \ ,
\ee
and $V = V(z^{I}, \bar{z}^{\bar{J}})$ denotes the scalar potential. The latter can be readily computed from (\ref{K_4}) and (\ref{W_SO(3)_R}) using the standard $\mathcal{N}=1$ formula 
\be
\label{V_N1_canonical}
V = 8 \, K^{I \bar{J}} \, \partial_{z^I} {\cal W} \, \partial_{\bar{z}^{\bar{J}}} {\cal W} - 12 \,  {\cal W}^2 \ ,
\ee
involving the gravitino mass term
\be
\label{eq.realsuperpotential}
{\cal W} = \frac{1}{2} e^{K/2} \left( W \overline W \right)^{1/2} \ ,
\ee
and the inverse K\"ahler metric $K^{I \bar{J}}$.

This $\mathcal{N}=1$ model suffices to capture all the known supersymmetric $\textrm{AdS}_{4}$ solutions of the $\textrm{ISO}(7)$ maximal supergravity with at least $\textrm{SU}(3)$ or $\textrm{SO}(4)$ gauge symmetry (see Table~\ref{tab.fixedpoints}). Moreover, and importantly for the purposes of this paper, these solutions also appear as supersymmetric within the $\mathcal{N}=1$ model presented here, thus satisfying the F-flatness conditions $F_{I}=\partial_{I}W+(\partial_{I}K) \, W=0$ that follow from the superpotential (\ref{W_SO(3)_R}). This fact will allow us to construct BPS domain-wall solutions that interpolate between the supersymmetric $\textrm{AdS}_{4}$ critical points of Table~\ref{tab.fixedpoints}. These domain-walls will describe, holographically, RG flows between the corresponding dual CFTs.

\begin{table}
\begin{center}
\begin{tabular}{|cccc|cccc|}
\hline
$\cal N$ & $G$ & $L_G$ & $V_G =-12/L_G^2$ & $z_1^{(G)}$ & $z_2^{(G)}$ & $z_3^{(G)}$ & $z_4^{(G)}$  \\
\hline\hline
1 & $\textrm{G}_2$ & $\frac{5 \cdot 15^{1/4}}{8\cdot 2^{2/3}}$ & $ -19.9871\cdots$ & $\frac{1 + i \sqrt{15}}{4 \cdot 2^{1/3}}$ & $\frac{1 + i \sqrt{15}}{4 \cdot 2^{1/3}}$ & $\frac{1 + i \sqrt{15}}{4 \cdot 2^{1/3}}$ & $\frac{1 + i \sqrt{15}}{4 \cdot 2^{1/3}}$ \\
2 & $\textrm{SU}(3)\times\textrm{U}(1)$ & $\frac{1}{3^{1/4}}$ & $ -20.7846\cdots$ & $\frac{i}{\sqrt{2}}$ & $\frac{i}{\sqrt{2}}$ & $\frac{1+i\sqrt{3}}{2}$ & $\frac{1+i\sqrt{3}}{2}$  \\
3 & $\textrm{SO}(4)$ & $\frac{3^{3/4}}{2 \cdot 2^{2/3}}$ & $ -23.2773\cdots$ & $\frac{-1+i\sqrt{3}}{2 \cdot 2^{1/3}}$ & $\frac{1+i\sqrt{3}}{2 \cdot 2^{1/3}}$ & $\frac{1+i\sqrt{3}}{2 \cdot 2^{1/3}}$ & $\frac{1+i\sqrt{3}}{2^{1/3}}$  \\
1 & $\textrm{SU}(3)$ & $\frac{5^{5/4}}{8 \cdot 3^{1/4}}$ & $ -23.7956\cdots$ & $\frac{\sqrt{3}+i\sqrt{5}}{4}$ & $\frac{\sqrt{3}+i\sqrt{5}}{4}$ & $\frac{-1+i\sqrt{15}}{4}$ & $\frac{-1+i\sqrt{15}}{4}$  \\
\hline
\end{tabular}
\caption{Supersymmetric $\textrm{AdS}_{4}$ solutions ordered by decreasing value of the scalar potential $V_{G}$. The first four columns show respectively the number of preserved supersymmetries in the maximal theory, the residual gauge symmetry preserved at the $\textrm{AdS}_{4}$ solution, the value of the $\textrm{AdS}_{4}$ radius, and a numeric approximation to the value of $V_{G}$. The last four columns give the position of the $\textrm{AdS}_{4}$ solutions in field space. We have set $g=m=1$.}
\label{tab.fixedpoints}
\end{center}
\end{table}

\subsection{Domain-wall setup}

In order to describe the three-dimensional RG flows holographically, we are interested in gravitational configurations that preserve $\textrm{SO}(1,2)$ Lorentz symmetry. This requirement is accommodated by a domain-wall Ansatz of the type
\be
d s^2 = e^{2A(\rho)} \eta_{\mu\nu} dx^\mu dx^\nu + d \rho^2 \ ,
\ee
where $\rho\in \mathbb{R}$ is the coordinate transverse to the domain-wall and holographically dual to the energy scale in the field theory, ${\eta_{\mu\nu}=\textrm{diag}(-1,1,1)}$ is the 2+1-dimensional Minkowski metric, and $A(\rho)$ is a function that we will refer to as the domain-wall function. This Ansatz enjoys two reparameterisation symmetries related to shifts in the holographic radial coordinate and re-scalings of the Minkowski coordinates
\be\label{eq.rescaling}
x^\mu \to \sigma\, x^\mu \ , \qquad A \to A - \log \sigma \ , \qquad \rho \to \rho + \rho_s \ .
\ee

The minimisation of the action \eqref{eq.action} gives rise to a set of second order ordinary differential equations. However, we are interested in BPS configurations preserving various amounts of supersymmetry. Such configurations are solutions of a set of BPS first-order differential equations
\be
\label{eq.BPSequations}
\partial_\rho A = 2\, {\cal W}
\hspace{5mm} , \hspace{5mm}
\partial_\rho z^I = - 4\, K^{I \bar{J}}\, \partial_{\bar{z}_{\bar{J}}}{\cal W} \ , 
\ee
where $\cal W$ is the gravitino mass term given in equation~\eqref{eq.realsuperpotential}. In (\ref{eq.BPSequations}), it is convenient to scale away all the dependence on the coupling constants $g$ and $m$ by considering the redefinition
\be
z^I \to z^I \left( \frac{m}{g} \right)^{1/3} \quad \Rightarrow \quad 
K_{I \bar{J}} \to K_{I \bar{J}} \left( \frac{m}{g} \right)^{-2/3} \ .
\ee
From \eqref{K_4} and \eqref{W_SO(3)_R}, it then follows that the gravitino mass (\ref{eq.realsuperpotential}) scales as
\be
{\cal W} \to {\cal W} \left( \frac{g^7}{m} \right)^{1/6} \ .
\ee
The quantity $\ell \equiv (g^7/m)^{-1/6}$ becomes the natural length scale with respect to which all the remaining dimensionful fields and functions are measured
\be
V \to V/\ell^2 \ , \quad\quad\quad \{x^\mu,\rho\} \to \{x^\mu,\rho\} \,\ell \ , \quad\quad\quad A \to A \ .
\ee
The $\textrm{AdS}_{4}$ radius $L_G$ at each critical point scales accordingly as
\be
L_G \to L_G \, \ell \ .
\ee
From now on, we will always use the dimensionless quantities just introduced. The explicit dependence of quantities on $g$ and $m$ is restored by simply applying the above rescalings.

\subsection{Modes and dual operators around $\textrm{AdS}_4$ solutions} \label{sec:Modes}

The BPS equations \eqref{eq.BPSequations} admit $\textrm{AdS}_4$ solutions where the scalar fields acquire a value $z_I^{(G)}$ that extremises the superpotential: the r.h.s. of the BPS equations for the scalars vanishes and the domain-wall function takes the linear form
\be
A = \frac{\rho}{L_G} \ .
\ee
The constant $L_G=1/2{\cal W}(z_I^{(G)})$ here is the corresponding $\textrm{AdS}_{4}$ radius, and the label $G$ refers to the residual gauge symmetry preserved at a given $\textrm{AdS}_4$ solution. All the $\textrm{AdS}_4$ solutions considered in the main body of this paper were previously known \cite{Gallerati:2014xra,Guarino:2015qaa} and are summarised in Table~\ref{tab.fixedpoints}.

\begin{table}
\begin{center}
\begin{tabular}{|c|c|c|c|c|c|c|c|c|}
\hline
$G$ & \multicolumn{8}{c|}{$\Delta_{G,a}$} \\
\hline\hline
$\textrm{G}_2$ & \cellcolor{gray!25}$1-\sqrt{6}$ & \multicolumn{3}{c|}{$1-\frac{1}{\sqrt{6}}$ ($\times$3)} & \multicolumn{3}{c|}{$1+\frac{1}{\sqrt{6}}$ ($\times$3)} & $1+\sqrt{6}$ \\[2pt]
\hline\hline
$\textrm{SU}(3)\times\textrm{U}(1)$ & \cellcolor{gray!25}$\frac{1-\sqrt{17}}{2}$ & \cellcolor{gray!25}$\frac{3-\sqrt{17}}{2}$ & $\frac{2}{3}$ & \multicolumn{2}{c|}{$1$ ($\times$2)} & $\frac{4}{3}$ & $\frac{1+\sqrt{17}}{2}$ & $\frac{3+\sqrt{17}}{2}$ \\
\hline\hline
$\textrm{SO}(4)$ & \cellcolor{gray!25}$-\sqrt{3}$ & \multicolumn{2}{c|}{\cellcolor{gray!25}$1-\sqrt{3}$ ($\times$2)} & $2-\sqrt{3}$ & $\sqrt{3}$ & \multicolumn{2}{c|}{$1+\sqrt{3}$ ($\times$2)} & $2+\sqrt{3}$ \\
\hline\hline
$\textrm{SU}(3)$ & \multicolumn{2}{c|}{\cellcolor{gray!25}$1-\sqrt{6}$ ($\times$2)} & $0$ & $\frac{1}{3}$ & $\frac{5}{3}$ & $2$ & \multicolumn{2}{c|}{$1+\sqrt{6}$ ($\times$2)} \\
\hline
\end{tabular}
\caption{Modes allowed by the BPS equations around each of the supersymmetric $\textrm{AdS}_{4}$ solutions labelled by their residual gauge symmetry $G$. We have ordered them in increasing magnitude and highlighted with a gray background the modes with $\Delta_{G,a}<0$ which parameterise regular domain-wall solutions ending in the IR ($\rho\to-\infty$).}
\label{tab.modes}
\end{center}
\end{table}

The scalar spectrum around each of the $\textrm{AdS}_4$ solutions can be obtained by considering fluctuations of the chiral fields
\be\label{eq.perturbation}
z_I(\rho) = z_I^{(G)} + \delta z_I^{(G)}(\rho) \ .
\ee
Linearising the BPS equations in the variables $\delta z_I^{(G)}$ one finds that a generic solution can be expressed as a linear superposition
\be\label{eq.perturbation2}
\delta z_I^{(G)}(\rho) = \sum_{a=1}^8 z_{I,a}^{(G)} \, \zeta_{a}^{(G)} \, e^{-\frac{\rho}{L_G}\Delta_{G,a}} \ .
\ee
We will refer to the exponents $\Delta_{G,a}$ as \emph{modes}. These modes and the constant matrix of coefficients $z_{I,a}^{(G)}$ are completely determined by the BPS equations, whereas the eight integration constants $\zeta_a^{(G)}$ remain arbitrary and specify the most general solution to the linearised BPS equations. We list the modes corresponding to each of the $\textrm{AdS}_4$ solutions in Table~\ref{tab.modes}. It is worth noticing that the radial shift introduced in \eqref{eq.rescaling} gets reflected in a non-trivial transformation of the integration constants
\be
\label{eq.radialshift}
\rho \to \rho + \rho_s  \ , \qquad \zeta_a^{(G)} \to e^{\frac{\rho_s}{L_G}\Delta_{G,a}}\zeta_a^{(G)} \ .
\ee
This symmetry will prove useful in Sections~\ref{section.GYflow} and \ref{section.N=1flows} to construct domain-wall solutions. It will also be helpful in appendix~\ref{sec.previousDWs}, where the flows  previously constructed in the SU(3)--invariant sector of $\textrm{ISO}(7)$ supergravity \cite{Guarino:2016ynd} are re-obtained as solutions of our present four-chiral model.

The normalised spectrum of scalar masses at the various $\textrm{AdS}_{4}$ solutions of Table~\ref{tab.fixedpoints} can be obtained from the potential (\ref{V_N1_canonical}), and is related to the independent modes in Table~\ref{tab.modes}. The relation between the two sets of data is given by
\be
\label{eq.dualscaling}
M_{G,a}^2 \, L_G^2 = \Delta_{G,a} \left( \Delta_{G,a} - 3 \right) \ .
\ee
By virtue of the gauge/gravity correspondence, the modes with $M_{G,a}^2\, L_G^2 > 0$ are dual to irrelevant scalar operators (with conformal dimension $\Delta>3$) whereas those modes with ${-\frac{9}{4} \leq M_{G,a}^2\, L_G^2<0}$ are dual to relevant scalar operators (with conformal dimension $\Delta<3$). Modes with a vanishing mass squared correspond to marginal scalar operators  (with conformal dimension $\Delta=3$). For the $\textrm{AdS}_{4}$ solutions of Tables \ref{tab.fixedpoints} and \ref{tab.modes} the number of dual operators of each type is given in Table~\ref{tab.modes_2}. 

Note that the counting of relevant and irrelevant operators dual to the scalar fields does not coincide, in general, with the counting of positive and negative modes around the $\textrm{AdS}_{4}$ solutions (the latter are highlighted in Table~\ref{tab.modes}). Recall that the reason lies in the existence of two solutions, denoted $\Delta_{\pm}$, to the equation \eqref{eq.dualscaling} \cite{Klebanov:1999tb}. On the one hand, for a positive mass squared, dual to an irrelevant operator, there are two solutions $\Delta_+>3$ and $\Delta_-<0$, and the BPS equations select one of them. If the negative root $\Delta_-$ is selected, the gauge/gravity correspondence establishes that the dual CFT Lagrangian is deformed by adding an irrelevant operator, which has an important effect in the UV of the theory. If, on the contrary, the positive root $\Delta_+$ is selected, the dual field theory possesses a non-trivial vacuum expectation value (vev) for the corresponding irrelevant operator, but the source of this operator is absent in the Lagrangian. On the other hand, for a negative mass squared, dual to a relevant operator, both solutions $\Delta_\pm$ are positive, and the BPS equations again selects only one of them. As before, if the negative root $\Delta_-$ is selected the dual CFT is deformed by adding a relevant operator, whereas if the positive root $\Delta_+$ is, the dual field theory possesses a vev with no source in the Lagrangian. However, the r\^ole of the roots $\Delta_\pm$ can be reversed in the case where the negative mass square lies in the range $-\frac{9}{4} < M_{G,a}^2 \, L_G^2 \leq -\frac{5}{4}$, when an alternative quantisation of the scalar field is possible \cite{Klebanov:1999tb}.

Whenever one of the modes in the expansion \eqref{eq.perturbation2} is activated, the maximally symmetric $\textrm{AdS}_4$ geometry ceases to be a solution of the BPS equations (\ref{eq.BPSequations}). The latter dictate a new and non-trivial domain-wall solution whose holographic interpretation corresponds to an RG flow of the dual field theory. When the active modes are the negative ones associated with irrelevant operators, as we will consider shortly, the $\textrm{AdS}_4$ solution provides the IR endpoint of the RG flow. In field theory language, the RG flow brings the dual field theory to a fixed point in the IR with appropriate deformations turned on. 

\begin{table}
\begin{center}
\begin{tabular}{|c|ccc|}
\hline
 $G$ & Relevant operators & Irrelevant operators & Marginal operators  \\
\hline\hline
$\textrm{G}_2$ & $6$ & $2$ & $0$ \\
$\textrm{SU}(3)\times\textrm{U}(1)$ & $5$ & $3$ & $0$ \\
$\textrm{SO}(4)$ & $4$ & $4$ & $0$ \\
$\textrm{SU}(3)$ & $3$ & $4$ & $1$ \\
\hline
\end{tabular}
\caption{Number of relevant, irrelevant and marginal operators dual to the scalar modes in the $\textrm{AdS}_{4}$ solutions.}
\label{tab.modes_2}
\end{center}
\end{table}

In \cite{Guarino:2016ynd} we presented a study of the domain-wall solutions involving the ${\cal N}=1$, $\textrm{G}_2$ point, the ${\cal N}=2$, $\textrm{SU}(3)\times\textrm{U}(1)$ point and the ${\cal N}=1$, $\textrm{SU}(3)$ point as IR endpoints. In that reference the negative modes in the IR are the same ones that appear in the truncation under scrutiny here, and therefore the structure of domain-wall solutions with these fixed points in the IR is the same as in \cite{Guarino:2016ynd}. See appendix~\ref{sec.previousDWs} for a summary. For this reason, in the rest of this paper we will focus on domain-wall solutions ending at the ${\cal N}=3$, $\textrm{SO}(4)$ solution in the IR. From equation \eqref{eq.perturbation2} and Table~\ref{tab.modes}, only three modes can be seen to allow for a regular solution in the IR ($\rho\to-\infty$). We will denote these as
\be
\label{eq.SO4modes}
\Delta_{SO(4),1} = -\sqrt{3} \ , \quad\quad \Delta_{SO(4),2} = \Delta_{SO(4),3}=1-\sqrt{3} \ .
\ee
For these modes, the coefficients are determined in terms of three integration constants $\zeta_{a}^{(SO(4))}$ (with $a=1,2,3$), which determine the chiral field fluctuations by means of the equation \eqref{eq.perturbation2} with a parameterisation of the coefficients of the form
\be
\label{eq.SO4coefficients}
z_{Ia}^{(SO(4))} = \begin{pmatrix}
  \frac{ 1 + ( 2+\sqrt{3})i}{2\cdot 2^{1/3}} &   \frac{ 1 - (2-\sqrt{3})i}{2\cdot 2^{1/3}}  &  0    \\ 
- \frac{1 - (2+\sqrt{3})i}{2\cdot 2^{1/3}}  &  -\frac{\sqrt{3} - (3-2\sqrt{3})i}{6\cdot 2^{1/3}}   &   \frac{1 + (2-\sqrt{3})i}{2\cdot 2^{1/3}}  \\
- \frac{1 - (2+\sqrt{3})i}{2\cdot 2^{1/3}} &   0  &  - \frac{1 + (2-\sqrt{3})i }{2\cdot 2^{1/3}}    \\
- \frac{(\sqrt{3} + 1) - (\sqrt{3} - 1)i }{2^{1/3}}  &   \frac{2^{2/3}}{3} \left( 3- 2\sqrt{3} \right) -  \frac{2^{2/3}}{\sqrt{3}} i  & 0     
\end{pmatrix} \ .
\ee
Using the shift symmetry (\ref{eq.radialshift}) we can set one of the integration constants to any desired value, for example, to one. Furthermore, the fact that two of the modes are equal, namely $\Delta_{SO(4),2} = \Delta_{SO(4),3}=1-\sqrt{3}$, implies that the dual field theory is perturbed by two operators of the same dimensionality. But there are particular combinations that result more convenient to study certain domain-walls, as we will show.

\section{The gravity dual of the Gaiotto--Yin flow}
\label{section.GYflow}

A concrete supersymmetric domain-wall solution of the flow equations derived in Section~\ref{section.minimal} connects the $\cN=2$ $\textrm{SU}(3)\times\textrm{U}(1)$-invariant fixed point in the UV to the  $\cN=3$ $\textrm{SO}(4)$ fixed point in the IR. We will argue that this domain-wall corresponds holographically to one of the field theory RG flows envisaged by GY in \cite{Gaiotto:2007qi}. We will review the boundary and bulk sides of the story in Sections~\ref{sec:FT} and \ref{sec:StSugra}, and will finally integrate the numerical domain-wall solution in Section~\ref{sec.DWflowdomainwall}.

\subsection{Field theory}
\label{sec:FT}

The SCFTs of interest arise as low-energy phases of the theory defined on the worldvolume of a stack of $N$ planar D2-branes in $\mathbb{R}^7$, namely, three-dimensional $\cN=8$ $\textrm{SU}(N)$ SYM, upon turning on supersymmetric CS terms at level $k$ for the $\textrm{SU}(N)$ gauge fields. At sufficiently high energies, the relevant field content thus includes $\bm{1}$ vector field $A_\mu$, $\bm{7}$ real scalars $X^I$, $I=1, \ldots , 7$, corresponding to the directions transverse to the D2-branes, and $\bm{8}$ Majorana fermions $\lambda^A$, $A=1 ,\ldots , 8$, all of them in the adjoint of the $\textrm{SU}(N)$ gauge group and in the indicated representations of the $\textrm{SO}(7)$ R-symmetry. The fields have canonical dimensions  $\Delta (X^I) = \frac12$, $\Delta (A_\mu) = \Delta( \lambda^A) = 1$. The CS terms overrule the (irrelevant) Yang--Mills contributions and dominate the low-energy physics. Additional couplings can be included among the matter fields $X^I$, $\lambda^A$ that render the resulting CS-matter models superconformal. 

Two such CS-matter SCFTs have $\cN=2$ and $\cN=3$ supersymmetry. In general, the on-shell field content of this type of theories includes, in $\cN=2$ language, a non-Abelian gauge field $A_\mu$ in a vector multiplet, along with a number $N_f$ (arbitrary for $\cN=2$ and $N_f=2$ for $\cN=3$) of complex scalars $Z^a$ and complex fermions $\chi^a$, $a = 1 , \ldots , N_f$. These are the on-shell components of chiral multiplets $\Phi^a$, and lie in a given representation of the gauge group. For the cases at hand, we chose gauge group $\textrm{SU}(N)$ and matter in the adjoint in order to make contact with the D2-brane description at sufficiently high energies. In these cases, the $Z^a$ and $\chi^a$ will respectively be complexifications of $X^I$ and $\lambda^A$, namely, $Z^1 = X^1 + i X^2$, etc. At weak coupling, these SCFTs admit the on-shell Lagrangian description \cite{Schwarz:2004yj,Gaiotto:2007qi} 
{\setlength\arraycolsep{2pt}
\begin{eqnarray} \label{eq:WeakCoupLag}
{\cal L} &=& \textrm{tr} \Big[ \tfrac{k}{4\pi} \,  \epsilon^{\mu \nu \rho} \big( A_\mu \partial A_\rho + \tfrac23 \, A_\mu A_\nu A_\rho \big) + D_\mu \bar{Z}_a \, D^\mu Z^a + i \, \bar{\chi}_a \, \gamma^\mu D_\mu \chi^a  \Big] \nonumber \\[4pt]
&& -\tfrac{4\pi}{k} \,\textrm{tr} ( \bar{Z}_a T_i Z^a ) \,  \textrm{tr} ( \bar{\chi}_b T^i \chi^b ) -\tfrac{8\pi}{k} \,\textrm{tr} ( \bar{\psi}_a T^i Z^a ) \, \textrm{tr} ( \bar{Z}_b T^i \chi^b ) \nonumber \\[4pt]
&&  -\tfrac{4\pi}{k} \, \textrm{tr} ( \bar{Z}_a T_i Z^a ) \, \textrm{tr} ( \bar{Z}_b T_j Z^b ) \, \textrm{tr} ( \bar{Z}_c T^i T^j Z^c ) + {\cal L}_W \; ,
\end{eqnarray}
}where the traces are taken in the adjoint and $T^i$ are the  $\textrm{SU}(N)$ generators. The Yukawa terms and the quartic scalar potential arise upon elimination of auxiliary fields. In addition, we have allowed for further interaction terms ${\cal L}_W$ governed by a superpotential $W$. This is a holomorphic function of $Z^a$, and arises as the lowest component of a chiral superfield ${\cal W}$ holomorphic in $\Phi^a$. Explicitly, these interaction terms read
\begin{align} \label{eq:LfromW}
{\cal L}_W  = 
& \textrm{tr} \left( \frac{\partial W(Z)}{\partial Z^a} \frac{\partial \overline{ W(Z)}}{\partial \bar{Z}_a} 
+   \frac{1}{2}\frac{\partial^2  W(Z)}{\partial Z^a \partial Z^b}\chi^a \chi^b+\frac{1}{2}\frac{\partial^2 \overline{ W(Z)}}{\partial \bar{Z}_a
\partial \bar{Z}_b}\bar{\chi}_a \bar{\chi}_b \right)\; , 
\end{align}
see {\it e.g.} (A.34) of \cite{Araujo:2017hvi}. The addition of a superpotential will typically break the manifest $\textrm{U}(N_f)$ flavour symmetry of the theory with no superpotential to a subgroup thereof.

The $\cN=3$ theory has $N_f = 2$, flavour symmetry $\textrm{SU}(2) \equiv \textrm{SO}(3)_\textrm{R}$, and quartic superpotential \cite{Gaiotto:2007qi} 
\begin{equation} \label{N=3SuperPot}
{\cal W}_{\cN=3} = \tfrac{2\pi}{k} \, \textrm{tr}\,  \big( [\Phi^1 ,  \Phi^2 ]  \big)^2 \; ,
\end{equation}
with (dimensionless) coefficient locked in terms of the Chern-Simons level $k$. With free-field assignments for the conformal dimensions, $\Delta (\Phi^a) \equiv \Delta (Z^a) = \tfrac12$,  $\Delta (\chi^a) = 1$, the superpotential (\ref{N=3SuperPot}) is marginal and the classical action (\ref{eq:WeakCoupLag}) is manifestly scale-invariant. A more general quartic superpotential of the type (\ref{N=3SuperPot}) but with a generic coupling $\alpha$ would only preserve $\cN=2$. GY argue, at weak coupling $k \gg 1$, that this more general $\cN=2$ $\alpha$-dependent theory flows into the theory with superpotential (\ref{N=3SuperPot}), therefore experiencing an $\cN=3$ supersymmetry enhancement at low energies \cite{Gaiotto:2007qi}. The superpotential is non-renormalised and the $\cN=3$ theory does not have R-charge or wave function renormalisation either. For this reason, the Lagrangian (\ref{eq:WeakCoupLag})--(\ref{N=3SuperPot}) can be expected to provide a good description of the $\cN=3$ theory also at strong 't Hooft coupling $\lambda \equiv N/k \gg 1$ with $k$ of order 1 \cite{Gaiotto:2007qi}. We will review evidence from the field theory later in this section and from supergravity in Section~\ref{sec:StSugra} that support this picture. The full symmetry of this theory is $\textrm{OSp}(4|3) \times \textrm{SO}(3)_\textrm{R}$. Denoting  by $\textrm{SO}(3)_\textrm{d}$ the $\cN=3$ R-symmetry group contained in $\textrm{OSp}(4|3)$, the full global bosonic symmetry of the $\cN=3$ SCFT is thus $\textrm{SO}(3)_\textrm{d} \times \textrm{SO}(3)_\textrm{R}$.

The $\cN=2$ theory (\ref{eq:WeakCoupLag}) with no superpotential, ${\cal W} = 0$, and free-field dimension assignments is also manifestly scale-invariant. In contrast to the $\cN=3$ case, however, the $\cN=2$ chirals may undergo both R-charge and wave function renormalisation \cite{Gaiotto:2007qi}. Based on holographic evidence at strong coupling \cite{Guarino:2015jca}, we claim that the $\cN=2$ theory with $N_f=3$, which we fix henceforth, and cubic superpotential 
\begin{equation} \label{N=2SuperPot}
{\cal W}_{\cN=2} = \tfrac16 \, \epsilon_{abc} \, \textrm{tr}\, \big( [\Phi^a ,  \Phi^b ] \, \Phi^c  \big) \; ,
\end{equation}
is in fact also conformal. This could not possibly happen without R-charge renormalisation. The coefficient of (\ref{N=2SuperPot}) is not fixed by $\cN=2$ supersymmetry (in particular, it is not fixed to the Chern-Simons level $k$) but is nevertheless dimensionless, consistent with conformal invariance. Thus, the Lagrangian (\ref{eq:WeakCoupLag}), (\ref{eq:LfromW}) does not provide a good description of the $\cN=2$ theory with superpotential (\ref{N=2SuperPot}), and it should be replaced by the Wilsonian effective action corresponding to the CS-driven flow from $\cN=8$ SYM. The latter may contain, for example, a K\"ahler potential for the kinetic terms of the chirals. The full symmetry of this strongly coupled $\cN=2$ SCFT is thus $\textrm{OSp}(4|2) \times \textrm{SU}(3)$, where the latter factor is the flavour symmetry preserved by the (non-renomalised) superpotential (\ref{N=2SuperPot}). The R-symmetry group contained in $\textrm{OSp}(4|2)$ will be denoted  $\textrm{U}(1)_\psi$, following the geometric conventions of Section~\ref{sec.uplift}. The full global bosonic symmetry of the $\cN=2$ SCFT is thus $\textrm{U}(1)_\psi \times \textrm{SU}(3)$.

With the benefit of hindsight, it is possible to argue purely in field-theoretical terms that a strongly coupled $\cN=2$ SCFT theory with flavour $\textrm{SU}(3)$ and superpotential (\ref{N=2SuperPot}) makes perfect sense. The free energy $F$ of this type of field theories on $S^3$ can be determined at strong coupling \cite{Jafferis:2010un,Jafferis:2011zi} using localisation techniques \cite{Kapustin:2009kz,Jafferis:2010un,Hama:2010av}. If the SCFT has a superpotential, then $F$ can be computed as a function of arbitrary dimension assignments $\Delta_a$ for the chirals $\Phi^a$, $a=1,2,3$, subject to the sole requirement that the (exact, non-renormalised) superpotential be marginal. For (\ref{N=2SuperPot}), this translates into the condition
\begin{equation} \label{eq:DeltaConstN=2}
\Delta_1 + \Delta_2 + \Delta_3  = 2 \; .
\end{equation}
The real part of the leading order free energy as a function of $\Delta_a$ is \cite{Fluder:2015eoa}
\begin{equation} \label{eq:GeneralF}
F = \frac{3\sqrt{3}\, \pi}{20 \cdot 2^{1/3} } \, \Big[ 1 + \sum_{a=1}^{N_f}  \big(1-\Delta_a \big) \big[1 - 2\, ( 1-\Delta_a)^2 \big] \Big]^{2/3} k^{1/3} \, N^{5/3} \; ,
\end{equation}
with $N_f = 3$. On the surface (\ref{eq:DeltaConstN=2}), the function (\ref{eq:GeneralF}) attains an extremum at
\begin{equation} \label{eq:DeltaConstSolN=2}
\Delta_1 = \Delta_2 = \Delta_3  = \tfrac23 \; ,
\end{equation}
consistent with $\textrm{SU}(3)$ symmetry and renormalised away from the free field values. These, and only these, must be the dimensions of the chirals at the $\cN=2$ fixed point \cite{Jafferis:2010un}, reproducing the assignments of \cite{Guarino:2015jca}. With (\ref{eq:DeltaConstSolN=2}), the free energy (\ref{eq:GeneralF}) evaluates to
\begin{equation} \label{eq:FN=2}
F_{\cN=2} = \frac{3^{13/6} \pi}{40} \left( \frac{32}{27} \right)^{2/3} k^{1/3} \, N^{5/3} \; ,
\end{equation}
thus reducing to the result of \cite{Guarino:2015jca}. Subleading corrections to this free energy  have been worked out in \cite{Liu:2018bac}.

Of course, the leading order free energy of the $\cN=3$ SCFT can be computed in the exact same way \cite{Pang:2015rwd}. Assigning arbitrary dimensions $\Delta_a$ to the two chirals consistent with the marginality of the (again, exact and non-renormalised) superpotential (\ref{N=3SuperPot}),
\begin{equation} 
\Delta_1 + \Delta_2  = 1 \; ,
\end{equation}
the free energy, (\ref{eq:GeneralF}) with $N_f = 2$, becomes extremal for the free-field values
\begin{equation} \label{eq:DeltaConstN=3}
\Delta_1 = \Delta_2  = \tfrac{1}{2} \; .
\end{equation}
These are now compatible with $\textrm{SU}(2) \equiv \textrm{SO}(3)_\textrm{R}$ symmetry. This provides evidence that the classical $\cN=3$ action (\ref{eq:WeakCoupLag})--(\ref{N=3SuperPot}) with $N_f=2$ is not renormalised at strong coupling. At the extremum (\ref{eq:DeltaConstN=3}), the leading contribution of the $\cN=3$ free energy becomes \cite{Pang:2015rwd}
\begin{equation} \label{eq:FN=3}
F_{\cN=3} = \frac{3^{13/6} \pi}{40}  \, k^{1/3} \, N^{5/3} \; .
\end{equation}

From (\ref{eq:FN=2}) and (\ref{eq:FN=3}), it straightforwardly follows that 
\begin{equation}
F_{\cN=2} \, > \, F_{\cN=3} \; .
\end{equation}
By the argument of \cite{Jafferis:2011zi}, these two theories could thus be connected by an RG flow, with the $\cN=2$ SCFT in the UV and the $\cN=3$ one in the IR. In fact, GY had previously argued that this flow is indeed generated upon deforming the $\cN=2$ theory by a mass term for one of the three chirals\footnote{Here we focus on a simplified version of the model in Section~4.2 of \cite{Gaiotto:2007qi} with no D6-branes, $N_f=0$ there, and consequently no fundamental matter, $Q^j = \tilde{Q}^j = 0$ there. It is this simplified flow that we dub GY after these authors, although we will argue slightly differently.}. Consider a deformation of the $\cN=2$ superpotential (\ref{N=2SuperPot}) quadratic in, say, the $\Phi^3$ superfield:
\begin{equation} \label{N=2SuperPotDef}
{\cal W}_{\cN=2 \, , \, \textrm{def}} =   \textrm{tr}\, \big( [\Phi^1 ,  \Phi^2 ] \, \Phi^3  + \tfrac12 \, \mu \, (\Phi^3)^2  \big) \; .
\end{equation}
A mass term must always be relevant. Indeed, for the $\cN=2$ assignment $\Delta_3 = \frac23$ in (\ref{eq:DeltaConstSolN=2}), the dimension of the operator $(\Phi^3)^2$ is $\frac43$, less than the marginal dimension $2$ of the superpotential. The dimensionful parameter $\mu$ introduces a scale, conformal invariance is lost, and the $\cN=2$ theory plunges down an RG flow. At sufficiently low energies, the massive field $\Phi^3$ is integrated out. From (\ref{N=2SuperPotDef}), the effective superpotential becomes
\begin{equation} \label{N=3SuperPotCoef}
{\cal W} = \tfrac{1}{2\mu} \, \textrm{tr}\,  \big( [\Phi^1 ,  \Phi^2 ]  \big)^2 \; .
\end{equation}
GY argue that this $\cN=2$ superpotential will finally end up flowing to the $\cN=3$ superconformal fixed point whose superpotential has coefficient fixed by the Chern-Simons level (see below equation (\ref{N=3SuperPot}) above). At long distances, conformal invariance is restored and supersymmetry is even enhanced.

\begin{table}[t!]
 \centering
\begin{tabular}{cccccc}
\multicolumn{2}{c}{Flavour} & 
 R-symmetry 
& 
&  Flavour
&  R-symmetry \\
\multicolumn{2}{c}{\cellcolor{blue!15} \hspace{1em}$\textrm{SU}(3)$\hspace{1.25em} $\times$} & 
\cellcolor{green!15}$\textrm{U}(1)_\psi$ 
& 
& \cellcolor{red!15}
& \cellcolor{orange!15} \\
\multicolumn{2}{c}{\cellcolor{blue!15} $\textrm{SO}(3)_\textrm{R} \times \textrm{U}(1)_\tau$}  
& \cellcolor{green!15}$ \hspace{-0.935em}\times \hspace{0.3em}\textrm{U}(1)_\psi$ 
&  
& \multirow{-2}{*}{\cellcolor{red!15} $\textrm{SO}(3)_\textrm{R}$}
& \multirow{-2}{*}{\cellcolor{orange!15} $\hspace{-1.275em}\times \hspace{0.3em} \textrm{SO}(3)_\textrm{d}$} \\
\cellcolor{blue!15}$\textrm{SO}(3)_\textrm{R}$ 
& \multicolumn{2}{c}{\cellcolor{teal!15}$\hspace{-1.5em}\times \hspace{1.25em}\textrm{U}(1)_\textrm{d}$} 
& $\xrightarrow{\phantom{\textrm{SO}(3)_\textrm{R} } \textrm{SO}(3)_\textrm{R} \times \textrm{U}(1)_\textrm{d} \phantom{\textrm{SO}(3)_\textrm{R} }}$
& \cellcolor{red!15} $\textrm{SO}(3)_\textrm{R}$
& \cellcolor{orange!15} $\hspace{-1.85em}\times \hspace{0.3em}  \textrm{U}(1)_\textrm{d}$  \\
\multicolumn{3}{c}{UV} &
RG flow &
\multicolumn{2}{c}{IR}
\end{tabular}
 \caption{Summary of bosonic global symmetry groups involved in the GY flow. The top lines correspond to the full symmetry enjoyed by the fixed points, with subsequent rows giving the explicit subgroups mentioned in the text.\label{tab:GYGroups}}
\end{table}

It is interesting to determine the symmetry groups preserved along the GY flow. See Table~\ref{tab:GYGroups} for a summary and Appendix~\ref{eq:SymGYFT} for further details. The mass deformation in (\ref{N=2SuperPotDef}) obviously breaks the $\textrm{SU}(3)$ UV flavour to the $\textrm{SU}(2)$ subgroup such that $\bm{3} \rightarrow \bm{2} + \bm{1}$. Here $\Phi^1$, $\Phi^2$ are the doublet and $\Phi^3$ the singlet. By construction, this $\textrm{SU}(2)$ is identified with the $\textrm{SO}(3)_\textrm{R}$ flavour symmetry of the IR SCFT. In addition, the GY flow preserves an extra U(1). This is a mixture of the $\textrm{U}(1)$ (call it $\textrm{U}(1)_\tau$ following again Section~\ref{sec.uplift}) that commutes  with $\textrm{SO}(3)_\textrm{R}$ inside $\textrm{SU}(3)$, and the UV R-symmetry $\textrm{U}(1)_\psi$. This mixing follows from a group theory argument whose implementation is cleaner if the parameter $\mu$ in (\ref{N=2SuperPotDef}) is thought as dimensionless. In this case, a reassignment of the dimensions of $\Phi^a$ is needed, as in {\it e.g.} \cite{Freedman:2013ryh,Bobev:2018uxk,Bobev:2018wbt}. Both terms in the superpotential (\ref{N=2SuperPotDef}) must now be separately requested to be marginal. This in turn leads to a split of the constraint (\ref{eq:DeltaConstN=2}) as $\Delta_1+ \Delta_2 = 1$ and $\Delta_3= 1$. The free energy (\ref{eq:GeneralF}) with $N_f=3$ is now extremal under these constraints when
\begin{equation} \label{eqAssign2}
\Delta_1 = \Delta_2 = \tfrac12 \; , \qquad \Delta_3 = 1 \; ,
\end{equation}
of course reproducing the $\textrm{SO}(3)_\textrm{R}$--symmetric assignments (\ref{eq:DeltaConstN=3}) for the doublet that survives in the IR. But by $\textrm{OSp}(4|2)$ representation theory, these dimensions are also the R-charges (with opposite sign in our conventions) preserved along the flow. The $\textrm{U}(1)$ charges (\ref{eqAssign2}) only branch appropriately from $\textrm{SU(3)} \times \textrm{U}(1)_\psi$ if 
this $\textrm{U}(1)$ is strictly contained in $\textrm{U}(1)_\tau \times \textrm{U}(1)_\psi$. This $\textrm{U}(1)$ can also be shown to be contained in the $\textrm{SO}(3)_\textrm{d}$ R-symmetry of the IR (it can thus be denoted $\textrm{U}(1)_\textrm{d}$). This follows by assuming that both UV, $\textrm{SU(3)} \times \textrm{U}(1)_\psi$, and IR, $\textrm{SO(3)}_\textrm{R} \times \textrm{SO(3)}_\textrm{d}$, global symmetry groups are contained in SO(7), as required for both $\cN=2$ and $\cN=3$ theories to arise as different CS-matter phases of the D2-brane field theory. To summarise, the global bosonic symmetry preserved by the GY flow is $\textrm{U}(1)_\textrm{d} \times \textrm{SO}(3)_\textrm{R}$, where $\textrm{SO}(3)_\textrm{R}$ is flavour and $\textrm{U}(1)_\textrm{d}$ is the R-symmetry. From (\ref{N=2SuperPotDef}), the GY flow is manifestly $\cN=2$ like the UV theory. However, the R-symmetry $\textrm{U}(1)_\textrm{d}$ that rotates the supercharges along the flow is different from the R-symmetry $\textrm{U}(1)_\psi$ of the UV. Instead, $\textrm{U}(1)_\textrm{d}$ corresponds to the precise mixture of UV R-symmetry $\textrm{U}(1)_\psi$ and UV flavour $\textrm{U}(1)_\tau$ that is contained in the IR R-symmetry group $\textrm{SO}(3)_\textrm{d}$ (see Table~\ref{tab:GYGroups} and Appendix~\ref{eq:SymGYFT}).

Finally, it is useful to elucidate the $\cN=2$ operators that drive the GY flow at the level of the Lagrangian. We have argued that the weak-coupling Lagrangian description (\ref{eq:WeakCoupLag}) breaks down at strong coupling. A K\"ahler potential might be generated, and the superpotential interaction terms (\ref{eq:LfromW}) should require modification accordingly. We may nevertheless make naive use of (\ref{eq:LfromW}) to find the schematic form for these operators. Plugging the superpotential (\ref{N=2SuperPotDef}) into (\ref{eq:LfromW}), cubic interaction terms and quadratic mass terms are generated for the component fields of $\Phi^3$. The latter are of the form
\begin{equation} 
\label{eq:GYMasses}
{\cal O}_B \sim \textrm{tr} \,  \bar{Z}_3 \, Z^3 \; , \qquad
{\cal O}_F \sim \textrm{tr} \,  \big( \chi^3 \chi^3 + \bar{\chi}_3 \bar{\chi}_3 \big) \; .
\end{equation}
These mass operators are clearly invariant under the $\textrm{SO}(3)_\textrm{R}$ flavour group of the GY flow. Moreover, as the group theory analysis of Appendix~\ref{eq:SymGYFT} shows, these operators are also neutral under the $\textrm{U}(1)_\textrm{d}$ R-symmetry along the flow.

\subsection{Supergravity and the field-operator map}
\label{sec:StSugra}

The CS-matter SCFTs described in Section~\ref{sec:FT} have gravity duals in massive type IIA string theory  \cite{Guarino:2015jca,Varela:2015uca,Pang:2015vna,DeLuca:2018buk}. In addition, these models enjoy a convenient four-dimensional description in terms of maximal, $\cN=8$, supergravity with a dyonic $\textrm{ISO}(7)$ gauging \cite{Guarino:2015qaa}. This is similar to the existence of holographic descriptions of four-dimensional $\cN=4$ SYM \cite{Brink:1976bc} and ABJM \cite{Aharony:2008ug} in terms of the maximal supergravities in five and four dimensions with SO(6) \cite{Gunaydin:1984qu} and SO(8) gauge groups \cite{deWit:1982ig}. In those cases, like in the present case, some of the supergravity fields are dual to mass terms for the boundary fields. Let us determine the map between supergravity fields and gauge-invariant operators of the boundary field theories.

We find it convenient to work in the SL(8) frame for the $\cN=8$ supergravity, the frame used in \cite{Guarino:2015qaa}, because the proper scalars can be straightforwardly identified with quadratic combinations of the vector representation where SL(8) acts. Identifying these as the coordinates transverse to the branes and ultimately as the adjoint scalars in the boundary theory, these quadratic combinations become related to mass terms for the latter. The supergravity pseudoscalars, in turn, are related to mass terms for the dual field theory fermions. The SL(8) frame is, however, rather inconvenient to identify these mass terms, as the pseudoscalars are parametrised in this frame as self-dual four-rank antisymmetric tensors in the vector representation. A triality rotation is needed to bring the parametrisation into quadratic combinations of spinor representations, for which the relation to the field theory's fermion mass terms then becomes obvious. In the following, we will assume that appropriate triality transformations on the supergravity pseudoscalars have been performed.

Although there is no supersymmetric AdS critical point with SO(7) symmetry, the 35 proper scalars of $\textrm{ISO}(7)$ supergravity can be nevertheless assigned to the $\bm{27} + \bm{7} + \bm{1}$ representations of SO(7), and the pseudoscalars to the $\bm{35}$. The latter correspond to gauge-invariant mass terms $\textrm{tr} \, \lambda^{(A} \lambda^{B)}$ for the $\bm{8}$ fermions $\lambda^A$ of $\cN=8$ SYM. The $\bm{27}$ scalars are dual to symmetric-traceless gauge-invariant mass terms $\textrm{tr} \,  X^{\{I } X^{ J\} }$ for the $\bm{7}$ scalars $X^I$. The singlet can be seen to be related to the second factor in the $D=4$ $\cN=8$ branching $\textrm{E}_{7(7)} \supset \textrm{SL}(7) \times \textrm{SO}(1,1)$ and thus to the IIA dilaton. For this reason, this bosonic operator can be assigned to the Yang--Mills Lagrangian, $\textrm{tr}\, F_{\mu \nu} F^{\mu \nu}$, in analogy with the $D=5$ $\cN=8$ situation where $\textrm{E}_{6(6)} \supset \textrm{SL}(6) \times \textrm{SL}(2)$, with the second factor associated to the IIB axion-dilaton. Finally, the $\bm{7}$ in the branching  $\bm{27} + \bm{7} + \bm{1}$ does not have a holographic interpretation because it corresponds to St\"uckelberg scalars that are eaten and disappear from the physical spectrum.

\begin{table}[t]
\centering
\resizebox{\textwidth}{!}{%
\begin{tabular}{|l|c|c|c|c|c|l|}
\hline
scalar/pseudoscalar                                                                                                                                                     & $\textrm{SU}(3) \times \textrm{U}(1)_\psi$  & $\textrm{SO}(3)_\textrm{R} \times \textrm{U}(1)_\tau \times \textrm{U}(1)_\psi$  & $\textrm{SO}(3)_\textrm{R} \times \textrm{U}(1)_\textrm{d} $   & $M^2L^2$        & $\Delta$                & Osp($4|2$) multiplet \\ \hline\hline
$Z^a\bar{Z}_b-\frac{1}{3}\delta^a_bZ^c\bar{Z}_c$                                                                                                            & $\mathbf{8}_0$    & $\bm{1}_{(0,0)} +\bm{2}_{(-\frac32,0)} +\bm{2}_{(\frac32,0)} +\bm{3}_{(0,0)}  $ & {\begin{tikzpicture}\draw[overlay, thin, rounded corners=1mm] (0.025,-0.15) rectangle (0.6,0.4); \end{tikzpicture}} 
$ \textcolor{cyan}{\bm{1}_{0}} +\bm{2}_{-\frac{1}{2}} +\bm{2}_{\frac{1}{2}} +\bm{3}_{0}  $    & $-2$              & 1                       & massless vector      \\ \hline
$Z^a\bar{Z}_4$                                                                                                                                              & $\mathbf{3}_{-2/3}$    & $\bm{1}_{(-1,-\frac23)} +\bm{2}_{(\frac12,-\frac23)}  $ & $\bm{1}_{-1} +\bm{2}_{-\frac12}  $     & $-\frac{14}{9}$ & $\frac{7}{3}$           & short gravitino      \\ \hline
$\bar{Z}_a Z^4$                                                                                                                                             & $\mathbf{\bar{3}}_{2/3}$  & $\bm{1}_{(1,\frac23)} +\bm{2}_{(-\frac12,\frac23)}  $ & $\bm{1}_{1} +\bm{2}_{\frac12}  $  & $-\frac{14}{9}$ & $\frac{7}{3}$           & short gravitino      \\ \hline
$Z^{(a}Z^{b)}$                                                                                                                                              & $\mathbf{6}_{-4/3}$     & $\bm{1}_{(-2,-\frac43)} +\bm{2}_{(-\frac12,-\frac43)} +\bm{3}_{(1,-\frac43)}  $ & $\bm{1}_{-2} +\bm{2}_{-\frac32} +\bm{3}_{-1}  $    & $-\frac{20}{9}$ & $\frac{4}{3}$           & hypermultiplet      \\ \hline
$\bar{Z}_{(a}\bar{Z}_{b)}$                                                                                                                                  & $\mathbf{\bar{6}}_{4/3}$   & $\bm{1}_{(2,\frac43)} +\bm{2}_{(\frac12,\frac43)} +\bm{3}_{(-1,\frac43)}  $ & {\begin{tikzpicture}\draw[overlay, thin, rounded corners=1mm] (-0.325,-0.15) rectangle (0.75,1.15); \end{tikzpicture}} $\bm{1}_{+2} +\bm{2}_{\frac32} +\bm{3}_{1}  $ & $-\frac{20}{9}$ & $\frac{4}{3}$           & hypermultiplet      \\ \hline
$Z^{a}\bar{Z}_a-3Z^4 \bar{Z}_4$                                                                                                                                    & $\mathbf{1}_0$       & $\bm{1}_{(0,0)} $ & {\begin{tikzpicture}\draw[overlay, thin, rounded corners=1mm] (0.025,-0.15) rectangle (0.6,0.4); \end{tikzpicture}} $\textcolor{cyan}{\bm{1}_{0}}  $       & $3-\sqrt{17}$   & $\frac{1+\sqrt{17}}{2}$ & long vector          \\ \hline
Re($Z^4Z^4$)                                                                                                                                                & $\mathbf{1}_0$        & $\bm{1}_{(0,0)} $ & {\begin{tikzpicture}\draw[overlay, thin, rounded corners=1mm] (0.025,-0.15) rectangle (0.6,0.4); \end{tikzpicture}} $\bm{1}_{0}  $        & $3+\sqrt{17}$   & $\frac{5+\sqrt{17}}{2}$ & long vector          \\ \hline

Im($Z^4Z^4$)                                                                                                                                                & $\mathbf{1}_0$         & $\bm{1}_{(0,0)} $ & $\bm{1}_{0}  $        & 0               & -                       & eaten                \\ \hline
$Z^aZ^4$                                                                                                                                                    & $\mathbf{3}_{-2/3}$   & $\bm{1}_{(-1,-\frac23)} +\bm{2}_{(\frac12,-\frac23)}  $ & $\bm{1}_{-1} +\bm{2}_{-\frac12}  $       & 0               & -                       & eaten                \\ \hline
$\bar{Z}_a\bar{Z}_4$                                                                                                                                        & $\mathbf{\bar{3}}_{2/3}$    & $\bm{1}_{(1,\frac23)} +\bm{2}_{(-\frac12,\frac23)}  $ & $\bm{1}_{1} +\bm{2}_{\frac12}  $    & 0               & -                       & eaten              
  \\ \hline\hline
$\chi^a\bar{\chi}_b-\frac{1}{3}\delta^a_b \chi^c\bar{\chi}_c$               & $\mathbf{8}_0$      & $\bm{1}_{(0,0)} +\bm{2}_{(-\frac32,0)} +\bm{2}_{(\frac32,0)} +\bm{3}_{(0,0)}  $ & {\begin{tikzpicture}\draw[overlay, thin, rounded corners=1mm] (0.025,-0.15) rectangle (0.6,0.4); \end{tikzpicture}} $\bm{1}_{0} +\bm{2}_{-\frac{1}{2}} +\bm{2}_{\frac{1}{2}} +\bm{3}_{0}  $       & $-2$              & 2                       & massless vector      \\ \hline
$\chi^{(a} \chi^{b)}$                                                                                                & $\mathbf{6}_{2/3}$   & $\bm{1}_{(-2,\frac23)} +\bm{2}_{(-\frac12,\frac23)} +\bm{3}_{(1,\frac23)}  $    &     $\textcolor{cyan}{\bm{1}_{0}} +\bm{2}_{\frac12} +\bm{3}_{1}  $ & $-\frac{14}{9}$  & $\frac{7}{3}$           & hypermultiplet      \\ \hline
$\bar{\chi}_{(a} \bar{\chi}_{b)}$                                                                             & $\mathbf{\bar{6}}_{-2/3}$  & $\bm{1}_{(2,-\frac23)} +\bm{2}_{(\frac12,-\frac23)} +\bm{3}_{(-1,-\frac23)}  $    &     {\begin{tikzpicture}\draw[overlay, thin, rounded corners=1mm] (-0.10,-0.15) rectangle (0.7,1.15); \end{tikzpicture}} $\textcolor{cyan}{\bm{1}_{0}} +\bm{2}_{-\frac12} +\bm{3}_{-1}  $ & $-\frac{14}{9}$  & $\frac{7}{3}$           & hypermultiplet      \\ \hline
Re($\chi^4\chi^4$)                                                                                                        & $\mathbf{1}_{-2}$     & $\bm{1}_{(0,-2 ) }  $ & {\begin{tikzpicture}\draw[overlay, thin, rounded corners=1mm] (0.025,-0.15) rectangle (0.8,0.4); \end{tikzpicture}} $\bm{1}_{-2}  $      & $2  $            & $\frac{3+\sqrt{17}}{2}$ & long vector          \\ \hline
Im($\chi^4\chi^4$)                                                                   & $\mathbf{1}_{2}$     & $\bm{1}_{(0,2 ) }  $ & {\begin{tikzpicture}\draw[overlay, thin, rounded corners=1mm] (0.025,-0.15) rectangle (0.7,0.4); \end{tikzpicture}} $\bm{1}_{2}  $     &  $2$           & $\frac{3+\sqrt{17}}{2}$ & long vector          \\ \hline
$\chi^{a}\bar{\chi}_a-3\chi^4 \bar{\chi}_4$                                                                                                  & $\mathbf{1}_0$    & $\bm{1}_{(0,0) }  $ & $\bm{1}_{0}  $        & $ 2  $            & $\frac{3+\sqrt{17}}{2}$ & long vector          \\ \hline
$\chi^a\bar{\chi}_4$                                                                                                            & $\mathbf{3}_{-2/3}$     & $\bm{1}_{(-1,-\frac23)} +\bm{2}_{(\frac12,-\frac23)}  $ & $\bm{1}_{-1} +\bm{2}_{-\frac12}  $      & 0               & -                       & eaten                \\ \hline
$\bar{\chi}_a \chi^4$                                                                                                       & $\mathbf{\bar{3}}_{2/3}$  & $\bm{1}_{(1,\frac23)} +\bm{2}_{(-\frac12,\frac23)}  $ & $\bm{1}_{1} +\bm{2}_{\frac12}  $  & 0               & -                       & eaten                \\ \hline
$\chi^a \chi^4$ & $\mathbf{3}_{4/3}$  & $\bm{1}_{(-1,\frac43)} +\bm{2}_{(\frac12,\frac43)}  $ & $\bm{1}_{1} +\bm{2}_{\frac32}  $       & 0               & -                       & eaten                \\ \hline
$\bar{\chi}_a \bar{\chi}_4$ &           $\mathbf{\bar{3}}_{-4/3}$  & $\bm{1}_{(1,-\frac43)} +\bm{2}_{(-\frac12,-\frac43)}  $ & $\bm{1}_{-1} +\bm{2}_{-\frac32}  $     & 0               & -                       & eaten                \\ \hline
\end{tabular}%
}
\caption{
The scalar spectrum at the $\cN=2$ point. }
\label{tab:n=0SU3U1D2masses}
\end{table}

The spectrum at the $\cN=2$ critical point of ISO(7) supergravity was given in \cite{Guarino:2015qaa} and allocated into representations of $\textrm{OSp}(4|2) \times \textrm{SU}(3)$ in \cite{Pang:2017omp}. See Table~\ref{tab:n=0SU3U1D2masses} for a summary. The $\textrm{SU}(3) \times \textrm{U}(1)_\psi$ representations branch from the SO(7) representations discussed above, see Appendix~\ref{eq:SymGYFT}. Further branchings under the various subgroups of $\textrm{SU}(3)$ introduced in Section~\ref{sec:FT} have also been included for convenience. From the table, chiral condensates $\textrm{tr} \, Z^{(a} Z^{b)}$ and real (traceless) mass terms $\textrm{tr} \,  (Z^a \bar{Z}_b -\textrm{trace})$ for the boundary scalars $Z^a$, $a=1,2,3$, can be seen to be included in the spectrum. The chiral condensates have dimension $\frac{4}{3}$, consistent with the dimensions (\ref{eq:DeltaConstSolN=2}) for $Z^a$. Curiously, the real traceless mass terms have dimension $1$, rather than twice the dimension of $Z^a$. This is perhaps not so surprising when one realises that these operators arise as the lowest component of the conserved SU(3) flavour supercurrent multiplet, and thus must have protected dimension 1. The $\textrm{AdS}_4$ mass squared of these fields is $M^2 L^2 = -2$. Alternative quantisation \cite{Klebanov:1999tb} is therefore needed for these supergravity fields to be dual to dimension 1 operators.

Two $\textrm{SU}(3) \times \textrm{U}(1)_\psi$ singlets are contained in the spectrum of proper scalars, both of them contained in the same long vector multiplet of $\textrm{OSp}(4|2)$. One of them, $\textrm{tr} \, \textrm{Re}( Z^4 Z^4)$, descends directly from the SO(7)  singlet identified above, and thus corresponds to $\textrm{tr}\, F_{\mu \nu} F^{\mu \nu}$. Indeed, it is this mode that drives holographically the $\cN=8$ SYM theory into the $\cN=2$ CS-matter SCFT \cite{Guarino:2016ynd}. It is fun to note that supergravity in this context gives a precise value, $(5+\sqrt{17})/2$, for the dimension of the  irrelevant Yang--Mills term in three-dimensions. The other singlet is related, up to a term proportional to the Konishi-like operator 
\be \label{eq:KonishilikeOp}
{\cal O}_0 = \textrm{tr} \left( Z^1 \bar{Z}_1 + Z^2 \bar{Z}_2 + Z^3 \bar{Z}_3 + Z^4 \bar{Z}_4 \right) \ , 
\ee
to the square of the 7th coordinate, $X^7$, transverse to the D2-branes. Indeed, $(X^7)^2$ can be identified with the auxiliary, SU(3)-singlet  scalar $\sigma$ in the vector multiplet that arises from splitting the $\cN=8$ SYM field content into $\cN=2$ multiplets. This auxiliary field turns out to be  integrated out as $\textrm{tr} \, \sigma \sim \textrm{tr} \,   Z^a \bar{Z}_a$ (with sum in $a=1,2,3$), see {\it e.g.} (A.33) of \cite{Araujo:2017hvi}, thus matching the table assignment\footnote{Up to terms in the Konishi-like operator (\ref{eq:KonishilikeOp}), which has a dual only in the full type IIA string theory and not in $D=4$ $\cN=8$ ISO(7) supergravity, the operator $\textrm{tr}(X^7)^2 \sim \textrm{tr}Z^a \bar{Z}_a$ is in turn akin to the Konishi operator of $\cN=4$ SYM in four-dimensions. Unlike in $D=5$ $\cN=8$ SO(6) supergravity, this operator does have a dual scalar in ISO(7) supergravity, again up to terms proportional to the actual Konishi-like operator (\ref{eq:KonishilikeOp}) in the present case. The fact that $\textrm{tr}(X^7)^2$ is integrated out of the (weakly coupled) on-shell $\cN=2$ Lagrangian (\ref{eq:WeakCoupLag}) does not mean that it becomes irrelevant in the $\cN=2$ CS-matter SCFT. On the contrary, supergravity predicts a relevant dimension $(1+\sqrt{17})/2$ for this operator.}. The three scalars $\textrm{tr} \, Z^a \bar{Z}_4$ and their complex conjugates are potentially related to minimal couplings in the covariant derivatives of $Z^a$. They belong to massive gravitino multiplets, and thus are dual to operators in the six $\cN=8$ SYM supersymmetry current multiplets broken by the $\cN=2$ SCFT. 

The pseudoscalar spectrum contains mass terms, $\textrm{tr} \, \big( \textrm{Re} (\chi^4 \chi^4) + i \, \textrm{Im} (\chi^4 \chi^4) \big) $, for the complex gaugino that enters the $\cN=2$ vector multiplet. Like the auxiliary scalar $\textrm{tr} \sigma \sim \textrm{tr} (X^7)^2$ in this multiplet, the complex gaugino is also integrated out from the weakly-coupled Lagrangian (\ref{eq:WeakCoupLag}). These fermionic mass terms belong to the same $\textrm{OSp}(4|2)$ long vector multiplet. Other pseudoscalars in the spectrum can be assigned to different quadratic fermionic operators in the boundary. Most importantly for our purposes, the spectrum contains gauge-invariant mass terms $\textrm{tr} \, \chi^{(a} \chi^{b)}$ for the boundary fermions $\chi^a$. Supergravity predicts that these mass terms should have a renormalised dimension of $\frac73$, consistent with the dimension $\frac76$ for $\chi^a$ that follows from (\ref{eq:DeltaConstSolN=2}), see Appendix~\ref{eq:SymGYFT}.

A proposal for the supergravity fields that should drive holographically the GY flow can be made upon inspection of Table~\ref{tab:n=0SU3U1D2masses}. The appropriate scalars should be singlets under the $\textrm{SO}(3)_\textrm{R} \subset \textrm{SU}(3)$ flavour symmetry preserved along the flow. Twenty such singlets can be found in the fourth column of the table. These are the scalars contained in the {$\textrm{SO}(3)_\textrm{R}$-invariant} model of \cite{Guarino:2019jef}. Further discrete symmetries and identifications can be imposed, along the lines of Appendix~\ref{App:7_chirals}, that allow one to retain only the boxed fields in that column (boxes containing two entries account for a single supergravity field). The boxed fields correspond to those of the eight-real-scalar model described in Section~\ref{sec:fourchiral}. In addition to be $\textrm{SO}(3)_\textrm{R}$ singlets, the supergravity fields driving the GY flow must also be invariant under the $\textrm{U}(1)_\textrm{d}$ R-symmetry along the flow. In particular, the pseudoscalar in the $\textbf{1}_{0} \subset \textbf{8}_{0}$ with $M^2 L^2 = -2$ is set to zero along the flow by virtue of the BPS compatibility condition \eqref{eq.extraDWflowconditions} below. Of course, they must be relevant in the UV as well. The only supergravity fields that satisfy all these requirements are the two scalars and the pseudoscalar marked in blue in Table~\ref{tab:n=0SU3U1D2masses}. The selected pseudoscalar indeed corresponds to a mass term ${\cal O}_F$ of the type discussed in (\ref{eq:GYMasses}). Up to terms in the Konishi-like operator (\ref{eq:KonishilikeOp}), the proper scalars are also of the form ${\cal O}_B$ argued in  (\ref{eq:GYMasses}). The numerical integration of Section~\ref{sec.DWflowdomainwall} will confirm that these are indeed the modes that drive the flow out of the $\cN=2$ UV phase. 

As the field theory flows into the $\cN=3$ fixed point at long distances, the dual supergravity spectrum allocates itself into $\textrm{OSp}(4|3) \times \textrm{SO}(3)_\textrm{R}$ representations. These were worked out in \cite{Gallerati:2014xra}, to which we refer for the details. Here we only note that the supergravity spectrum at the $\cN=3$ point contains an $\textrm{SO}(3)_\textrm{R}$ triplet of massless $\textrm{OSp}(4|3)$ vector multiplets, each one containing in turn an $\textrm{SO}(3)_\textrm{d}$ triplet of scalars and an $\textrm{SO}(3)_\textrm{d}$ triplet of pseudoscalars, along with the adjoint $\textrm{SO}(3)_\textrm{d}$ massless R-symmetry vectors. All of these scalars and pseudoscalars have AdS mass $M^2L^2 = -2$. The 9 proper scalars have AdS mass $M^2L^2 = -2$ and are dual to $\Delta = 1$ operators in alternative quantisation. More concretely, each vector multiplet contains the $\textrm{SO}(3)_\textrm{d}$ triplet 
\begin{equation} \label{eq:tripletsSO3dSO3R}
\big(  Z^{(a} Z^{b)} \; , \; \bar{Z}_{(a} \bar{Z}_{b)} \; , \; Z^a\bar{Z}_b-\tfrac{1}{2}\delta^a_bZ^c\bar{Z}_c  \, \big)  \; , \quad a=1,2 \; ,
\end{equation}
where each entry is itself an $\textrm{SO}(3)_{\textrm{R}}$ triplet. The first two entries are condensates of the chiral fields $Z^a$, $a=1,2$, and their conjugates, that remain massless on the GY flow. The field theory operators dual to the supergravity scalars (\ref{eq:tripletsSO3dSO3R}) have now dimension 1, protected by the conservation of the $\textrm{SO}(3)_\textrm{R}$ flavour current. Moreover, the dimension 1 of all these operators is now consistent with the free-field dimension assignment (\ref{eq:DeltaConstN=3}) for $Z^a$, $a=1,2$, in the IR. This is in contrast with the situation at the $\cN=2$ fixed point. This provides a holographic argument that the $\cN=3$ Lagrangian (\ref{eq:WeakCoupLag})--(\ref{N=3SuperPot}) is not renormalised at strong coupling. A similar analysis for the pseudoscalars can be made. These are dual to dimension 2 fermionic mass terms with a structure analogue to (\ref{eq:tripletsSO3dSO3R}). Finally, we note that the supergravity modes with irrelevant dimensions (\ref{eq.SO4modes}) that were shown in Section~\ref{sec:Modes} to possibly drive flows into the $\cN=3$ IR fixed point, belong to the long $\textrm{OSp}(4|3)$ gravitino multiplet retained in the $\textrm{SO}(3)_\textrm{R}$-invariant model of \cite{Guarino:2019jef}. We do not have a concrete proposal for the dual operators.


\subsection{The numerical four-dimensional domain-wall}
\label{sec.DWflowdomainwall}

The intricacy of the BPS equations \eqref{eq.BPSequations} forces us to consider a numeric strategy to integrate them. The strategy we follow consists in performing an IR shooting from a convenient radial value $\rho=\rho_{IR}\ll 0$, with the perturbations in \eqref{eq.perturbation2} used as a boundary condition. This allows us to set the starting point in $\rho_{IR}$ instead of the deep IR $\rho\to-\infty$. As previously discussed, regularity of the flow permits only the negative modes in \eqref{eq.SO4modes}, thus corresponding to turning on irrelevant operators in the dual field theory. This strategy has been implemented successfully in similar contexts (see e.g. \cite{Freedman:1999gp} for an early example).

Let us first note that in both the UV and IR endpoints of the domain-wall the $z_1$ and $z_2$ chiral scalars are identified as
\be
\label{eq.DWflowcondition}
z_1 = - \overline z_2 \ .
\ee
Using the parameterisation (\ref{chiralsRealNotation_1}), this condition amounts to identifying $\phi_1 = \phi_2$ and $b_{11} = b_{22}$ in the larger (half-maximal) theory constructed in \cite{Guarino:2019jef}. It is in this half-maximal context where \eqref{eq.DWflowcondition} appears as the requirement that the $\textrm{U}(1)_\textrm{d} \equiv \textrm{SO}(2) \subset \textrm{SO(3)}_{\textrm{d}}$ subgroup of the R-symmetry group in the IR is preserved also in the UV  and, more generally, along any domain-wall solution where \eqref{eq.DWflowcondition} holds.

At the level of the fluctuations described by the coefficients in \eqref{eq.SO4coefficients}, we observe that the first column therein already satisfies \eqref{eq.DWflowcondition} whereas only a combination of the second and the third columns does  (recall that the modes to which they are associated have the same value). Alternatively, we can re-express this as a condition on the two corresponding integration constants
\be
\label{eq.DWflowcondition_2}
\zeta^{(SO(4))}_3 = \left( \frac{1}{\sqrt{3}} - 1 \right) \zeta^{(SO(4))}_2 \ .
\ee
As a result, any deviation from the $\textrm{AdS}_4$ solution with $\textrm{SO}(4)$ symmetry that is regular towards $\rho\to-\infty$ and satisfies the condition \eqref{eq.DWflowcondition} for $\textrm{U}(1)_\textrm{d}$ invariance takes the form
\begin{eqnarray}
z_1(\rho) \,\,=\,\, -\bar z_2(\rho) & = & z_1^{(SO(4))} + \left(  \frac{ 1 + ( 2+\sqrt{3})i}{2\cdot 2^{1/3}}   \right) \, \zeta_1^{(SO(4))}  \,  e^{ \frac{\sqrt{3}\,\rho}{L_{SO(4)}} } \nonumber \\[4pt]
&& \qquad \qquad \quad \!\!\! +\left(  \frac{1 - (2-\sqrt{3})i}{2\cdot 2^{1/3}}  \right) \, \zeta_2^{(SO(4))} \,  e^{ \frac{ \left( \sqrt{3} - 1 \right) \rho}{L_{SO(4)}}} + \cdots  \ ,\nonumber  \\[10pt]
z_3(\rho) & = & z_3^{(SO(4))} + \left( - \frac{1 - (2+\sqrt{3})i}{2\cdot 2^{1/3}} \right) \,\zeta_1^{(SO(4))}  \,  e^{ \frac{\sqrt{3}\,\rho}{L_{SO(4)}} } \label{eq.DWflowIRexpansion} \\[4pt]
&& \qquad \qquad \quad \!\!\! + \left( \frac{3 - \sqrt{3}}{ 6 \cdot 2^{1/3}} + \frac{9-5\sqrt{3}}{6 \cdot 2^{1/3}}   i \right) \, \zeta_2^{(SO(4))} \,  e^{ \frac{ \left( \sqrt{3} - 1 \right) \rho}{L_{SO(4)}}} + \cdots  \ , \nonumber \\[10pt]
z_4(\rho) & = & z_4^{(SO(4))} + \left( - \frac{(\sqrt{3} + 1) - (\sqrt{3} - 1)i }{2 \cdot 2^{1/3}} \right) \, \zeta_1^{(SO(4))}  \, e^{ \frac{\sqrt{3}\,\rho}{L_{SO(4)}} } \nonumber \\[4pt]
&& \qquad \qquad \quad \!\!\! + \left( \frac{2^{2/3}}{3}(3- 2\sqrt{3}) - \frac{2^{2/3}}{\sqrt{3}}  i \right) \, \zeta_2^{(SO(4))} \,  e^{ \frac{ \left( \sqrt{3} - 1 \right) \rho}{L_{SO(4)}}} + \cdots\nonumber 
\end{eqnarray}
where the ellipsis represent terms with a dependence of the form $e^{ \left( m_1 \sqrt{3} + m_2 (\sqrt{3}-1) \right) \, \rho/L_{SO(4)}}$ with $m_1+m_2>1$. Truncating at order $m_1+m_2=k$ corresponds to an expansion that keeps all the correct radial dependence only up to $e^{k\, \sqrt{3}\, \rho/L_{SO(4)}}$.

Despite being consistent at the level of the equations of motion that derive from the action (\ref{eq.action}), the identification \eqref{eq.DWflowcondition} poses additional constraints on the BPS equations \eqref{eq.BPSequations}. Plugging \eqref{eq.DWflowcondition} into \eqref{eq.BPSequations} yields two algebraic relations for the imaginary parts of $z_3$ and $z_4$ of the form\footnote{In the parameterisation (\ref{chiralsRealNotation_1}), (\ref{chiralsRealNotation_2}), the algebraic relations \eqref{eq.extraDWflowconditions} become the two conditions
\begin{equation}
m \, e^{\sqrt{2} \phi_3 } +2g \, \chi \,  \big(1 +e^{2\varphi} \chi^2 \big) = 0 \; , \quad\quad \tfrac{1}{\sqrt{2}} \, b_{33} \, e^{\frac{1}{\sqrt{2}} \phi_3} = \chi \, e^{\varphi} \ ,
\end{equation}
where the supergravity couplings $g$ and $m$ have been restored.}
\be
\label{eq.extraDWflowconditions}
\frac{\textrm{Re} (z_3) }{\textrm{Im} (z_3)} = \frac{\textrm{Re} (z_4) }{\textrm{Im} (z_4) } =  \frac{\sqrt{2 \,  \textrm{Re} (z_3)^2 \, \textrm{Re} (z_4) }}{\sqrt{1 - 2  \,  \textrm{Re} (z_3)^2 \, \textrm{Re} (z_4) }} \ .
\ee
These are in turn consistent with the flow equations \eqref{eq.BPSequations}, in the sense that the latter are identically satisfied on the combinations (\ref{eq.extraDWflowconditions}). The remaining undetermined components,  $\textrm{Re}(z_1)$, $\textrm{Im}(z_1)$, $\textrm{Re}(z_3)$ and $\textrm{Re}(z_4)$, then satisfy a set of coupled differential equations
\be
\bal\label{eq.4BPSs}
 \textrm{Re}(z_1)' & =  \frac{ \textrm{Re}(z_1)\left( 2\, \textrm{Re}(z_3) - \textrm{Re}(z_4) \right) }{ \left(2\,  \textrm{Re} (z_3)^2 \, \textrm{Re} (z_4) \right)^{1/4} \left( 1 - 2 \,  \textrm{Re} (z_3)^2 \, \textrm{Re} (z_4)  \right)^{3/4}} \ , \\[4pt]
 \textrm{Im}(z_1)' & = - \frac{\textrm{Im}(z_1)^3}{4} \,   \frac{1 -8 \, \textrm{Re}(z_3)^2 \textrm{Re}(z_4) +  \textrm{Re}(z_1)^2 \left(  2\textrm{Re}(z_3) - \textrm{Re}(z_4) \right) }{ \left(2 \,  \textrm{Re} (z_3)^2 \, \textrm{Re} (z_4) \right)^{1/4} \left( 1 - 2  \,  \textrm{Re} (z_3)^2 \, \textrm{Re} (z_4)  \right)^{3/4}} \ , \\[4pt]
 \textrm{Re} (z_3)' & = - \frac{\textrm{Re} (z_3)}{2} \left( \frac{1- 2  \,  \textrm{Re} (z_3)^2 \, \textrm{Re} (z_4) }{2 \,  \textrm{Re} (z_3)^2 \, \textrm{Re} (z_4) } \right)^{1/4} \left( \textrm{Im}(z_1)^2 \left( 1 - \textrm{Re} (z_1)^2 \textrm{Re} (z_3) \right) -4 \textrm{Re}(z_3) \right) \ , \\[4pt]
 \textrm{Re} (z_4)' & = -\frac{\textrm{Re} (z_4)}{2} \left( \frac{1- 2  \,  \textrm{Re} (z_3)^2 \, \textrm{Re} (z_4) }{2 \,  \textrm{Re} (z_3)^2 \, \textrm{Re} (z_4) } \right)^{1/4} \left( \textrm{Im}(z_1)^2 \left( 1 + \textrm{Re} (z_1)^2 \textrm{Re} (z_4) \right) -4 \textrm{Re}(z_4) \right)  \ ,
\eal
\ee
where the prime represents a derivative with respect to $\rho$. A numeric domain-wall solution to the BPS equations can now be readily constructed by performing a shooting from the IR using \eqref{eq.DWflowIRexpansion} as boundary conditions.

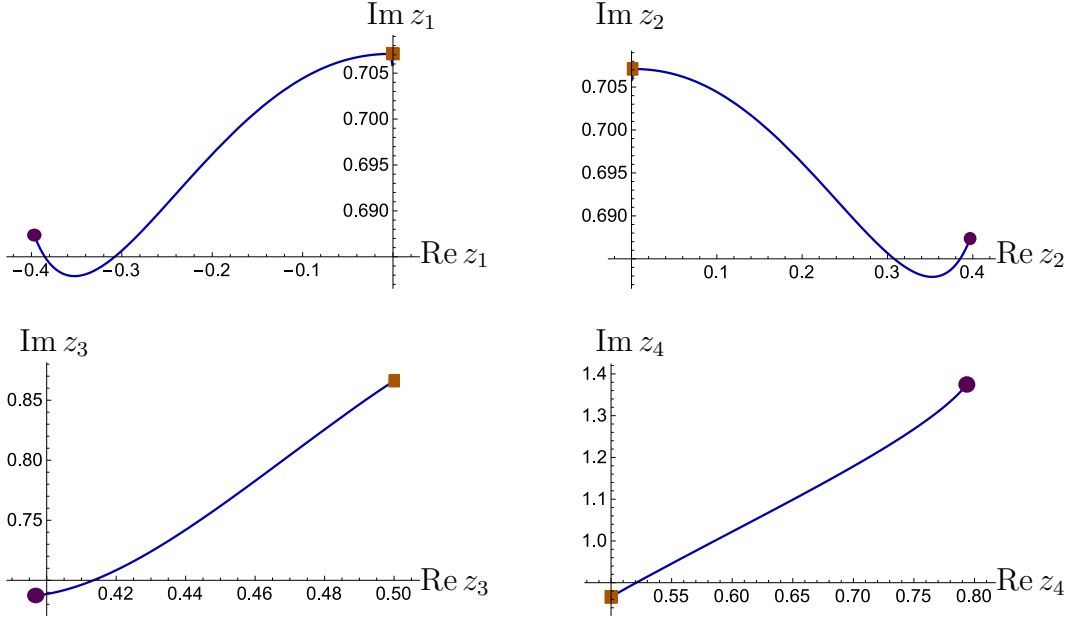
\begin{figure}[t]
\begin{center}
\includegraphics[width=0.35\textwidth]{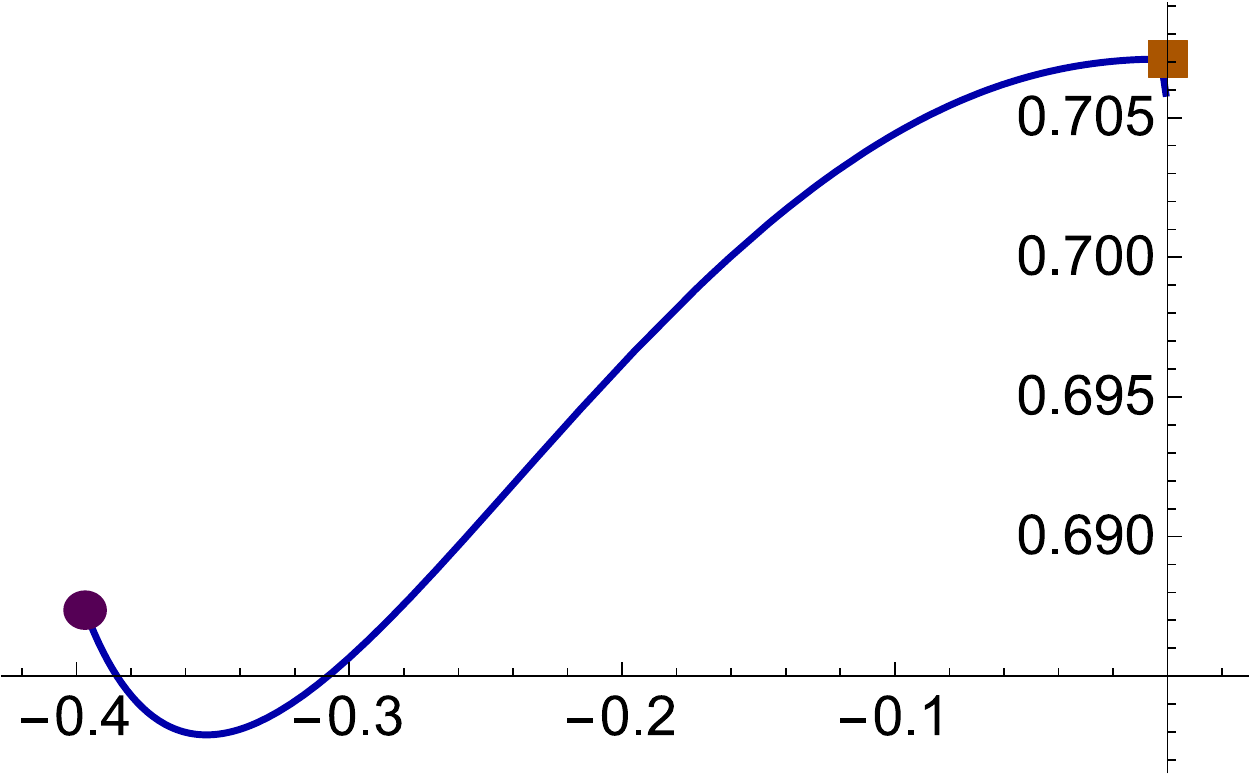}
\put(0,10){\large Re$z_1$}
\put(-20,100){\large Im$z_1$}
\hspace{20mm}
\includegraphics[width=0.35\textwidth]{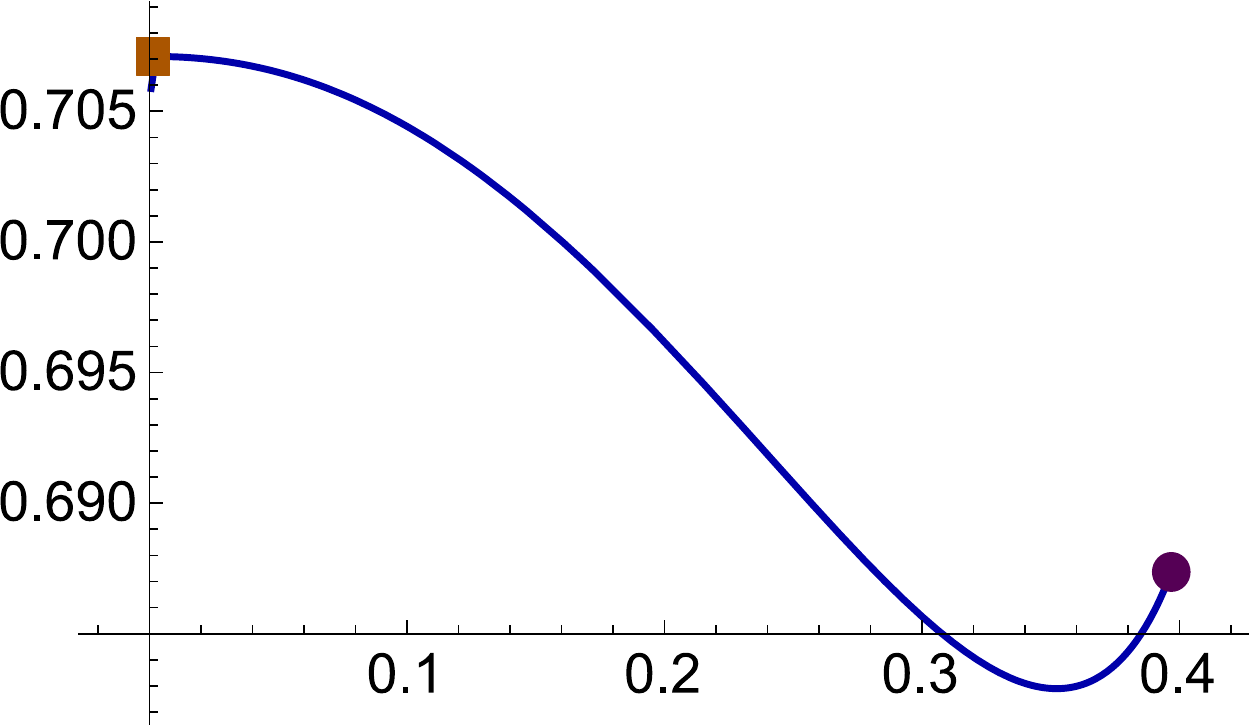}
\put(0,10){\large Re$z_2$}
\put(-150,100){\large Im$z_2$}
\\[5mm]
\includegraphics[width=0.35\textwidth]{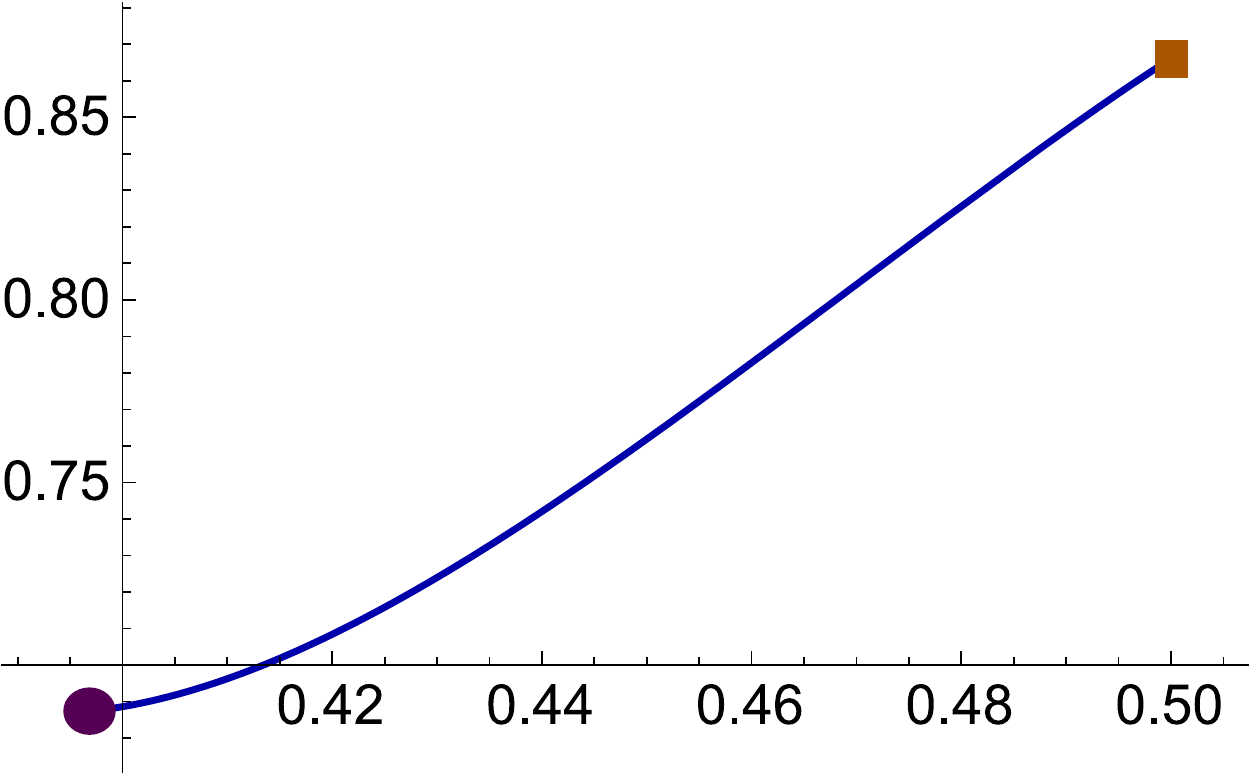}
\put(0,10){\large Re$z_3$}
\put(-150,100){\large Im$z_3$}
\hspace{20mm}
\includegraphics[width=0.35\textwidth]{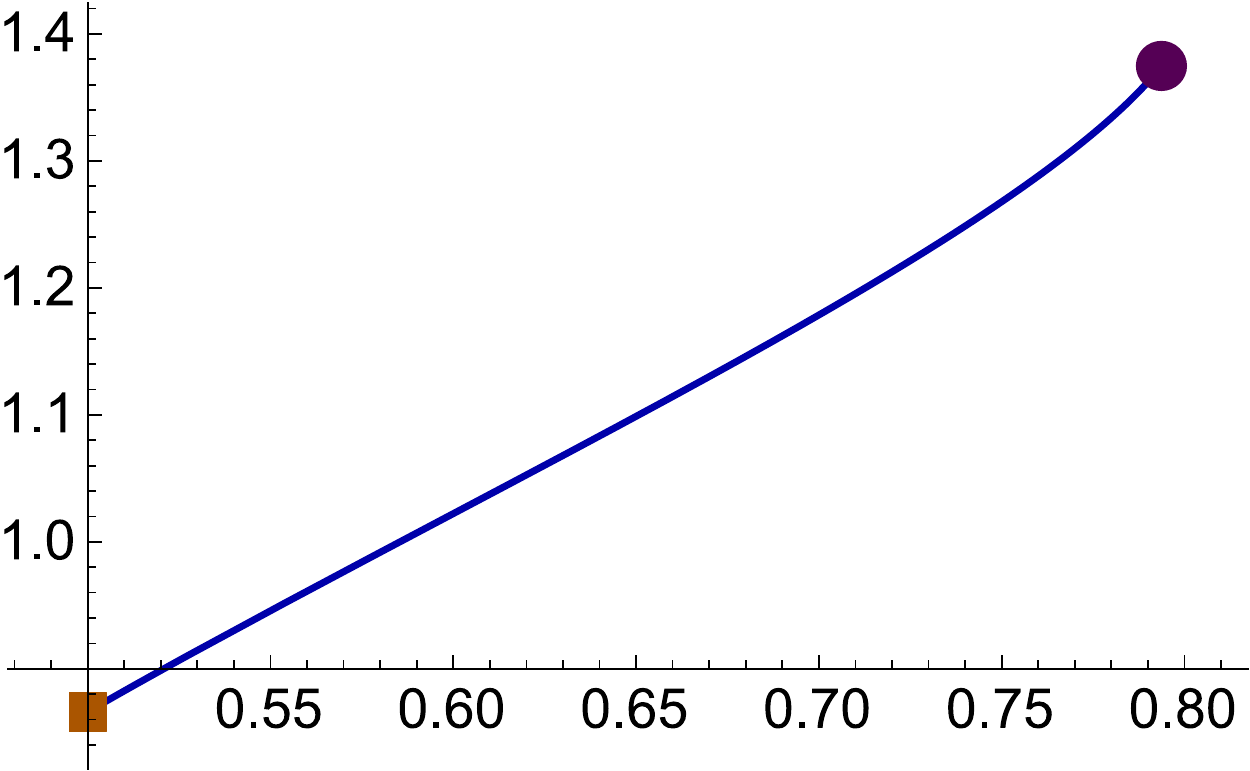}
\put(0,10){\large Re$z_4$}
\put(-150,100){\large Im$z_4$}
\caption{Numerically integrated domain-wall with $z_1 = -\bar{z}_2$ that interpolates between $\textrm{AdS}_{4}$ solutions with $\textrm{SU}(3)\times\textrm{U}(1)$ symmetry in the UV (brown square) and $\textrm{SO}(4)$ symmetry in the IR (purple circle). This domain-wall is dual to the GY flow reviewed in Section~\ref{sec:FT}.}
\label{fig.DWflowdomainwall}
\end{center}
\end{figure}

We have performed the numeric integration after using the shift symmetry in \eqref{eq.radialshift} to set $\zeta_1^{SO(4)}=1$ and choosing a particular value for $\zeta_2^{SO(4)}$. The value of $\zeta_2^{SO(4)}$ turns to be restricted to a finite range for the domain-wall solution to admit a non-singular UV behaviour. A generic value within this range produces a domain-wall that reaches the scaling behaviour of the D2-brane solution (deformed by Roman's mass \cite{Guarino:2016ynd}) in the UV. However, there is a critical value for $\zeta_2^{SO(4)}$ at the edge of the permitted range that gives rise to the special domain-wall displayed in Figure~\ref{fig.DWflowdomainwall}. The UV endpoint of this particular domain-wall corresponds to the $\cN=2$ AdS fixed point. This critical value for $\zeta_2^{SO(4)}$ lacks physical significance since, by repeated use of the arbitrary shift symmetry \eqref{eq.radialshift}, it can be set to any value upon changing the value of $\zeta_1^{SO(4)}$ accordingly. Still we can provide a radial shift-independent relation between the two IR parameters that determines the domain-wall depicted in Figure~\ref{fig.DWflowdomainwall} uniquely. This relation reads
\be
\label{IR_coeff}
\left( \zeta_2^{SO(4)} \right)^\frac{1}{1-\sqrt{3}}  \simeq 15.54 \left(\zeta_1^{SO(4)}\right)^\frac{1}{-\sqrt{3}} \ .
\ee

As we have just shown, there is a unique domain-wall solution that is $\textrm{SO}(3)_\textrm{R}$-invariant by construction and is also subject to the relations \eqref{eq.DWflowcondition} that ensure that $\textrm{U}(1)_\textrm{d}$-invariance is also preserved. This domain-wall is $\cN=2$ all along and interpolates between the $\cN=2$ AdS critical point in the UV and the $\cN=3$ one in the IR. In Section~\ref{sec:Modes} we discussed in detail the allowed deformations around the IR ($\rho\to-\infty$) as well as the relations \eqref{eq.DWflowcondition_2} and \eqref{IR_coeff}. Here we analyse the UV ($\rho\to\infty$) regime of the domain-wall and perform a characterisation of deformations around the ${\cal N}=2$, $\textrm{SU}(3)\times\textrm{U}(1)$ solution.

Amongst the modes listed in Table~\ref{tab.modes}, only the six that are positive correspond to regular solutions in the UV. However, not all of them are compatible with the conditions \eqref{eq.DWflowcondition} and \eqref{eq.extraDWflowconditions} that need to be imposed when constructing the domain-wall of Figure~\ref{fig.DWflowdomainwall}. Out of the six positive modes, only three are compatible with these conditions. These are:
\be
\label{eq.SU(3)modesintheUV}
\Delta_{(SU(3)\times U(1)),3}= \frac{2}{3} \ , \quad \Delta_{(SU(3)\times U(1)),4}= 1 \ , \quad \Delta_{(SU(3)\times U(1)),7}= \frac{1+\sqrt{17}}{2} \ .
\ee
The fluctuations around the $\textrm{AdS}_{4}$ UV solution are determined by the matrix of coefficients $z_{I,a}^{(SU(3)\times U(1))}$, where $a=3,4,7$ in the current notation, with the label specifying the position in Table~\ref{tab.modes}. It is illuminating to provide higher-order terms in the near-UV solution,  corresponding to the complementary modes to those listed in \eqref{eq.SU(3)modesintheUV}. These can be calculated as $3-\Delta_{(SU(3)\times U(1)),a}$, and appear as free coefficients when integrating the second order equations of motion but are completely determined in terms of the UV coefficients\footnote{In the remaining of this section we omit the label $^{(SU(3)\times U(1))}$ in the UV coefficients and the AdS radius to avoid excessive cluttering.} $\zeta_{3,4,7}$ when considering the BPS equations subject to the conditions \eqref{eq.DWflowcondition} and \eqref{eq.extraDWflowconditions}. This analysis gives
\be\label{eq.UVexpansionSU3U1}
\begin{split}
\begin{pmatrix} z_1 \\ z_2 \\ z_3 \\ z_4 \end{pmatrix}
= &
\begin{pmatrix} \frac{1}{\sqrt{2}} \\ - \frac{1}{\sqrt{2}} \\ 0 \\ 0 \end{pmatrix}\zeta_3\, e^{-\frac{2}{3} \frac{\rho}{L}}
+
\begin{pmatrix} 0 \\ 0 \\ -\frac{1+\sqrt{3}i}{4} \\ \frac{1+\sqrt{3}i}{2} \end{pmatrix}\zeta_4\, e^{-\frac{\rho}{L}}
+
\begin{pmatrix} \frac{i}{2} \\ \frac{i}{2} \\ -\frac{1+\sqrt{17}}{8\sqrt{2}} (1-\sqrt{3}i) \\ -\frac{1+\sqrt{17}}{8\sqrt{2}} (1-\sqrt{3}i) \end{pmatrix}\zeta_7\, e^{-\frac{1+\sqrt{17}}{2} \frac{\rho}{L}} \\
& 
+
\begin{pmatrix} -\frac{5\,i}{3\sqrt{2}} \\ -\frac{5\,i}{3\sqrt{2}} \\ \frac{3}{32}-\frac{11}{8\sqrt{3}}i \\ - \frac{1}{16}-\frac{17}{8\sqrt{3}}i \end{pmatrix}\zeta_3^2\zeta_4\, e^{-\frac{7}{3} \frac{\rho}{L}}
+
\left[ \begin{pmatrix} \frac{33}{32\sqrt{2}} \\ -\frac{33}{32\sqrt{2}} \\ 0 \\ 0 \end{pmatrix}\zeta_3^3 + \begin{pmatrix} \frac{i}{2\sqrt{2}} \\ \frac{i}{2\sqrt{2}} \\ - \frac{1-\sqrt{3}i}{8} \\  \frac{1 + 2\sqrt{3}i}{4} \end{pmatrix}\zeta_4^2 \right] e^{-2 \frac{\rho}{L}} + \cdots \ ,
\end{split}
\ee
with the ellipsis denoting other terms that we are not interested in. I.e., the expression \eqref{eq.UVexpansionSU3U1}  is not a UV expansion since we are omitting terms, that in particular scale as $e^{-\frac{4}{3} \frac{\rho}{L}}$ and $e^{-\frac{5}{3} \frac{\rho}{L}}$, which are more important in the UV limit $\rho\to\infty$ than some of the terms shown here. Notice also that a non-normalisable contribution $e^{-\frac{5-\sqrt{17}}{2} \frac{\rho}{L}}$ does not appear in Table~\ref{tab.modes} and is thus not allowed by the BPS equations. For the numerical domain-wall of Figure~\ref{fig.DWflowdomainwall}, the coefficients $\zeta_{3,4,7}$ in (\ref{eq.UVexpansionSU3U1}) become functions of the IR parameters (\ref{IR_coeff}).

Armed with the expansion \eqref{eq.UVexpansionSU3U1} we can make contact with the field theory picture reviewed in Section~\ref{sec:FT}, using the field-operator map discussed in Section~\ref{sec:StSugra}. The constant $\zeta_3$ corresponds to the source for the $\textrm{SO}(3)_\textrm{R} \times \textrm{U}(1)_\textrm{d} $-invariant fermion bilinear operator ${\cal O}_F \sim \textrm{tr} \,  \big( \chi^3 \chi^3 + \bar{\chi}_3 \bar{\chi}_3 \big)$ with dimension $\Delta=\frac73$ that belongs to an $\textrm{OSp}(4|2)$ hypermultiplet in Table~\ref{tab:n=0SU3U1D2masses}. This source is the only dimensionful parameter in the field theory side, and therefore its exact value carries no physical significance. The corresponding property in the gravitational side is, again, the use of the shift symmetry \eqref{eq.radialshift} that allows to set $\zeta_3$ to any value without changing the physics. From our numerical integration we can find the UV constants of integration in terms of the IR parameter $\zeta_2^{(\textrm{SO}(4))}$. For our discussion it suffices to give the physically-meaningful relations between the UV ones
\be
\label{eq.UVvaluesrelations}
\zeta_4 \simeq 0.97722 \, \zeta_3^{3/2} \ , \qquad  \zeta_7 \simeq 1.04957 \, \zeta_3^\frac{3(1+\sqrt{17})}{4}  \ .
\ee
The constants $\zeta_4$ and $\zeta_7$, related to $\zeta_3$ via (\ref{eq.UVvaluesrelations}), correspond in principle to vevs for the boson bilinears with conformal dimensions $\Delta=1$ and $\Delta=\tfrac{1+\sqrt{17}}{2}$ that belong to the massless and massive (long) vector multiplets in Table~\ref{tab:n=0SU3U1D2masses}, respectively. 

The coefficients in the second line of Equation~\eqref{eq.UVexpansionSU3U1} also carry information about the behaviour of the dual operators. The first one scales like $e^{-\frac{7}{3} \frac{\rho}{L}} $ and corresponds to a vev for ${\cal O}_F$. This term, proportional to $\zeta_3^4 \, \zeta_4$, takes the exact value that allows to kill the normalizable mode associated to the field of mass square $M^2L^2=-\tfrac{14}{9}$, i.e., there is no condensate for this fermion bilinear. To see this explicitly we constructed the asymptotic solution to the second order equations of motion for the relevant fields and compared it to the expansion in \eqref{eq.UVexpansionSU3U1}. The matching between both expressions determined that the constant of integration associated to the fermion bilinear condensate, via holographic renormalisation \cite{Skenderis:2002wp}, has to vanish. A similar holographic renormalisation analysis should be performed to assess whether the naive vevs mentioned above for the boson bilinears actually hold up as actual vevs for operators that turn out to condense along the flow. This is, however, immaterial for our discussion. More important is the term in \eqref{eq.UVexpansionSU3U1} that scales like $e^{-2\frac{\rho}{L}} $. This term corresponds to a source for the dimension 1 operator that requires alternative quantisation and sits in the massless vector multiplet of Table~\ref{tab:n=0SU3U1D2masses}. Crucially, from \eqref{eq.UVexpansionSU3U1} and (\ref{eq.UVvaluesrelations}), the coefficient of this term is completely determined from the fermion mass parameter as $\zeta_3^3$.

This analysis confirms that the $\textrm{SO}(3)_\textrm{R} \times \textrm{U}(1)_\textrm{d} $-invariant domain-wall plotted in Figure~\ref{fig.DWflowdomainwall} approaches the $\cN=2$ UV fixed point with sources, governed by a unique parameter, for the field theory operators ${\cal O}_F$ and ${\cal O}_B$ defined in (\ref{eq:GYMasses}), up to contributions of the Konishi-like operator (\ref{eq:KonishilikeOp}). This domain-wall is thus dual to the GY flow.

\section{A family of holographic RG flows with $\mathcal{N}=1  \rightarrow \mathcal{N}=3$}
\label{section.N=1flows}

Relaxing the field identification in \eqref{eq.DWflowcondition}, the unique domain-wall depicted in Figure~\ref{fig.DWflowdomainwall} gets generalised to a full family of domain-walls labelled by a single parameter. This family consists of new domain-wall solutions that connect the ${\cal N}=3$, $\textrm{SO}(4) $ solution in the IR and the ${\cal N}=1$, $\textrm{G}_2$ solution in the UV. The generic flows in the family are $\cN=1$ and preserve only the $\textrm{SO}(3)_\textrm{R}$ flavour of the IR fixed point. The family is limited on one side by the $\cN=2$, $\textrm{SO}(3)_\textrm{R} \times \textrm{U}(1)_\textrm{d}$-invariant GY domain-wall of Section~\ref{sec.DWflowdomainwall}, and bounded on the other side by a new $\cN=1$ $\textrm{SO}(3)_\textrm{R} \times \textrm{SO}(2)$-invariant domain-wall. The supersymmetry of the latter is also $\cN=1$, and $\textrm{SO}(2)$ corresponds to an additional flavour symmetry.

\subsection{Generic flows}

A numerical study reveals that the range of the parameter describing different members of this family is delimited by two special cases. On one side of the allowed range there is the \textit{limiting} domain-wall that passes arbitrarily close to the ${\cal N}=2$, $\textrm{SU}(3) \times \textrm{U}(1)$ solution before reaching the ${\cal N}=1$, $\textrm{G}_2$ in the UV. This limiting domain-wall eventually disappears in favour of the $\mathcal{N}=2  \rightarrow \mathcal{N}=3$ domain-wall studied in Section~\ref{section.GYflow} for which the scalar identification in \eqref{eq.DWflowcondition} holds and we concluded that
\begin{equation}
\label{cone_bound_1}
\frac{\zeta^{(SO(4))}_3}{\zeta^{(SO(4))}_2} = \left( \frac{1}{\sqrt{3}} - 1 \right) \ ,
\end{equation}
see \eqref{eq.DWflowcondition_2}. The $\mathcal{N}=2  \rightarrow \mathcal{N}=3$ domain-wall of Figure~\ref{fig.DWflowdomainwall} can be ``glued" to the $\mathcal{N}=1  \rightarrow \mathcal{N}=2$ domain-wall connecting the $\textrm{G}_{2}$ and $\textrm{SU}(3) \times \textrm{U}(1)$ solutions \cite{Guarino:2016ynd} (see appendix~\ref{sec.previousDWs}).

\begin{figure}[t]
\begin{center}
\includegraphics[width=0.35\textwidth]{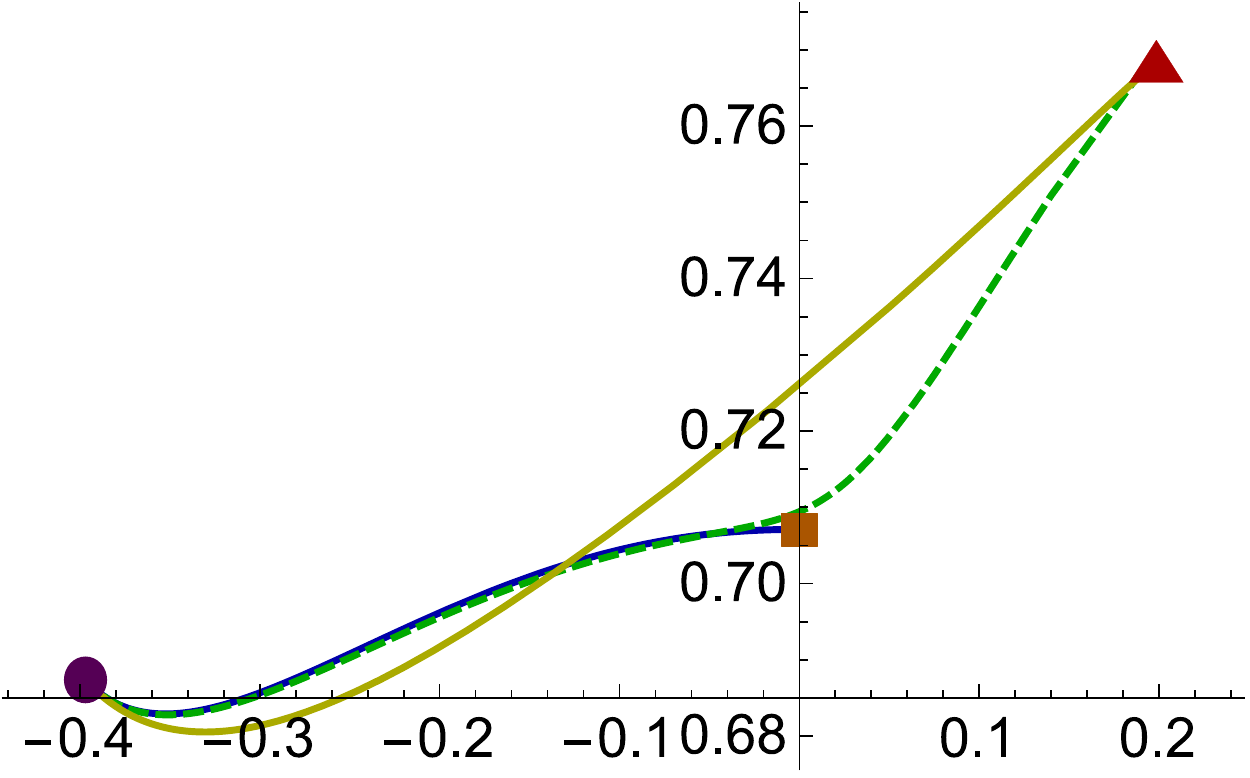}
\put(0,10){\large Re$z_1$}
\put(-65,100){\large Im$z_1$}
\hspace{20mm}
\includegraphics[width=0.35\textwidth]{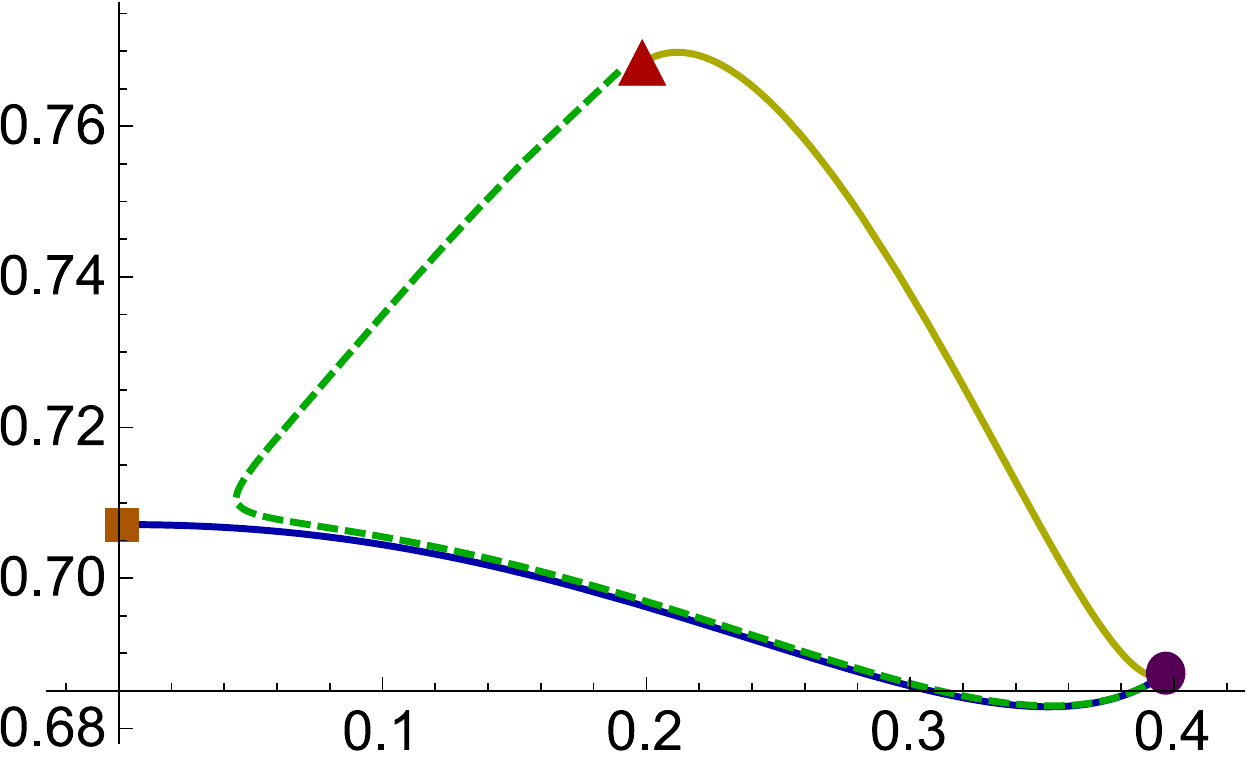}
\put(0,10){\large Re$z_2$}
\put(-150,100){\large Im$z_2$}
\\[5pt]
\includegraphics[width=0.35\textwidth]{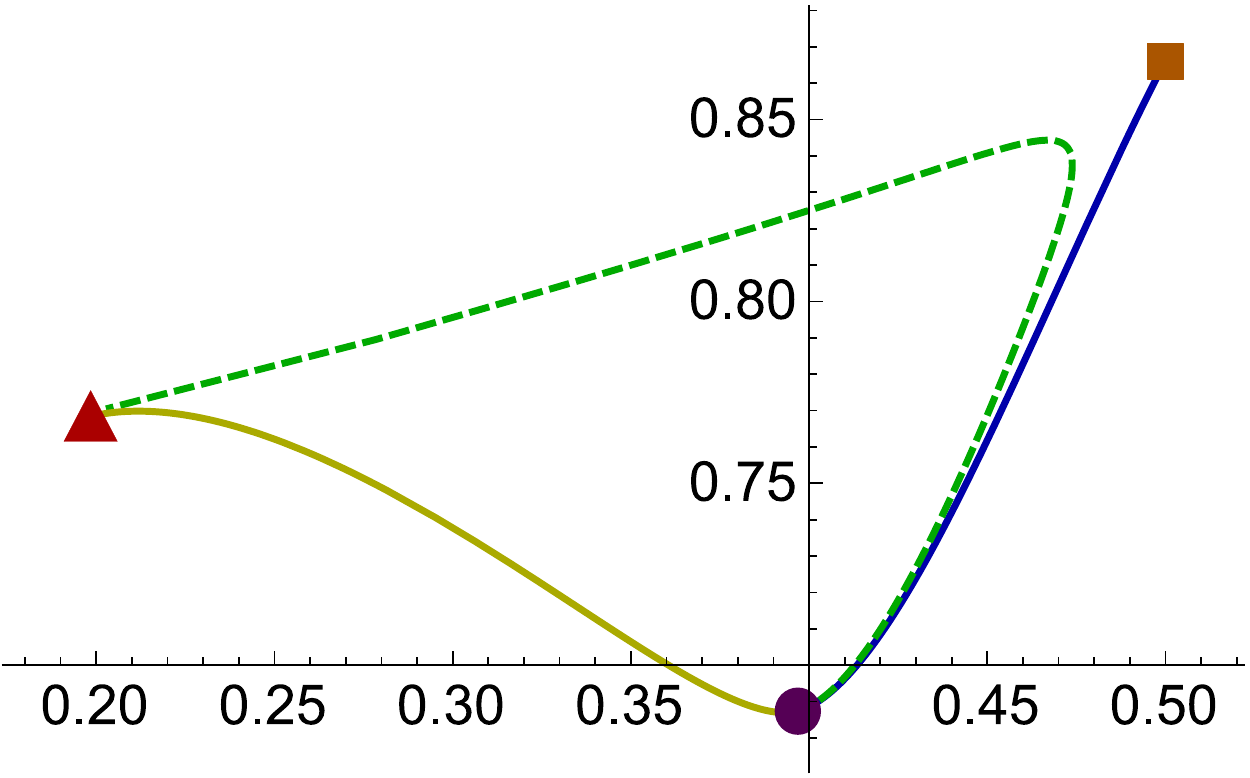}
\put(0,10){\large Re$z_3$}
\put(-65,100){\large Im$z_3$}
\hspace{20mm}
\includegraphics[width=0.35\textwidth]{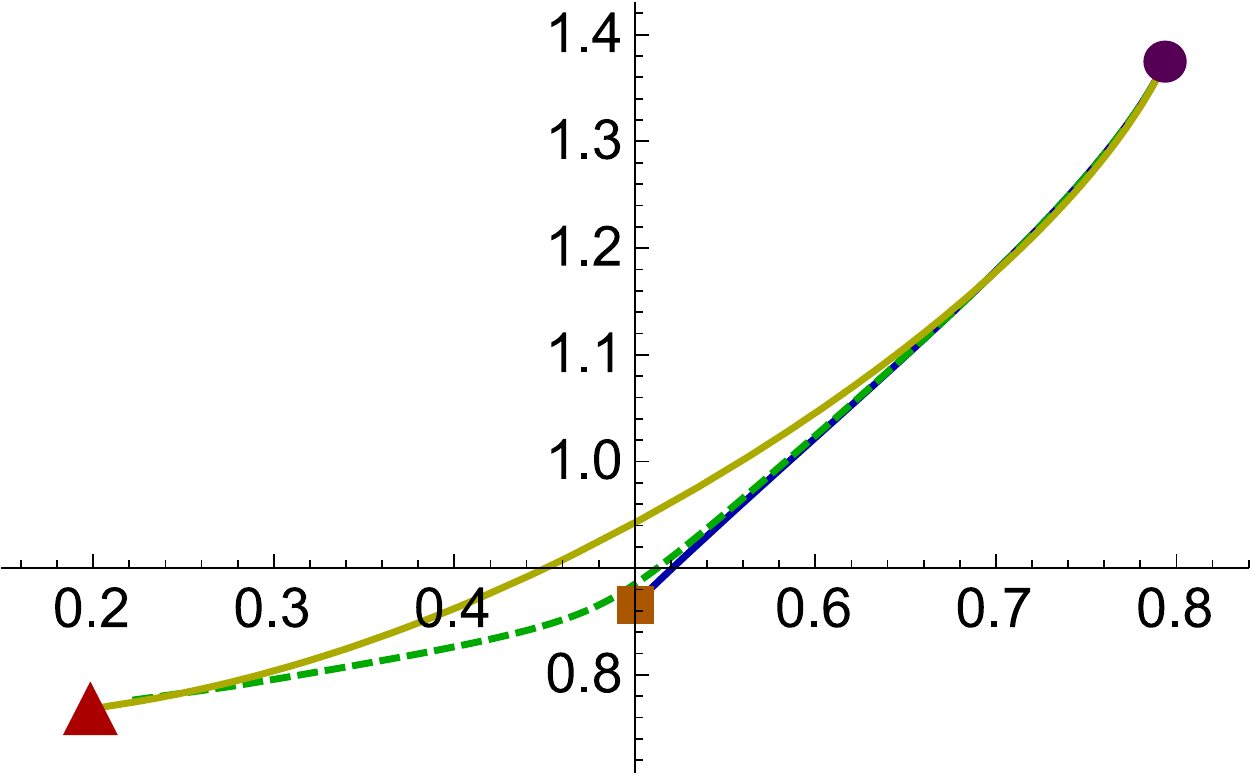}
\put(0,10){\large Re$z_4$}
\put(-85,100){\large Im$z_4$}
\caption{Numerically integrated domain-walls that interpolate between the $\textrm{G}_2$ solution (red triangle), the $\textrm{SU(3)}\times\textrm{U(1)}$ solution (brown square) and the $\textrm{SO}(4)$ solution (purple circle). They form a one-parameter family of domain-walls delimited by two solid lines: the limiting domain-wall dual to the GY flow (blue solid line) and the bounding domain-wall (yellow solid line). In both cases additional scalar identifications occur which translate into symmetry-enhanced domain-walls. The dashed, green line corresponds to a generic domain-wall with $\zeta^{(SO(4))}_3 = -\frac{2}{5} \, \zeta^{(SO(4))}_2$ that passes close to the $\cN=2$ $\textrm{SU(3)}\times\textrm{U}(1)_\psi$ fixed point without reaching it.}
\label{fig.domainwallcone}
\end{center}
\end{figure}
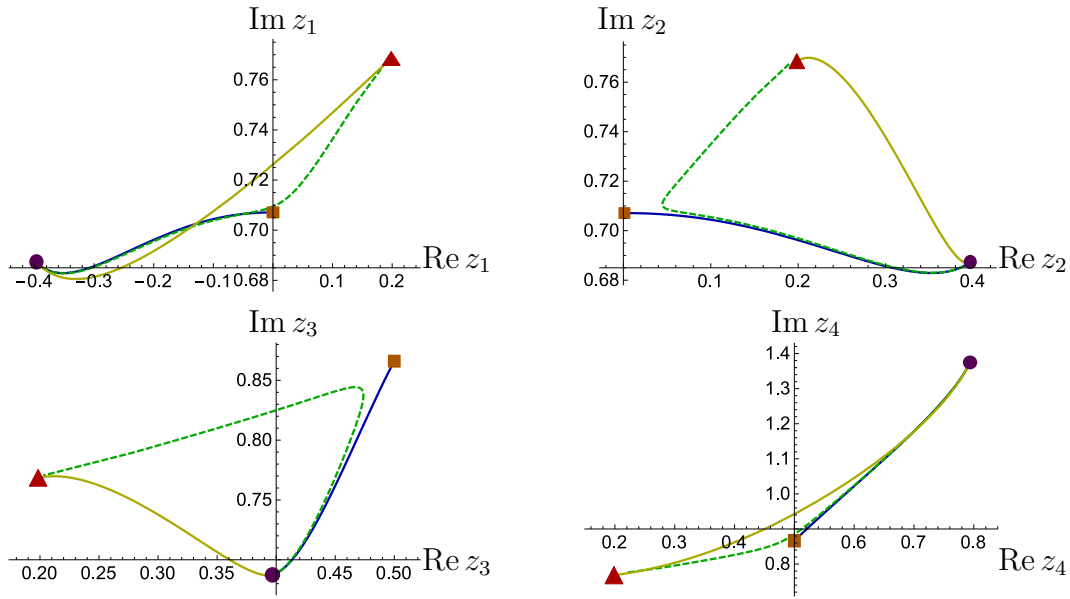

On the other side of the allowed range for the parameter there is a \textit{bounding} domain-wall that does reach the ${\cal N}=1$, $\textrm{G}_2$ point in the UV. For this bounding domain-wall a different scalar identification of the form $z_{2}=z_{3}$ holds, which translates into the condition
\begin{equation}
\label{cone_bound_2}
\frac{\zeta^{(SO(4))}_3}{\zeta^{(SO(4))}_2} = \frac{1}{2\sqrt{3}}  \ .
\end{equation}
As a result, a family of domain-wall solutions exists and is given by a parameter delimited by the radial shift-independent values in \eqref{cone_bound_1} and \eqref{cone_bound_2}, namely,
\begin{equation}
\left( \frac{1}{\sqrt{3}} - 1 \right) \,\, < \,\,  \frac{\zeta^{(SO(4))}_3}{\zeta^{(SO(4))}_2} \,\, \leq \,\, \frac{1}{2\sqrt{3}}   \ .
\end{equation}
An example of a member of this family of BPS domain-walls is displayed in Figure~\ref{fig.domainwallcone}, together with the limiting (blue solid line) and bounding (yellow solid line) domain-walls.

\subsection{The bounding symmetry-enhanced domain-wall}

Similarly to Section~\ref{sec.DWflowdomainwall}, let us note here that in both the UV and IR endpoints of the bounding domain-wall, the $z_{2}$ and $z_{3}$ chiral scalars are identified as
\be
\label{eq.G2condition}
z_2 = z_3 \ .
\ee
Using the parameterisation \eqref{chiralsRealNotation_1}, this identification translates into ${b_{22}=b_{33}}$ and ${\phi_2=\phi_3}$ in the larger (half-maximal) theory constructed in \cite{Guarino:2019jef}. As a result, an $\textrm{SO}(2) \subset \textrm{SO}(3)_{\textrm{d}}$ subgroup of the R-symmetry in the IR is again preserved in the UV. The identification (\ref{eq.G2condition}) is this time consistent with the BPS equations (\ref{eq.BPSequations}) and holds all along the bounding domain-wall. Its effect in the linearised solution around the $\textrm{AdS}_{4}$ endpoint in the IR reduces to the identification of the integration constants
\be
\label{eq.G2parametercondition}
\zeta^{(SO(4))}_3 = \frac{1}{2\sqrt{3}} \zeta^{(SO(4))}_2 \ ,
\ee
since the first column in \eqref{eq.SO4coefficients} is already compatible with \eqref{eq.G2condition}.

\begin{figure}[t]
\begin{center}
\includegraphics[width=0.35\textwidth]{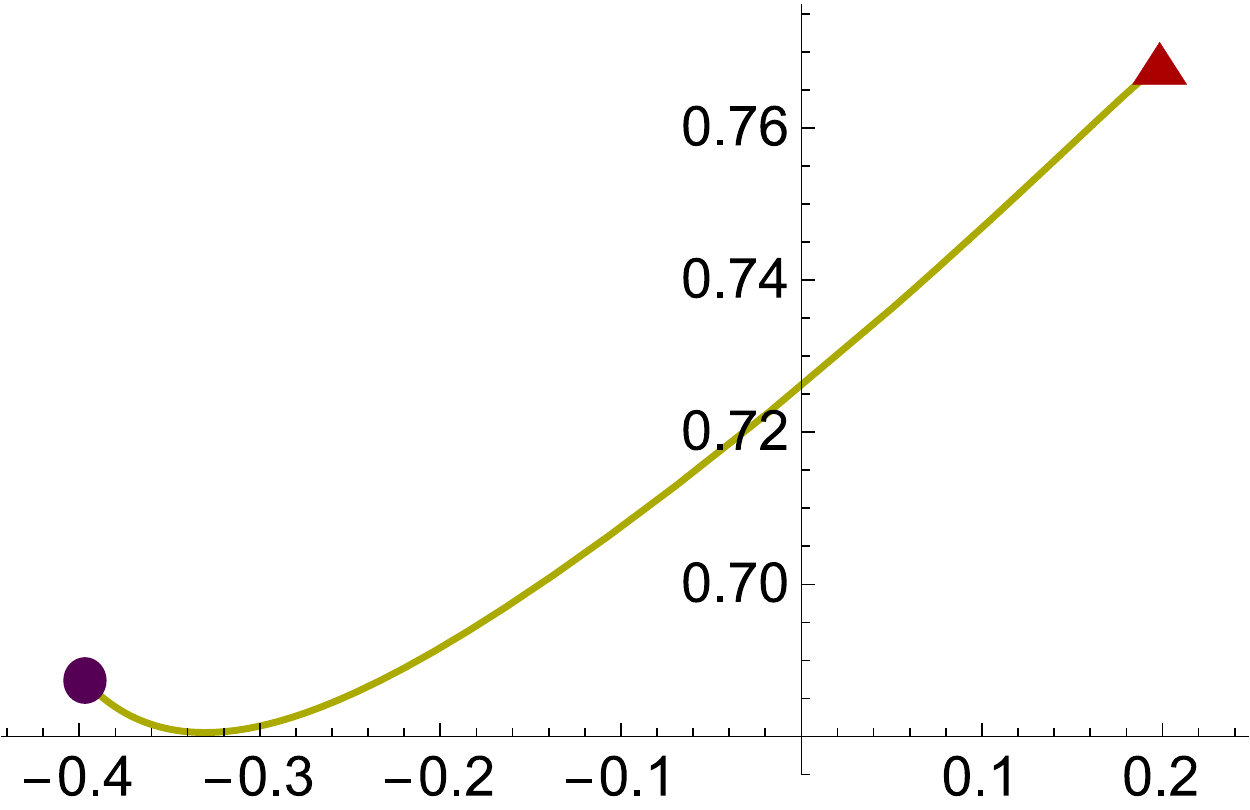}
\put(0,10){\large Re$z_1$}
\put(-65,100){\large Im$z_1$}
\hspace{20mm}
\includegraphics[width=0.35\textwidth]{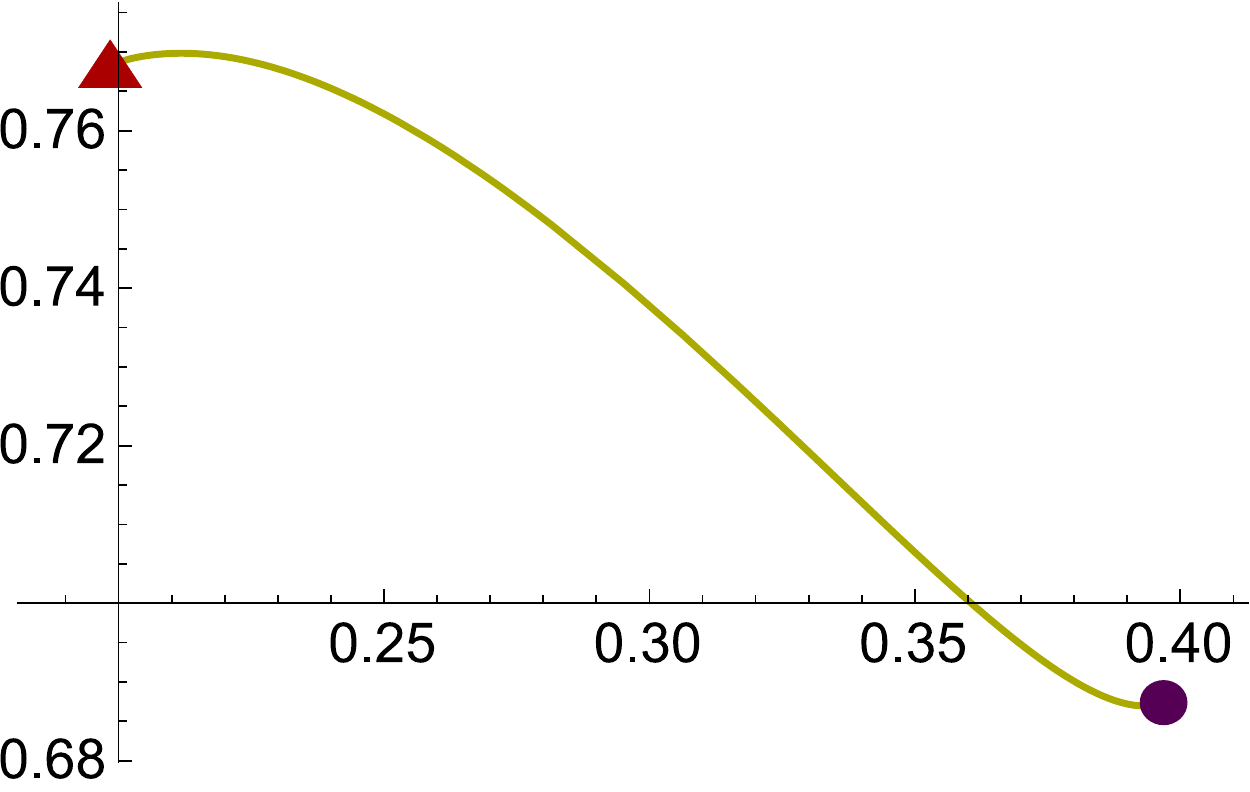}
\put(0,10){\large Re$z_2$}
\put(-150,100){\large Im$z_2$}
\\[5mm]
\includegraphics[width=0.35\textwidth]{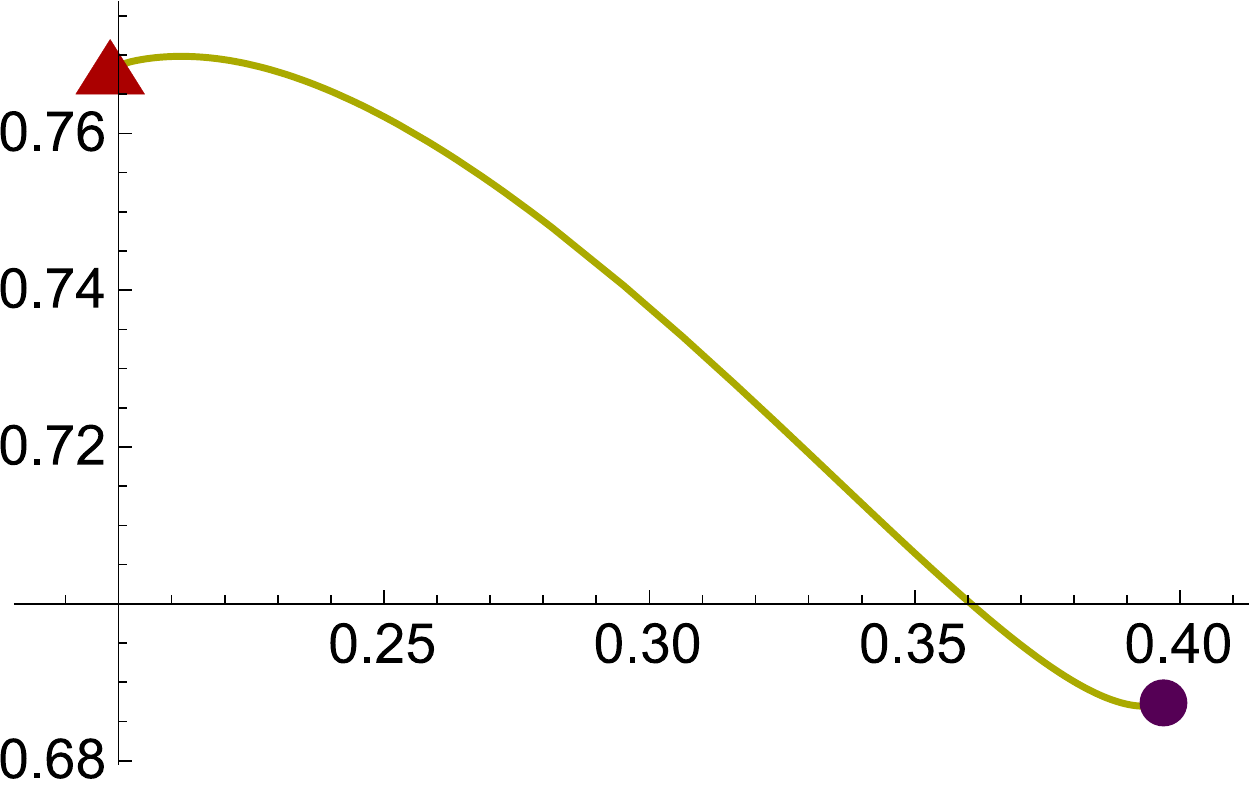}
\put(0,10){\large Re$z_3$}
\put(-150,100){\large Im$z_3$}
\hspace{20mm}
\includegraphics[width=0.35\textwidth]{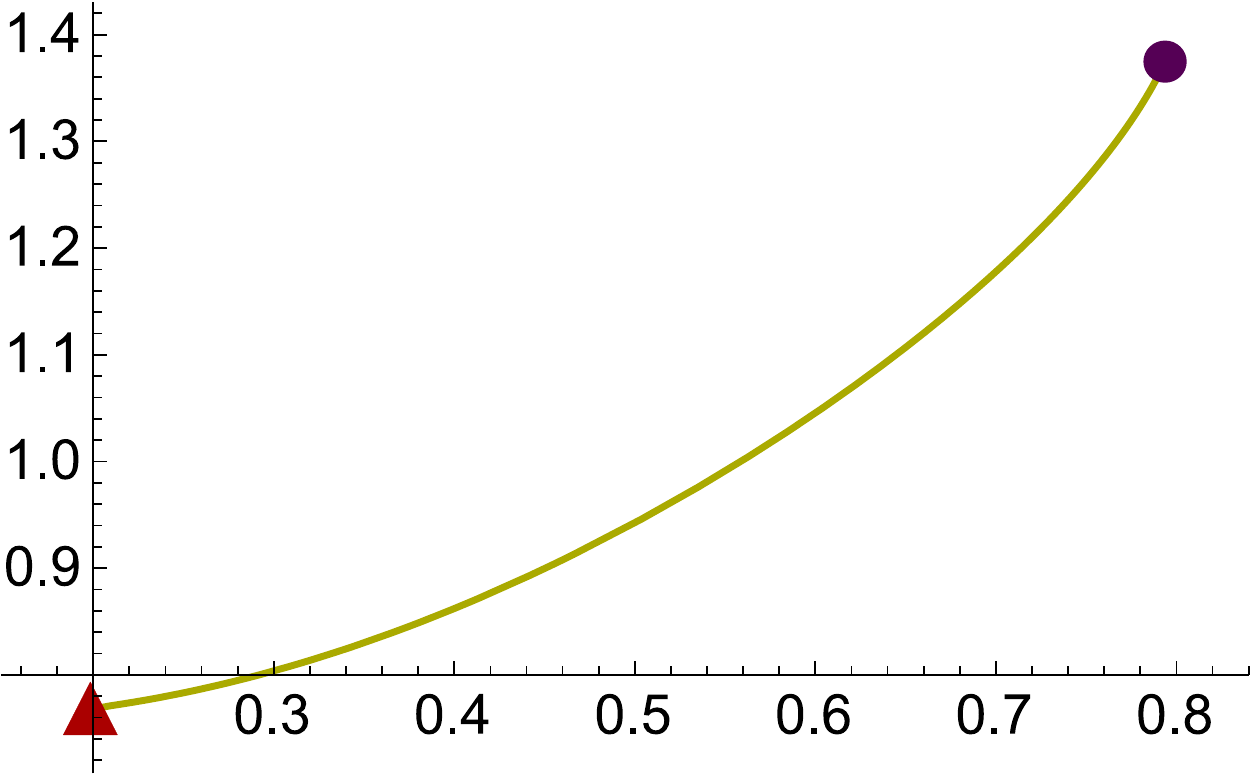}
\put(0,10){\large Re$z_4$}
\put(-150,100){\large Im$z_4$}
\caption{Numerically integrated bounding domain-wall with $z_2 = z_3$ that interpolates between $\textrm{AdS}_{4}$ solutions with $\cN=1$, $\textrm{G}_{2}$ symmetry in the UV (red triangle) and $\cN=3$ $\textrm{SO}(4)$ symmetry in the IR (purple circle). This domain-wall holographically describes an RG flow between three-dimensional SCFTs with $\mathcal{N}=1 \rightarrow \mathcal{N}=3$ supersymmetry enhancement.}
\label{fig.G2domainwall}
\end{center}
\end{figure}
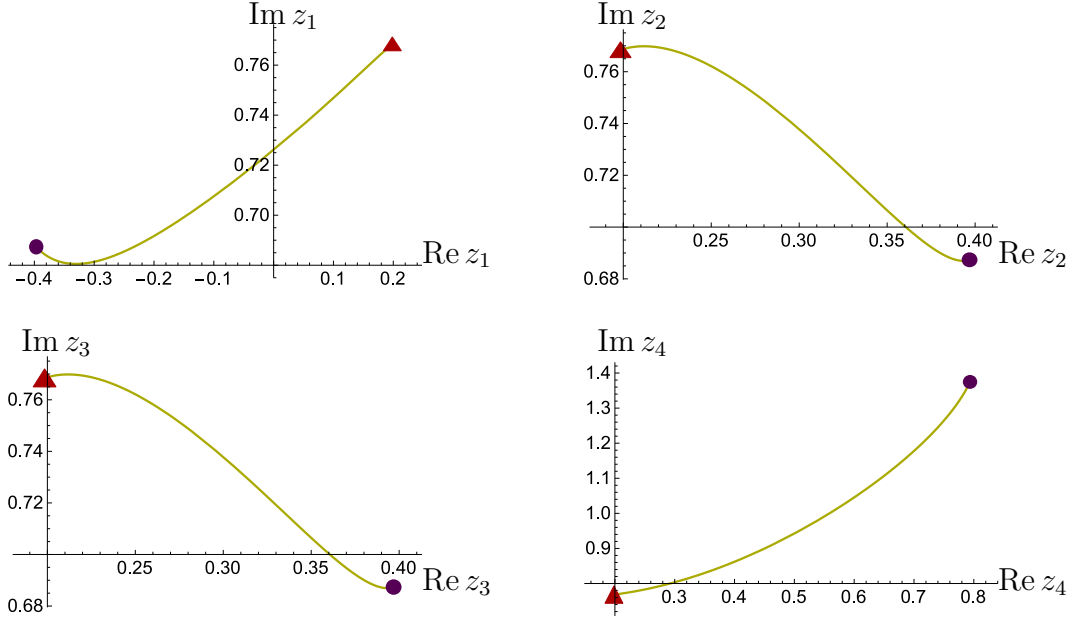

As before, we can use of the radial shift symmetry \eqref{eq.radialshift} to set $\zeta^{(SO(4))}_1=1$ without loss of generality. It is then straightforward to solve the first-order BPS equations by shooting from the IR with the boundary conditions being determined by the expansions \eqref{eq.perturbation} and \eqref{eq.perturbation2}, and the parameters \eqref{eq.SO4coefficients} being constrained by \eqref{eq.G2parametercondition}. There is again a critical value of $\zeta^{(SO(4))}_2 $ such that the domain-wall depicted by the yellow line in Figure~\ref{fig.domainwallcone} occurs. This unique domain-wall appears when the relation between the IR parameters reads
\be
\label{IR_coeff_2}
\left( \zeta_2^{SO(4)} \right)^\frac{1}{1-\sqrt{3}}  \simeq 11.26 \left(\zeta_1^{SO(4)}\right)^\frac{1}{-\sqrt{3}} \ .
\ee
This domain-wall is plotted in Figure~\ref{fig.G2domainwall}. 

\section{The flows in ten dimensions}
\label{sec.uplift}

We conclude with some considerations about the type IIA uplift of the flows that we have constructed in this paper, focusing on the GY flow of Section~\ref{section.GYflow}. All of the above four-dimensional domain-walls give rise to solutions of massive type IIA supergravity upon uplift on deformed six-spheres. In order to obtain the ten-dimensional geometries, we have particularised the formulae for the consistent truncation \cite{Guarino:2015jca,Guarino:2015vca} of massive type IIA supergravity \cite{Romans:1985tz} on $S^6$ to the eight-scalar sector of $D=4$ $\textrm{ISO}(7)$ supergravity \cite{Guarino:2015qaa} that was identified in Section~\ref{section.minimal}. The resulting uplifting formulae are quite complicated and can be consulted in appendix\footnote{As discussed in the appendix, we have in fact obtained the uplift of {\it the  entire} $\cN=4$ $\textrm{SO}(3)_\textrm{R}$-invariant sector of $\textrm{ISO}(7)$ supergravity to the ten-dimensional type IIA metric, dilaton and Ramond-Ramond one-form.} \ref{sec.SO3Ruplift}. 

Here we will simply analyse some qualitative features of the six-dimensional internal geometry corresponding to the GY flow of Section~\ref{section.GYflow}. The geometry corresponding to the UV $\cN=2$ fixed point \cite{Guarino:2015jca} can be understood as an $S^2$ bundle over $\mathbb{CP}^2$, with $S^6$ topology for the total space. The $S^2$ fibre is deformed and displays a $\textrm{U}(1)_\psi$ isometry only. The effect on this geometry of the field theory deformation (\ref{N=2SuperPotDef}) that triggers the GY flow is to realign the deformed $S^2$ fibres inside of the ambient $\mathbb{R}^7$ in which $S^6$ is defined. This realignment selects a $\textrm{U}(1)_\textrm{d}$ symmetry group, preserved along the entire flow, as a certain combination of $\textrm{U}(1)_\psi$ and another $\textrm{U}(1)_\tau$ that acts on the UV $\mathbb{CP}^2$ base. As the flow proceeds towards lower energies, the total $S^6$ internal space undergoes, in turn, a combination of two types of deformations. On the one hand, ellipsoidal deformations are inflicted by the proper scalars $\varphi$, $\phi_1$, $\phi_2$, $\phi_3$. On the other hand, the pseudoscalars $\chi$, $b_{11}$, $b_{22}$, $b_{33}$ together with the proper scalars squash inhomogeneously the deformed, $\textrm{U}(1)_\textrm{d}$-invariant, $S^2$ fibres against the four-dimensional base. At the IR $\cN=3$ fixed point \cite{Pang:2015vna,DeLuca:2018buk} the deformation of the fibre disappears. The fibre becomes a round $S^2$, and its $\textrm{U}(1)_\textrm{d}$ symmetry blows up into the full $\textrm{SO}(3)_\textrm{d}$ IR R-symmetry group. A flavour $\textrm{SO}(3)_\textrm{R}$ symmetry is also preserved along the flow.

Let us see how this works in more detail, referring to Appendix~\ref{sec:GYsymmetriesGeom} for the technicalities. The six-dimensional geometry corresponding to the $\cN=2$ UV fixed point is \cite{Guarino:2015jca}
\begin{eqnarray} \label{eq:N=2UVGeom}
ds_6^2 &=& \frac32 \,  d\alpha^2 +  \frac{ 9 \sin^2 \alpha}{ 5 + \cos 2\alpha } \, \big(d\psi + \frac12 \sin^2\xi \, (d\tau + \sigma) \big)^2
  \nonumber \\
&& +  \frac{ 6 \sin^2 \alpha}{ 3 + \cos 2\alpha } \, \Big[ d \xi^2 + \cos^2 \xi \,  ds^2 ( \mathbb{CP}^1) +  \frac14 \cos^2 \xi \,  \sin^2 \xi \,  (d\tau + \sigma)^2 \Big] \; .
\end{eqnarray}
The angles $\alpha$, $\psi$, parametrise a (globally defined, see \cite{Varela:2015uca}) $S^2$, fibred over the four-dim\-en\-sio\-nal geometry within brackets. In this case, this base is also  globally defined and corresponds to the complex projective plane, $\mathbb{CP}^2$, equipped with the Fubini-Study metric. In (\ref{eq:N=2UVGeom}), the latter has been written out in a standard parametrisation that exhibits a manifest ${\textrm{SO}(3)_\textrm{R} \times \textrm{U}(1)_\tau}$ symmetry, with $\textrm{SO}(3)_\textrm{R} \sim \textrm{SU}(2)$ acting homogeneously on the $\mathbb{CP}^1$ factor and $\textrm{U}(1)_\tau$ generated by the Killing vector $\partial_\tau$. Of course, the full symmetry of the term within brackets is the $\textrm{SU}(3)$ that acts homogeneously on $\mathbb{CP}^2$, in agreement with the flavour symmetry of the dual $\cN=2$ field theory. This  $\textrm{SU}(3)$ rotates the adjoint chirals $\Phi^a$, $a=1,2,3$, of the dual $\cN=2$ SCFT, leaving the superpotential (\ref{N=2SuperPot}) invariant. The UV configuration (\ref{eq:N=2UVGeom}) also has a manifest $\textrm{U}(1)_\psi$ symmetry generated by $\partial_\psi$. This is dual to the R-symmetry of the $\cN=2$ UV field theory. The metric on the topological $S^2$ parametrised by $\alpha$, $\psi$ is deformed away from the standard round form by a function of $\alpha$, and thus displays only $\textrm{U}(1)_\psi$ symmetry. At the UV configuration (\ref{eq:N=2UVGeom}) we thus have 
\begin{equation} \label{eq:UVSymGY}
\cN=2 \; \textrm{UV symmetry} \; : \textrm{U}(1)_\psi \times \textrm{U}(1)_\tau \times \textrm{SO}(3)_\textrm{R} \; \; \textrm{manifest \; (actually $\textrm{U}(1)_\psi \times \textrm{SU}(3))$}  .
\end{equation}

In the field theory, the $\textrm{SO}(3)_\textrm{R} \sim \textrm{SU}(2)$ subgroup of the full UV $\textrm{SU}(3)$ flavour symmetry group acts on the first two adjoint chirals, $\Phi^a$, $a=1,2$, as a doublet and leaves $\Phi^3$ invariant. The  mass deformation (\ref{N=2SuperPotDef}) that generates the GY flow therefore breaks the $\textrm{SU}(3)$ flavour of the UV field theory to $\textrm{SO}(3)_\textrm{R}$. This mass deformation is also   invariant under a combination $\textrm{U}(1)_\textrm{d}$ of the residual flavour symmetry   $\textrm{U}(1)_\tau$ that commutes with $\textrm{SO}(3)_\textrm{R}$ inside $\textrm{SU}(3)$, and the UV R-symmetry $\textrm{U}(1)_\psi$. Thus, along the $\cN=2$ GY flow we have 
\begin{equation} \label{eq:SymGY}
\cN=2 \; \textrm{GY flow symmetry} \; : \textrm{U}(1)_\textrm{d} \times \textrm{SO}(3)_\textrm{R} \; ,
\end{equation}
where the first factor is R-symmetry and the second is flavour. Geometrically, $\textrm{U}(1)_\textrm{d}$ is generated by the Killing vector (\ref{eq:U1dinUVCoords}) of the UV geometry (\ref{eq:N=2UVGeom}), and still acts on the $S^2$ fibres. However, these get realigned inside the ambient $\mathbb{R}^7$ that contains the topological $S^6$ as a consequence of the mixing of $\partial_\tau$ and $\partial_\psi$. Also as a consequence of this mixing, the four-dimensional base of the $S^2$ fibration is no longer globally defined. Nevertheless, the base still displays an intact $\mathbb{CP}^1$ upon which $\textrm{SO}(3)_\textrm{R}$ acts. The scalars and pseudoscalars squash the $\textrm{U}(1)_\textrm{d}$-invariant $S^2$ fibres against the $\textrm{SO}(3)_\textrm{R}$-invariant base as they run along the flow. The proper scalars by themselves tend to deform the metric on $S^6$ inherited by the constraint (\ref{eq:S6constraint}) to the ellipse $M_{IJ} \mu^I \mu^J =1$ in $\mathbb{R}^7$, for some scalar-dependent matrix $M_{IJ}$. This can be seen by setting the pseudoscalars to zero (formally, as they are non-vanishing along the flow) in the internal geometry (\ref{eq:10Dmetric}), (\ref{eq:10DmetricComps}). In any case, the symmetry group (\ref{eq:SymGY}) at intermediate stages of the GY flow is a subgroup of both the manifest and the full symmetries, (\ref{eq:UVSymGY}), of the UV geometry (\ref{eq:N=2UVGeom}). 

At the $\cN=3$ fixed point, the  internal six-dimensional geometry \cite{Pang:2015vna,DeLuca:2018buk} becomes, in a notation close to \cite{DeLuca:2018buk},
\begin{equation} \label{eq:N=3IRGeom}
ds_6^2 = \frac{2 \big( 3 + \cos 2\beta  \big) \cos^2 \beta}{ 3 \cos^4 \beta +3  \cos^2 \beta +2 } \, \delta_{ij}    D \tilde{\mu}^i D \tilde{\mu}^j  + 2 \,  \Big[ d\beta^2+ \frac{8 \sin^2 \beta }{3 + \cos 2\beta} \big( ds^2 (\mathbb{CP}^1) + \tfrac14 \, (\rho^3)^2 \big) \Big] \; .
\end{equation}
Here, $\beta$ is an angle related to $\alpha$, $\xi$ in (\ref{eq:N=2UVGeom}) through (\ref{S6CoordsChange}), and $\tilde{\mu}^i$, $i=1,2,3$, are constrained coordinates on $\mathbb{R}^3$ defining a round $S^2$, $\delta_{ij} \, \tilde{\mu}^i \tilde{\mu}^j = 1$. This $S^2$ is fibred on the (local) geometry within brackets in (\ref{eq:N=3IRGeom}) via 
\begin{eqnarray}
D \tilde{\mu}^i \equiv  d\tilde{\mu}^i + \epsilon^i{}_{jk} {\cal A}^j \tilde{\mu}^k \; , \qquad  \textrm{with} \qquad 
{\cal A}^i =  \frac{\sin^2 \beta }{3 + \cos 2\beta} \, \rho^i \; ,
\end{eqnarray}
where $\rho^i$, $i=1,2,3$, are the right-invariant one-forms on an $S^3$ within the local four-dimensional  base. This $S^3$ should be regarded as the Hopf fibration over $\mathbb{CP}^1$, with $\rho^3$ the one-form along the Hopf fibre. The $\textrm{SO}(3)_\textrm{R} \sim \textrm{SU}(2)$ flavour symmetry acts homogeneously on the (global) $\mathbb{CP}^1$ factor within the $S^3$ in the base. This $\mathbb{CP}^1$ factor is inherited by the IR geometry (\ref{eq:N=3IRGeom}) from its UV counterpart (\ref{eq:N=2UVGeom}), and survives the GY flow unscathed. More interestingly, an enhanced $\textrm{SO}(3)_\textrm{d}$ R-symmetry emerges in the IR, as the metric on the $S^2$ fibres becomes round. At the $\cN=3$ IR fixed point, we get
\begin{equation} \label{eq:IRSymGY}
\cN=3 \; \textrm{IR symmetry} \; : \textrm{SO}(3)_\textrm{d} \times \textrm{SO}(3)_\textrm{R} \; ,
\end{equation}
where the first factor, which contains the $\textrm{U}(1)_\textrm{d}$ R-symmetry along the GY flow, is the R-symmetry and the second factor is flavour. The IR symmetry group (\ref{eq:IRSymGY}) contains the GY flow symmetry (\ref{eq:SymGY}) but, interestingly, {\it is not} contained in the UV symmetry (\ref{eq:UVSymGY}). All three, UV, intermediate, and IR, symmetry groups are nevertheless contained in the $\textrm{SO}(7)$ group that rotates the undeformed internal $S^6$. This $\textrm{SO}(7)$ is also the R-symmetry of the three-dimensional $\cN=8$  super-Yang--Mills (SYM) theory defined on a stack of D2-branes, prior to deforming with Chern--Simons terms, see \cite{Guarino:2016ynd}.

With some differences, this behaviour is qualitatively analogue to that of similar solutions in type IIB \cite{Freedman:1999gp} and M-theory \cite{Benna:2008zy,Ahn:2000aq,Corrado:2001nv,Bobev:2009ms} that are respectively dual to mass deformations of four-dimensional $\cN=4$ SYM \cite{Brink:1976bc} and ABJM \cite{Aharony:2008ug}. The mass deformation of $\cN=4$ SYM considered in  \cite{Freedman:1999gp} breaks the UV $\textrm{SU}(4)$ R-symmetry to $\cN=1$, and preserves $\textrm{SU}(2) \times \textrm{U}(1)$ symmetry along the flow. Here, the first factor is flavour and the second is the R-symmetry. A similar flow on the M2-brane \cite{Benna:2008zy,Ahn:2000aq,Corrado:2001nv,Bobev:2009ms} breaks the manifest $\cN=6$ supersymmetry and $\textrm{SU}(4) \times \textrm{U}(1)$ global symmetry of ABJM to $\cN=2$ and $\textrm{SU}(3) \times \textrm{U}(1)$ with, again, the first factor corresponding to flavour and the second to R-symmetry. The internal geometries dual to these flows correspond to odd-dimensional sheres $S^{2n+1}$, with $n=2$ in type IIB and $n=3$ in M-theory. In both cases, the $\textrm{SU}(n)$ flavour group acts homogeneously on a $\mathbb{CP}^{n-1}$ submanifold of the $\mathbb{CP}^{n}$ base of $S^{2n+1}$ \cite{Corrado:2001nv}. The $\textrm{U}(1)$ R-symmetry is a combination of the $\textrm{U}(1)_\psi$ that acts on the Hopf fiber and the $\textrm{U}(1)_\tau$ that acts on $\mathbb{CP}^{n}$ and commutes with $\textrm{SU}(n)$ inside $\textrm{SU}(n+1)$ \cite{Corrado:2001nv}. As the flow proceeds, the $S^{2n+1}$ is squashed ellipsoidally by the proper scalars, and the Hopf fiber is squashed inhomogeneously against the base by the running scalars and pseudoscalars \cite{Corrado:2001nv}. An intact $\mathbb{CP}^{n-1}$ is preserved all along. Except for the family of M2-brane flows of \cite{Bobev:2009ms} and unlike the GY flow, these flows do not exhibit supersymmetry enhancement in the IR.

\section*{Acknowledgements}

We are grateful to Carlos Hoyos and Jes\'us Moreno for useful discussions. AG is partially supported by the Spanish government grant PGC2018-096894-B-100 and by the Principado de Asturias through the grant FC-GRUPIN-IDI/2018/000174.  JT is supported  by the European Research Council, grant no. 725369.  OV is supported by the NSF grant PHY-1720364 and, partially, by grants SEV-2016-0597, FPA2015-65480-P and PGC2018-095976-B-C21 from MCIU/AEI/FEDER, UE. Finally, AG and JT thank the Instituto de F\'isica Te\'orica (IFT) UAM-CSIC for hospitality during the completion of this work. JT would like to thank the University of Oviedo for providing support under the Grant for Visiting Docent Personnel funded by the Banco Santander.

\appendix

\addtocontents{toc}{\setcounter{tocdepth}{1}}


\section{${\cal N}=1$ truncation of $\textrm{ISO}(7)$ supergravity with seven chirals and new supersymmetric vacua}
\label{App:7_chirals}

The four-chiral model described in Section~\ref{section.minimal} was argued to arise as a further subtruncation of the $\textrm{SO}(3)_\textrm{R}$-invariant sector \cite{Guarino:2019jef} of $\cN=8$ $\textrm{ISO}(7)$ supergravity. We have  verified that the equations of motion, (2.6)--(2.8) of \cite{Guarino:2019jef}, of the $\textrm{SO}(3)_\textrm{R}$-invariant sector reduce consistently to the equations of motion for the model of this paper. Here, we provide an alternative derivation of the four-chiral model directly from the full $\cN=8$ $\textrm{ISO}(7)$ supergravity.

More precisely, the starting point here is a consistent $\cN=1$ subsector of $\cN=8$ $\textrm{ISO}(7)$ supergravity that retains seven chiral fields. As we will now show, this seven-chiral model arises as a $\mathbb{Z}^{*}_{2} \,\times \, \mathbb{Z}^{(1)}_{2} \times \mathbb{Z}^{(2)}_{2} $ invariant sector of the dyonic $\textrm{ISO}(7)$ maximal supergravity. Here,
\begin{equation}
\label{Z2xZ2}
\mathbb{Z}^{(1)}_{2} \, \times \, \mathbb{Z}^{(2)}_{2} \,\,\subset\,\, \textrm{SL}(8) \,\, \subset \,\, \textrm{E}_{7(7)} \ ,
\end{equation}
is a four-element Klein subgroup whose action on the $\textrm{SL}(8)$ fundamental index $A=1,\ldots, 8$ is given by
\begin{equation}
\label{Z2xZ2xZ2_action}
\begin{array}{rl}
\mathbb{I} : &  (x_{1}\,,\,x_{2}\,,\,x_{3}\,,\,x_{4}\,,\,x_{5}\,,\,x_{6}\,,\,x_{7}\,,\,x_{8}) \rightarrow  (x_{1}\,,\,x_{2}\,,\,x_{3}\,,\,x_{4}\,,\,x_{5}\,,\,x_{6}\,,\,x_{7}\,,\,x_{8}) \ ,\\[2mm]
\mathbb{Z}^{(1)}_{2} : &  (x_{1}\,,\,x_{2}\,,\,x_{3}\,,\,x_{4}\,,\,x_{5}\,,\,x_{6}\,,\,x_{7}\,,\,x_{8}) \rightarrow  (x_{1}\,,\,x_{2}\,,\,x_{3}\,,\,-x_{4}\,,\,-x_{5}\,,\,-x_{6}\,,\,-x_{7}\,,\,x_{8}) \ ,\\[2mm]
\mathbb{Z}^{(2)}_{2}  : &  (x_{1}\,,\,x_{2}\,,\,x_{3}\,,\,x_{4}\,,\,x_{5}\,,\,x_{6}\,,\,x_{7}\,,\,x_{8}) \rightarrow  (x_{1}\,,\,-x_{2}\,,\,-x_{3}\,,\,x_{4}\,,\,x_{5}\,,\,-x_{6}\,,\,-x_{7}\,,\,x_{8}) \ ,\\[2mm]
\mathbb{Z}^{(1)}_{2} \, \mathbb{Z}^{(2)}_{2} :&  (x_{1}\,,\,x_{2}\,,\,x_{3}\,,\,x_{4}\,,\,x_{5}\,,\,x_{6}\,,\,x_{7}\,,\,x_{8}) \rightarrow  (x_{1}\,,\,-x_{2}\,,\,-x_{3}\,,\,-x_{4}\,,\,-x_{5}\,,\,x_{6}\,,\,x_{7}\,,\,x_{8}) \ .
\end{array}
\end{equation}
In addition, we will also require invariance under an additional $\,\mathbb{Z}^{*}_{2}\,$ acting on the coordinates as
\begin{equation}
\label{Z2*}
\begin{array}{rl}
\mathbb{Z}^{*}_{2} : &  (x_{1}\,,\,x_{2}\,,\,x_{3}\,,\,x_{4}\,,\,x_{5}\,,\,x_{6}\,,\,x_{7}\,,\,x_{8}) \rightarrow  (x_{1}\,,\,-x_{2}\,,\,x_{3}\,,\,-x_{4}\,,\,x_{5}\,,\,-x_{6}\,,\,x_{7}\,,\,-x_{8}) \ .
\end{array}
\end{equation}
This truncation retains the seven dilatons of $\textrm{E}_{7(7)}$ together with seven axions in order to furnish seven $\mathcal{N}=1$ chiral multiplets. This sector has been extensively considered in the past when exploring $\mathcal{N}=1$ flux compactifications in the presence of generalised flux backgrounds \cite{Aldazabal:2006up,Aldazabal:2008zza,Aldazabal:2010ef}. In this context, the $\mathbb{Z}^{(1)}_{2} \, \times \, \mathbb{Z}^{(2)}_{2}$ factors in (\ref{Z2xZ2}) are associated with a (toroidal) orbifold action on $\textrm{T}^6$, whereas the $\mathbb{Z}^{*}_{2}$ factor in (\ref{Z2*}) is identified with an orientifold projection halving the number of supersymmetries \cite{Dibitetto:2011eu}. Recently, this sector has also been studied within the context of the SO(8) maximal supergravity \cite{Bobev:2019dik}.

Following the conventions of \cite{Guarino:2015tja}, the fourteen real scalars in the truncation are associated with the following E$_{7(7)}$ generators in the SL(8) basis. The dilatons have associated generators of the form
\begin{equation}
\begin{array}{llll}
g_{\varphi_{1}}  & = & -t_{1}{}^{1} -  t_{2}{}^{2} -  t_{3}{}^{3} +  t_{4}{}^{4} +  t_{5}{}^{5} +  t_{6}{}^{6} +  t_{7}{}^{7} -  t_{8}{}^{8} & , \\[2mm]
g_{\varphi_{2}}  & = & -t_{1}{}^{1} +  t_{2}{}^{2} +  t_{3}{}^{3} -  t_{4}{}^{4} -  t_{5}{}^{5} +  t_{6}{}^{6} +  t_{7}{}^{7} -  t_{8}{}^{8} & , \\[2mm]
g_{\varphi_{3}}  & = & -t_{1}{}^{1} +  t_{2}{}^{2} +  t_{3}{}^{3} +  t_{4}{}^{4} +  t_{5}{}^{5} -  t_{6}{}^{6} -  t_{7}{}^{7} -  t_{8}{}^{8} & , \\[4mm]
g_{\varphi_{5}}  & = & t_{1}{}^{1} -  t_{2}{}^{2} +  t_{3}{}^{3} +  t_{4}{}^{4} -  t_{5}{}^{5} +  t_{6}{}^{6} -  t_{7}{}^{7} -  t_{8}{}^{8} & , \\[2mm]
g_{\varphi_{6}}  & = & t_{1}{}^{1} +  t_{2}{}^{2} -  t_{3}{}^{3} -  t_{4}{}^{4} +  t_{5}{}^{5} +  t_{6}{}^{6} -  t_{7}{}^{7} -  t_{8}{}^{8} & , \\[2mm]
g_{\varphi_{7}}  & = & t_{1}{}^{1} +  t_{2}{}^{2} -  t_{3}{}^{3} +  t_{4}{}^{4} -  t_{5}{}^{5} -  t_{6}{}^{6} +  t_{7}{}^{7} -  t_{8}{}^{8} & , \\[4mm]
g_{\varphi_{4}}  & = & t_{1}{}^{1} -  t_{2}{}^{2} +  t_{3}{}^{3} -  t_{4}{}^{4} +  t_{5}{}^{5} -  t_{6}{}^{6} +  t_{7}{}^{7} -  t_{8}{}^{8} & ,
\end{array}
\end{equation}
whereas the axions correspond to generators of the form
\begin{equation}
\begin{array}{lllllllllll}
g_{\chi_{1}}  & = &   t_{1238}   & \hspace{5mm},\hspace{5mm}  & \,\,\,  g_{\chi_{5}}  & = &    t_{2578}  & \hspace{5mm},\hspace{5mm} \\[2mm]
g_{\chi_{2}}  & = &    t_{1458}  & \hspace{5mm},\hspace{5mm} & \,\,\, g_{\chi_{6}}  & = &    t_{4738}  & \hspace{5mm},\hspace{5mm} & \,\,\,\, g_{\chi_{4}}  & = &    t_{8246}  \  .\\[2mm]
g_{\chi_{3}}  & = &    t_{1678}  & \hspace{5mm},\hspace{5mm} & \,\,\, g_{\chi_{7}}  & = &    t_{6358}  & \hspace{5mm},\hspace{5mm}
\end{array}
\end{equation}
Exponentiating the above generators as
\begin{equation}
\mathcal{V} = \textrm{Exp} \left[ {-12 \, \displaystyle\sum_{I=1}^7 \chi_{I}\, g_{\chi_{I}}} \right]
\,
\textrm{Exp} \left[ {\frac{1}{4} \, \displaystyle\sum_{I=1}^7 \varphi_{I}\, g_{\varphi_{I}}} \right] \ ,
\end{equation}
yields a parameterisation of an $[\textrm{SL}(2)/\textrm{SO}(2)]^7$ coset space. The scalar kinetic terms in the resulting $\mathcal{N}=1$ supergravity model are of the form 
\begin{equation}
\mathcal{L}_{\textrm{kin}}=-\frac{1}{4} \displaystyle\sum_{I=1}^7  \left[  (\partial \varphi_{I})^2 + e^{2 \varphi_{I}} \,  (\partial \chi_{I})^2 \right] \ .
\end{equation}
These are the kinetic terms for a set of seven chiral fields $z_{I}=-\chi_{I}+ i \, e^{-\varphi_{I}}$ with K\"ahler potential
\begin{equation}\label{eq.kahlerseven}
K = - \displaystyle\sum_{I=1}^7   \log[-i(z_{I}-\bar{z}_{I})] \ .
\end{equation}
A direct computation shows that the scalar potential in this sector of the theory can be re-expressed in terms of a holomorphic superpotential of the form
\begin{equation}
\label{W_Z2xZ2}
W = 2 \, m  + 2 \,g \, \left[ \, 
z_{1} \, z_{2} \, z_{3} + 
z_{1} \, z_{6} \, z_{7} + 
z_{2} \, z_{7} \, z_{5} + 
z_{3} \, z_{5} \, z_{6} + 
(z_{1} \, z_{5} + z_{2} \, z_{6} + z_{3} \, z_{7}) \, z_{4} \, \right] \ ,
\end{equation}
using standard $\mathcal{N}=1$ formulas.

\subsubsection*{New $\mathcal{N}=1$ $\textrm{AdS}_{4}$ vacua}
\label{sec.new_vacua}

We have performed a numerical scan of supersymmetric extrema by solving the F-flatness conditions that derive from \eqref{W_Z2xZ2} and found six such $\textrm{AdS}_4$ vacua. Four of them were known previously, and their location in field space was discussed in the main text (see Table~\ref{tab.fixedpoints} with the identifications in Equation~\eqref{eq.z_identification_SO(3)_R} below). The other two vacua are new and preserve ${\cal N}=1$ supersymmetry together with only a $\textrm{U}(1)$ symmetry within the maximal $\textrm{ISO}(7)$ gauged supergravity. This $\textrm{U}(1)$ turns out to be the Cartan subgroup, $\textrm{U}(1)_\textrm{R}$, of the $\textrm{SO}(3)_\textrm{R}$ subgroup of SO(7) discussed in the main text.

The new vacua have a smaller value of the potential compared to the ones in Table~\ref{tab.fixedpoints}. In particular, setting $g=m=1$, we find:

\begin{itemize} 

\item The first one has a value of the potential
\be
V = - 25.6971 \ ,
\ee
and its location in field space is given by\footnote{There are additional (but equivalent) discrete realisations of this vacuum.\label{fn.position}}
\be\bal \label{eq:newpoint1}
z_1  = z_5 = 0.4874 + 0.5961\, i \ , \quad
z_2  = z_6 =  0.1082 + 1.1728 \, i \ , \quad \\[2pt]
z_3 =  -0.2178 + 0.5098\, i \ , \quad 
z_4  =  -0.5989 + 0.5894 \, i \ , \quad 
z_7 = 1.2101 + 0.8849 \, i \ .
\eal\ee
In terms of the $\textrm{AdS}_4$ radius, the spectrum of normalised scalar masses around this solution, within this seven-chiral sector, is given by
\be\bal
M^2 L^2  = &  \{ 8.1644 \ , \; 8.0986 \ , \; 4.2223 \ , \; 2.7101 \ , \; 2.6648 \ , \; 0.7839 \ , \; 0.1342 \ , \\
& \quad  -1.6232 \ , \; -1.6997 \ , \; -1.8625 \ , \; -1.8766 \ , \; -2.0988 \ , \; -2.1075 \ , \; -2.2066 \} \ ,
\eal\ee
and the corresponding values $\Delta$ of the modes that are selected by the BPS equations originating from \eqref{W_Z2xZ2} read
\be\bal
\Delta = & \{ -1.7271 \ , \; -1.7169 \ , \; -1.0441 \ , \; 3.7271 \ , \; 3.7169 \ , \; -0.2418 \ , \; 3.0441 \ , \\
& \quad  0.7083 \ , \; 2.2418 \ , \; 0.8775 \ , \; 0.8889 \ , \; 1.1111 \ , \; 1.1225 \ , \; 1.2917 \} \ .
\eal\ee
These modes arrange themselves into seven chiral multiplets of $\textrm{OSp}(4|1)$.

\item The second one has a smaller value of the potential
\be
V = - 35.6102 \ ,
\ee
and is located at (see footnote~\ref{fn.position})
\be\bal \label{eq:newpoint2} 
z_1  = z_5 = - 0.1103 + 0.7629\, i \ , \quad
z_2  = z_6 =  0.8364 + 0.3907 \, i \ , \quad \\[2pt]
z_3 =  -0.4021 + 0.3120\, i \ , \quad 
z_4  =  - 0.9449 + 1.4406 \, i \ , \quad 
z_7 = 0.7402 + 1.1526 \, i \ .
\eal\ee
The spectrum of normalised scalar masses around this solution is given by
\be\bal
M^2 L^2  = &  \{ 10.8555 \ , \; 9.8092 \ , \; 7.5707 \ , \; 4.6152 \ , \; 4.1131 \ , \; 4.0254 \ , \; 3.8639 \ , \\
& \quad  2.3031 \ , \; 0.0681 \ , \; 0.0152 \ , \; -1.1885 \ , \; -1.4465 \ , \; -2.2393 \ , \; -2.2491 \} \ ,
\eal\ee
and the corresponding values $\Delta$ of the modes that are selected by the BPS equations that follow from \eqref{W_Z2xZ2} read
\be\bal
\Delta = & \{ -2.1202 \ , \; -1.9726 \ , \; -1.6338 \ , \; 4.1202 \ , \; -1.0225 \ , \; -1.0051 \ , \; 3.9726 \ , \\
& \quad  3.6338 \ , \; 3.0225 \ , \; 3.0051 \ , \; 0.4697 \ , \; 0.6036 \ , \; 1.3964 \ , \; 1.5303 \} \ .
\eal\ee
Again, these modes arrange themselves into seven chiral multiplets of $\textrm{OSp}(4|1)$.

\end{itemize}

\subsubsection*{Enhancements to $\textrm{SO}(3)$ symmetry}
\label{sec.SO(3)sectors}

Denoting the chiral fields as $z_{i}=(z_{1},z_{2},z_{3})$ and $z_{\hat{i}}=(z_{5},z_{6},z_{7})$, two cases of symmetry enhancement are then immediately envisaged:
\begin{itemize}

\item A continuous $\textrm{SO}(3)$ invariance is recovered upon the identifications 
\begin{equation}
\label{identification_Z2xSO(3)}
z_{i} \equiv \Phi_{1} 
\hspace{5mm} , \hspace{5mm} 
z_{4} \equiv \Phi_{2}
\hspace{5mm} , \hspace{5mm} 
z_{\hat{i}}\equiv \Phi_{3} \ ,
\end{equation}
thus yielding the superpotential of the three-chiral model of Appendix~A of \cite{Guarino:2015qaa}
\begin{equation}
\label{W_Z2xSO(3)}
W = 2 \, m  + 2 \,g \, \left[ \, 
\Phi_{1}^3 + 
3 \, \Phi_{1} \, \Phi_{3}^2 + 
3 \, \Phi_{1} \, \Phi_{2} \, \Phi_{3} \, \right] \ .
\end{equation}
The identifications in (\ref{identification_Z2xSO(3)}) reduce the seven-chiral sector to the $\mathbb{Z}_{2} \times \textrm{SO}(3)$ invariant sector studied in \cite{Guarino:2015tja} for general CSO gaugings of maximal supergravity.

\item A different continuous $\textrm{SO}(3)$ invariance is restored upon the identification
\begin{equation}\label{eq.z_identification_SO(3)_R}
z_{i} = z_{\hat{i}} \ ,
\end{equation}
which connects with the $\textrm{SO}(3)_{\textrm{R}}$ invariance of \cite{Guarino:2019jef}. This yields a four-chiral model with a superpotential, from (\ref{W_Z2xZ2}), of the form
\begin{equation}
\label{W_SO(3)_R_app}
W = 2 \, m + 2\, g\, \left[ \, 4 \, z_1 \, z_2 \, z_3 +  \left( z_1^2 + z_2^2 + z_3^2 \right) \, z_{4}  \, \right] \ .
\end{equation}
This coincides with (\ref{W_SO(3)_R}) of the main text and is the superpotential that we have used in this paper. If (\ref{eq.z_identification_SO(3)_R}) is relaxed as $z_1 = z_5$, $z_2 = z_6$ while leaving $z_3$ and $z_7$ unidentified, as in the new $\cN=1$ critical points (\ref{eq:newpoint1}) and (\ref{eq:newpoint2}), the continuous symmetry is only the Cartan subgroup $\textrm{U}(1)_{\textrm{R}}$ of $\textrm{SO}(3)_{\textrm{R}}$.

\end{itemize}

Let us conclude with an observation. When analysed within the seven-chiral sector of this appendix, the $\mathcal{N}=3$ solution with $\textrm{SO(4)}$ symmetry has an enlarged scalar mass spectrum with no additional irrelevant operators apart from those already listed in Table~\ref{tab.modes}. Therefore, no additional domain-walls exist in the seven-chiral model ending at the $\mathcal{N}=3$ and $\textrm{SO(4)}$ symmetric solution in the IR.

\section{Previously known domain-wall solutions\label{sec.previousDWs}}

For completeness, we give here some details on the domain-wall solutions contained in the minimal model of Section~\ref{section.minimal} that coincide with flows discussed previously in \cite{Guarino:2016ynd}.

\begin{itemize}
\item The ${\cal N}=1$, $\textrm{G}_2$ fixed point has a unique negative (and therefore regular) mode in the IR, $\Delta_{G_2,1}=1-\sqrt{6}$, with the matrix of coefficients
\be
z_{I,1}^{(G_2)} = \frac{1}{4 \cdot 2^{1/3}} \left(\frac{ 9 -4\sqrt{6} }{\sqrt{2}} +\sqrt{\frac{15}{2}} i \right)
\ee
for $I=1,2,3,4$. The constant of integration $\zeta_1^{(G_2)}$ is the only dimensionful scale in the system, and the freedom to perform the radial shift \eqref{eq.radialshift} allows to re-scale it to any convenient value, for example $\zeta_1^{(G_2)}=1$; the field theory counterpart consists in noticing that the CFT is perturbed by a source that sets the only scale of the theory. Such scale can be always set to a convenient value by a redefinition of the energy units. 

The domain-wall describes a solution interpolating between the D2-brane geometry asymptotics in the UV and the $\textrm{G}_2$ fixed point in the IR. Holographically this corresponds to a $\textrm{G}_2$-preserving deformation in the UV  of the SYM theory that lives in the worldvolume of the D2-branes, to a CS-matter theory in the IR governed by a conformal fixed point \cite{Guarino:2016ynd}.

\item There are two regular deformations of the  ${\cal N}=2$, $\textrm{SU}(3)\times\textrm{U}(1)$ solution as the IR endpoint of a domain-wall solution: $\Delta_{SU(3)\times U(1),1}=\frac{1-\sqrt{17}}{2}$ and $\Delta_{SU(3)\times U(1),2}=\frac{3-\sqrt{17}}{2}$. The linearized BPS equations are solved by the matrix of coefficients
\be
z_{I,a}^{(SU(3)\times U(1))} = \begin{pmatrix}
 \frac{i}{2} & -\frac{i}{2} \\
 -\frac{i}{2} & -\frac{i}{2} \\
 0 & -\frac{1-\sqrt{17}}{8 \cdot 2^{5/6}}(1-\sqrt{3}i) \\
 0 & -\frac{1-\sqrt{17}}{8 \cdot 2^{5/6}}(1-\sqrt{3}i)
\end{pmatrix}
\ee
which preserves $\textrm{SU}(3)$ symmetry since the perturbation maintains $z_1=z_2$ and $z_3=z_4$. Once again, the arbitrary radial shift \eqref{eq.radialshift} allows to set the constant of integration $\zeta_1^{(SU(3)\times U(1))}=1$ without loss of generality. Then $\zeta_2^{(SU(3)\times U(1))}$ can take values in a compact range, parameterising a continuous family of solutions whose UV is given by the D2-brane geometry. In the field theory side there is a continuous deformation of the SYM Lagrangian that makes the theory flow to a CS-matter theory with $\textrm{SU}(3)\times\textrm{U}(1)$ symmetry. At one of the boundaries of this allowed range of values for $\zeta_2^{(SU(3)\times U(1))}$ the UV description of the domain-wall is dominated by the $\textrm{G}_2$ fixed point described above. In this case, as it corresponds to a UV description of the fixed point ($\rho\to\infty$), only positive modes (and therefore regular in the UV) are active around the $\textrm{G}_2$ solution \cite{Guarino:2016ynd}.

\item Finally, \cite{Guarino:2016ynd} also considers domain-walls where the IR endpoint of the solution is given by the ${\cal N}=1$, $\textrm{SU}(3)$ fixed point, with $\Delta_{SU(3),1}=\Delta_{SU(3),2}=1-\sqrt{6}$. The linearized BPS equations imply a matrix of coefficients
\be
z_{I,a}^{(SU(3))} = \begin{pmatrix}
 \frac{\sqrt{3}}{4} & \frac{\sqrt{5}}{4\sqrt{2}}i \\
 \frac{\sqrt{3}}{4} & \frac{\sqrt{5}}{4\sqrt{2}}i \\
 \frac{-3(4+3\sqrt{6}) + (3\sqrt{10}+4\sqrt{15})i}{16} & -\frac{5 \left( 2\sqrt{2}-\sqrt{3} \right) - \sqrt{5}(9-2\sqrt{6})i}{16} \\
 \frac{-3(4+3\sqrt{6}) + (3\sqrt{10}+4\sqrt{15})i}{16} & -\frac{5\left( 2\sqrt{2}-\sqrt{3} \right)  - \sqrt{5}(9-2\sqrt{6})i}{16}
\end{pmatrix}
\ee
which preserves $\textrm{SU}(3)$ symmetry since the perturbation maintains $z_1=z_2$ and $z_3=z_4$. By using the arbitrary radial shift of equation \eqref{eq.radialshift} one can set the constant of integration $\zeta_2^{(SU(3))}=1$ without loss of generality. Then $\zeta_1^{(SU(3))}$ can take values in a compact range, parameterising a continuous family of solutions whose UV is given by the D2-brane geometry, i.e., in the field theory side there is a continuous deformation of the SYM Lagrangian that makes the theory flow to a CS-matter theory with $\textrm{SU}(3)$ symmetry. At one of the limits of this set of allowed values for the constant $\zeta_1^{(SU(3))}$ the UV description of the domain-wall is dominated by the $\textrm{G}_2$ fixed point described above, with only positive modes turned on, as it corresponds to a regular UV description.
\end{itemize}


\section{Symmetries along the GY flow} \label{eq:SymGYFT}

Both $\cN=2$ and $\cN=3$ SCFTs connected by the GY flow can be regarded as different CS-matter phases of the D2-brane field theory: three-dimensional $\cN=8$ SYM. Accordingly, the symmetry groups, summarised in Table~\ref{tab:GYGroups}, at both endpoints and along the flow should be regarded as subgroups of the $\cN=8$ SYM R-symmetry group, SO(7). 

The $\textrm{SU}(3) \times \textrm{U}(1)_\psi$ global symmetry of the UV SCFT is embedded into SO(7) through 
\begin{equation} \label{eq:SO7intoSU3U1}
\textrm{SO}(7) \supset \textrm{SO}(6) \supset \textrm{SU}(3) \times \textrm{U}(1)_\psi \; .
\end{equation}
The $\bm{7}$ real scalars $X^I$ and $\bm{8}$ Majorana fermions $\chi^A$ of $\cN=8$ SYM accordingly branch as 
\begin{eqnarray}
  \label{eq:7ofSO7toSO6toU3branching}
  && \textbf{7} \;
  \xrightarrow{\scriptscriptstyle \mathrm{SO}(6)} \;
   \textbf{6} + \textbf{1} 
  \xrightarrow{\scriptscriptstyle \mathrm{SU}(3)\times\mathrm{U}(1)_\psi} \;
   \big( \textbf{3}_{-\frac23} + \overline{\textbf{3}}_{+\frac23} \big) +  \textbf{1}_0 \; , \\
  \label{eq:8ofSO7toSO6toU3branching}
  && \textbf{8} \;
  \xrightarrow{\scriptscriptstyle \mathrm{SO}(6)} \;
   \textbf{4} + \overline{\textbf{4}} 
 \xrightarrow{\scriptscriptstyle \mathrm{SU}(3)\times\mathrm{U}(1)_\psi} \;
   \big( \textbf{3}_{\frac13} + \textbf{1}_{-1} \big) + \big( \overline{\textbf{3}}_{-\frac13} +  \textbf{1}_{+1} \big) \; .
\end{eqnarray}
The triplets here correspond to the bosonic, $Z^a$, and fermionic, $\chi^a$, on-shell components of the $\cN=2$ SCFT chirals $\Phi^a$, $a=1,2,3$, via complexification of $X^1 , \ldots ,X^6$ and $\chi^1 , \ldots , \chi^6$. The singlets correspond to the real auxiliary scalar, $\sigma \sim ( X^7)^2$, and to the complex gaugino, $\lambda \sim \chi^7 + i \chi^8$, of the $\cN=2$ vector multiplet. Interestingly, the group theory branching (\ref{eq:7ofSO7toSO6toU3branching}) fixes the $\mathrm{U}(1)_\psi$ R-symmetry of $Z^a$ as $R(Z^a) = -\frac23$, in agreement with the independent field theory result (\ref{eq:DeltaConstSolN=2}). For reasons to be explained below, we use conventions where the conformal dimension $\Delta$ and the R-charge $R$ for the lower components of short $\textrm{OSp}(2|4)$ hypermultiplets are related via 
\begin{equation} \label{eq:DeltaR}
\Delta = -R \; , 
\end{equation}
rather than with the more familiar $+$ sign. Also, $X^7$ comes out R-neutral under the branching (\ref{eq:7ofSO7toSO6toU3branching}), consistent with the field theory result $\textrm{tr} \,  (X^7)^2 \sim  \textrm{tr} \, Z^a \bar{Z}_a$, see Section~\ref{sec:StSugra}. Finally, the branching (\ref{eq:8ofSO7toSO6toU3branching}) correctly reproduces the UV field theory R-charge assignment of $+\frac13$ for the fermions $\chi^a$. This follows independently in the field theory by writing out the superfield 
\begin{equation} \label{eq:superfield}
\Phi^a  = Z^a + \sqrt{2} \, \theta^\alpha \chi_\alpha^a + F^a
\end{equation}
in components and assigning R-charge $R(\theta^\alpha) = -1$ in agreement with the sign convention (\ref{eq:DeltaR}). Incidentally, the usual dimension assignment $[\theta^\alpha] = -\frac12$ leads to $\Delta(\chi^a) = \frac76$, so that the fermion mass terms $\textrm{tr} \, \chi^{(a} \chi^{b)}$  have  dimension $\frac73$ in agreement with the supergravity result of Table~\ref{tab:n=0SU3U1D2masses}. As noted in the text, the dimension $\frac23$ for $Z^a$ leads to dimension $\frac43$ for the condensates $\textrm{tr} \, Z^{(a} Z^{b)}$, but does not fix the dimension of the real SU(3)-traceless mass terms $\textrm{tr} \,  (Z^a \bar{Z}_b -\textrm{traces})$. 

The superpotential mass deformation in (\ref{N=2SuperPotDef}) that triggers the GY flow breaks the UV global symmetry $\textrm{SU}(3) \times \textrm{U}(1)_\psi$ down to $\textrm{SO}(3)_\textrm{R} \times \textrm{U}(1)_\textrm{d}$. Here, $\textrm{SO}(3)_\textrm{R} \sim \textrm{SU}(2)$ is the subgroup of SU(3) such that $\bm{3} \rightarrow \bm{2} + \bm{1}$, with $Z^a$, $a= 1,2$,  the doublet and $Z^3$ the singlet. This $\textrm{SO}(3)_\textrm{R}$ is the flavour symmetry group along the flow. The R-symmetry $\textrm{U}(1)_\textrm{d}$ along the flow is the subgroup of $\textrm{SU}(3) \times \textrm{U}(1)_\psi$ that leads to the R-charge assignments (\ref{eqAssign2}) (with opposite sign in our conventions, as in (\ref{eq:DeltaR})) along the flow. In order to determine how $\textrm{U}(1)_\textrm{d}$ is embedded in $\textrm{SU}(3) \times \textrm{U}(1)_\psi$, assume that $\textrm{U}(1)_\textrm{d} = p \, \textrm{U}(1)_\tau + q \, \textrm{U}(1)_\psi$, where $\textrm{U}(1)_\tau$ commutes with $\textrm{SO}(3)_\textrm{R}$ inside SU(3) and $p$, $q$ are constants to be determined. Under 
\begin{equation} \label{eq:BranchingU1d}
\textrm{SU}(3) \times \textrm{U}(1)_\psi \supset \textrm{SO}(3)_\textrm{R} \times \textrm{U}(1)_\tau \times \textrm{U}(1)_\psi \supset \textrm{SO}(3)_\textrm{R} \times \textrm{U}(1)_\textrm{d} \; ,
\end{equation}
the chiral bosons $Z^a$ and fermions $\chi^a$ branch as 
\begin{eqnarray}
  \label{eq:3branching}
  && \textbf{3}_{-\frac23} \;
  \xrightarrow{\scriptscriptstyle \textrm{SO}(3)_\textrm{R} \times \textrm{U}(1)_\tau \times \textrm{U}(1)_\psi} \;
  \textbf{2}_{(\frac12,-\frac23)} +  \textbf{1}_{(-1,-\frac23)}
  \xrightarrow{\scriptscriptstyle \textrm{SO}(3)_\textrm{R} \times \textrm{U}(1)_\textrm{d}} \;
 \textbf{2}_{\frac12 p -\frac23q} +  \textbf{1}_{-p-\frac23 q } \; , 
\\
  \label{eq:3branchingFermion}
  && \textbf{3}_{\frac13} \; 
  \xrightarrow{\scriptscriptstyle \textrm{SO}(3)_\textrm{R} \times \textrm{U}(1)_\tau \times \textrm{U}(1)_\psi} \;
  \textbf{2}_{(\frac12,\frac13)} +  \textbf{1}_{(-1,\frac13)}
  \xrightarrow{\scriptscriptstyle \textrm{SO}(3)_\textrm{R} \times \textrm{U}(1)_\textrm{d}} \;
 \textbf{2}_{\frac12 p +\frac13q} +  \textbf{1}_{-p+\frac13 q } \; .
\end{eqnarray}
The branching (\ref{eq:3branching}) reproduces the R-charge assignments (\ref{eqAssign2}) with the sign conventions of (\ref{eq:DeltaR}) for 
\begin{eqnarray}
\label{eq:U1charges}
\left.
\begin{array}{l}
\tfrac12 \, p - \tfrac23 q = -\tfrac12 \\
 -p - \tfrac23 q = -1 
\end{array} 
\right\} \; \Longrightarrow \; p = \tfrac13 \; , \; q = 1 \; .
\end{eqnarray} 
Thus, $\textrm{U}(1)_\textrm{d}$ is a proper mixture of $\textrm{U}(1)_\tau$ and $\textrm{U}(1)_\psi$, as advertised in Section~\ref{sec:FT}. Tensoring the branchings (\ref{eq:3branching}), (\ref{eq:3branchingFermion}) reproduces the charge assignments in Table~\ref{tab:n=0SU3U1D2masses} for the supergravity fields under $\textrm{SO}(3)_\textrm{R} \times \textrm{U}(1)_\tau \times \textrm{U}(1)_\psi$ and $\textrm{SO}(3)_\textrm{R} \times \textrm{U}(1)_\textrm{d}$ with $p,q$ in (\ref{eq:U1charges}).

In the $\cN=3$ SCFT fixed point, the surviving bosons, $Z^a$, and fermions, $\chi^a$, $a =1,2$, transform as doublets under the $\textrm{SO}(3)_\textrm{R} \sim \textrm{SU}(2)$ IR flavour symmetry. These are also charged under the R-symmetry $\textrm{U}(1)_\textrm{d}$: from (\ref{eq:3branching})--(\ref{eq:U1charges}), $R(Z^a) = -\frac12$ and $R(\chi^a) = \frac12$. These R-charge assignments are consistent with these fields being components of a superfield $\Phi^a$, as in (\ref{eq:superfield}) with now $a=1,2$. In the IR fixed point, the R-symmetry is in fact enlarged to a full $\textrm{SO}(3)_\textrm{d}$, consistent with $\cN=3$ supersymmetry. Thus, we still need to show that the  $\textrm{U}(1)_\textrm{d}$ subgroup of the UV global symmetry group $\textrm{SU}(3) \times \textrm{U}(1)_\psi$ defined via (\ref{eq:BranchingU1d})--(\ref{eq:U1charges}) is also a subgroup of the full $\textrm{SO}(3)_\textrm{d}$ R-symmetry group of the IR.

In order to show this, we resort to the embedding of both UV and IR global symmetries into the common SO(7) R-symmetry of the parent $\cN=8$ SYM theory. On the one hand, further branching (\ref{eq:SO7intoSU3U1}) via (\ref{eq:BranchingU1d}) by combining (\ref{eq:7ofSO7toSO6toU3branching}), (\ref{eq:8ofSO7toSO6toU3branching}) with (\ref{eq:3branching}), (\ref{eq:3branchingFermion}), the $\cN=8$ SYM bosons and fermions $X^I$, $\chi^A$ can be checked to split as
\begin{eqnarray}
  \label{eq:7SO7underSU3U1d}
  && \textbf{7} \;
  \xrightarrow{\scriptscriptstyle \textrm{SO}(3)_\textrm{R} \times \textrm{U}(1)_\textrm{d}} \;
\big(  \textbf{2}_{-\frac12 } +  \textbf{1}_{-1} \big) + \big(  \textbf{2}_{\frac12 } +  \textbf{1}_{1} \big) +\bm{1}_0  \; , 
\\
  \label{eq:8SO7underSU3U1d}
  && \textbf{8} \;
  \xrightarrow{\scriptscriptstyle \textrm{SO}(3)_\textrm{R} \times \textrm{U}(1)_\textrm{d}} \;
\big(  \textbf{2}_{\frac12 } +  \textbf{1}_{0} +  \textbf{1}_{-1} \big) + \big(  \textbf{2}_{-\frac12 } +  \textbf{1}_{0} +  \textbf{1}_{1} \big)  \; .
\end{eqnarray}
On the other hand, the global IR symmetry $\textrm{SO}(3)_{\textrm{d}} \times \textrm{SO}(3)_{\textrm{R}} $ is embedded into SO(7) via 
\begin{equation} \label{eq:embeddingSO3dSOrR}
\textrm{SO}(7) \, \supset \, 
  \textrm{SO}(3)^\prime \times \textrm{SO}(4)^\prime \equiv  \textrm{SO}(3)^\prime \times \textrm{SO}(3)_{\textrm{L}} \times \textrm{SO}(3)_{\textrm{R}} 
 \supset
  \textrm{SO}(3)_{\textrm{d}} \times \textrm{SO}(3)_{\textrm{R}} \; ,
\end{equation}
with $\textrm{SO}(3)_{\textrm{R}}$ the right-handed component of $\textrm{SO}(4)^\prime$ and $\textrm{SO}(3)_{\textrm{d}}$ the diagonal of $\textrm{SO}(3)^\prime \times \textrm{SO}(3)_{\textrm{L}}$, hence the labels employed in the main text. Under (\ref{eq:embeddingSO3dSOrR}), the $\cN=8$ SYM bosons and fermions branch as 
\begin{eqnarray}
  \label{eq:7SO7underSO3dSO3R}
  && \textbf{7} \;
  \xrightarrow{\scriptscriptstyle \textrm{SO}(3)^\prime \times \textrm{SO}(3)_{\textrm{L}} \times \textrm{SO}(3)_{\textrm{R}}} \;
( \bm{1} , \bm{2} , \bm{2} ) + ( \bm{3} , \bm{1} , \bm{1} ) 
  \xrightarrow{\scriptscriptstyle \textrm{SO}(3)_{\textrm{d}} \times \textrm{SO}(3)_{\textrm{R}}} \;
( \bm{2} , \bm{2} ) + ( \bm{3} , \bm{1}  )  \; , 
\\
  \label{eq:8SO7underSO3dSO3R}
  && \textbf{8} \;
  \xrightarrow{\scriptscriptstyle \textrm{SO}(3)^\prime \times \textrm{SO}(3)_{\textrm{L}} \times \textrm{SO}(3)_{\textrm{R}}} \;
( \bm{2} , \bm{2} , \bm{1} ) + ( \bm{2} , \bm{1} , \bm{2} ) 
  \xrightarrow{\scriptscriptstyle \textrm{SO}(3)_{\textrm{d}} \times \textrm{SO}(3)_{\textrm{R}}} \;
( \bm{2} , \bm{2} ) + ( \bm{3} , \bm{1}  ) + ( \bm{1} , \bm{1}  )  \; .\qquad
\end{eqnarray}
Further splitting these under the Cartan of $\textrm{SO}(3)_{\textrm{d}}$, we finally find the branchings
\begin{eqnarray}
  \label{eq:7SO7underSO3dSO3RCartan}
  && \textbf{7} \;
 \longrightarrow \; 
( \bm{2}_{\frac12} +  \bm{2}_{-\frac12} )  + ( \bm{1}_{1} +  \bm{1}_{0} +  \bm{1}_{-1} ) 
    \; , 
\\
  \label{eq:8SO7underSO3dSO3RCartan}
  && \textbf{8} \;
 \longrightarrow \; 
( \bm{2}_{\frac12} +  \bm{2}_{-\frac12} )  + ( \bm{1}_{1} +  \bm{1}_{0} +  \bm{1}_{-1} ) + \bm{1}_0 \; ,
\end{eqnarray}
into representations of $\textrm{SO}(3)_{\textrm{R}}$ and the U(1) Cartan of $\textrm{SO}(3)_{\textrm{d}}$. The branchings (\ref{eq:7SO7underSO3dSO3RCartan}), (\ref{eq:8SO7underSO3dSO3RCartan}) coincide with (\ref{eq:7SO7underSU3U1d}), (\ref{eq:8SO7underSU3U1d}). This proves that the $\textrm{U}(1)_\textrm{d}$ subgroup of the UV symmetry group $\textrm{SU}(3) \times \textrm{U}(1)_\psi$ defined via (\ref{eq:BranchingU1d})--(\ref{eq:U1charges}) is indeed the Cartan subgroup of the $\textrm{SO}(3)_\textrm{d}$ R-symmetry group of the IR.
 
We conclude with a justification of our unusual sign choice in the shortening relation (\ref{eq:DeltaR}) that relates the conformal dimension and the R-charge of $\textrm{OSp}(4|2)$ hypermultiplets. The reason is that, with this sign convention, the IR R-charge assignments for the mass deformed \cite{Benna:2008zy,Ahn:2000aq,Corrado:2001nv} ABJM chirals transverse to the M2-branes (see (3.16) of \cite{Klebanov:2008vq}),
\begin{equation}
\textrm{M2}: \qquad R(Z^a) = +\tfrac13 \; , \;  a=1,2,3 \; , \qquad R(Z^4) = +1 \; , 
\end{equation}
and the R-symmetry assignments in our case, given by equation (\ref{eq:7ofSO7toSO6toU3branching}),
\begin{equation}
\textrm{D2}: \qquad R(Z^a) = -\tfrac23 \; , \;  a=1,2,3 \; , \qquad R(X^7) = 0 \; , 
\end{equation}
are related by an SO(8) triality rotation. See \cite{Pang:2017omp} for the details.


\section{Ten-dimensional geometries}

\subsection{IIA uplift of the $D=4$ $\textrm{SO}(3)_\textrm{R}$-invariant sector}
\label{sec.SO3Ruplift}

Here we present the uplift the four-dimensional flows to massive IIA supergravity. In order to do this, we use the formulae of \cite{Guarino:2015jca,Guarino:2015vca} for the consistent truncation of massive type IIA down to $D=4$ $\cN=8$ dyonically gauged $\textrm{ISO}(7)$ supergravity \cite{Guarino:2015qaa}, and particularise them to the eight-scalar sector defined in Section~\ref{section.minimal}. We find it convenient to pack the scalars $\phi_1$, $\phi_2$, $\phi_3$ and the pseudoscalars $b_{11}$, $b_{22}$, $b_{33}$ into the $3 \times 3$ matrices
\begin{equation} \label{eq:defmats}
\bm{m} = \textrm{diag} \, \big( e^{-\sqrt{2} \phi_1} , \, e^{-\sqrt{2} \phi_2} , \, e^{-\sqrt{2} \phi_3} \big) \; ,
\qquad
\bm{b} = \textrm{diag} \, \big( b_{11} , \, b_{22} , \, b_{33} \big) \; .
\end{equation}
Whenever needed, the individual components of these matrices will be denoted $m_{ij}$ and $b^{\hat{\imath}}{}_j$, following the notation of \cite{Guarino:2019jef} with the index conventions of \cite{Varela:2019vyd}. In fact, the uplifting formulae below take on {\it the exact same form} for the entire  $\cN=4$, $\textrm{SO}(3)_\textrm{R}$-invariant scalar sector \cite{Guarino:2019jef}  of $\textrm{ISO}(7)$ supergravity, when $\bm{m}$ and $\bm{b}$ in (\ref{eq:defmats}) are replaced with the general $\textrm{SO}(3)_\textrm{R}$-invariant expressions given in \cite{Guarino:2019jef}.

We will give the uplifting formulae in $S^6$ embedding coordinates $\mu^I$, $I=1, \ldots, 7$, subject to the constraint
\begin{equation} \label{eq:S6constraint}
\delta_{IJ} \, \mu^I \mu^J = 1 \; .
\end{equation}
It is convenient to split the index $ I= (i, a)$, with $i=1,2,3$ and $a=4,5,6,7$ respectively labelling the fundamental representations of the $\textrm{SO}(3)^\prime$ and $\textrm{SO}(4)^\prime$ subgroups of $\textrm{SO}(7)$ defined in equation \eqref{eq:embeddingSO3dSOrR}. The $S^6$ embedding coordinates thus split as $\mu^I \equiv (\mu^i , \nu^a )$. Sometimes we will suppress the indices on these and will write $\bm{\mu}$ and $\bm{\nu}$, in line with the notation employed for the $D=4$ fields. Incidentally, the $S^6$ coordinates $\bm{\nu}$ should not be confused with the $D=4$ coset representative given in (2.9) of \cite{Guarino:2019jef}.
 
 It is helpful to introduce the following functions of the $D=4$ scalars and the $S^6$ embedding coordinates:
\begin{eqnarray} \label{Delta1}
\Delta_1 = e^{\sqrt{2} (\phi_1+\phi_2+\phi_3 ) } \, \bm{\mu}^\textrm{T} \bm{m} \bm{\mu} + e^{\varphi + \frac{1}{\sqrt{2}}  (\phi_1+\phi_2+\phi_3 )} \, \bm{\nu}^\textrm{T} \bm{\nu} \; , 
\end{eqnarray}
and
{\setlength\arraycolsep{2pt}
\begin{eqnarray} \label{Delta2}
\Delta_2 &=& e^{-\varphi+\sqrt{2} (\phi_1+\phi_2+\phi_3 ) } \, \big(1+ e^{2\varphi} \chi^2 \big) \big[  \bm{\mu}^\textrm{T} \big( \bm{m} + \tfrac12 \bm{b}^\textrm{T} \bm{b} \big)  \bm{\mu} \big]^2 \nonumber \\[4pt]
&& + e^\varphi \Big[ 1 + \tfrac12 \, \textrm{tr} \big( \bm{b}^\textrm{T} \bm{b} \bm{m}^{-1} \big) + \tfrac18 \big[ \textrm{tr} \big( \bm{b}^\textrm{T} \bm{b} \bm{m}^{-1} \big) \big]^2 -
 \tfrac18  \textrm{tr} \big( \bm{b}^\textrm{T} \bm{b} \bm{m}^{-1} \bm{b}^\textrm{T} \bm{b} \bm{m}^{-1} \big) \nonumber \\[4pt]
&& \qquad \; +\tfrac18 \, e^{\sqrt{2} (\phi_1+\phi_2+\phi_3 ) } \big( \textrm{det} \, \bm{b} \big)^2 \Big]  \big( \bm{\nu}^\textrm{T} \bm{\nu} \big)^2 \nonumber \\[4pt]
&& + \Big[ e^{\frac{1}{\sqrt{2}} (\phi_1 +\phi_2 +\phi_3 )} \big[ 2+ e^{2\varphi}\chi^2 + \tfrac12 \textrm{tr}  \big( \bm{b}^\textrm{T} \bm{b} \bm{m}^{-1} \big) \big] \nonumber \\[4pt]
&& \qquad \;  -\tfrac{1}{\sqrt{2}}  \, e^{\varphi+ \sqrt{2} (\phi_1 +\phi_2 +\phi_3 ) } \, \chi  \,  \textrm{det} \, \bm{b}   \Big] \big[  \bm{\mu}^\textrm{T} \big( \bm{m} + \tfrac12 \bm{b}^\textrm{T} \bm{b} \big)  \bm{\mu} \big] \, \bm{\nu}^\textrm{T} \bm{\nu}  \; .
\end{eqnarray}
}With these definitions, the uplift of the $D=4$ metric, $ds_4^2$, and scalars into the $D=10$ Einstein-frame metric reads
\begin{equation} \label{eq:10Dmetric}
d\hat{s}^2_{10} = \Delta_1^{1/8} \Delta_2^{1/4} \,  \big( ds_4^2 + g^{-2} \, \Delta_2^{-1} \, d\bar{s}_6^2 \big) \; ,
\end{equation}
where
{\setlength\arraycolsep{2pt}
\begin{eqnarray} \label{eq:10DmetricComps}
d\bar{s}^2_6 &=& M_{ij} \, d\mu^i \, d\mu^j + M_i \, d\mu^i \, \bm{\nu}^\textrm{T} d\bm{\nu} +M^{i \hat{\jmath}} \, d\mu_i \big( \bm{\nu} J_{\hat{\jmath}_-} d \bm{\nu} \big)  + M^{\hat{\imath} \hat{\jmath}} \, \big( \bm{\nu} J_{\hat{\imath} - } d \bm{\nu} \big) \big( \bm{\nu} J_{\hat{\jmath} - } d \bm{\nu} \big)   \nonumber \\[4pt]
&& + M \big( \bm{\nu}^\textrm{T} d\bm{\nu}  \big)^2  + P \, d \bm{\nu}^\textrm{T} d\bm{\nu} \; .
\end{eqnarray}
}Here, $J_{\hat{\imath}_-}$, $\hat{\imath} =1,2,3$, is the triplet of constant antiself-dual two-forms introduced in Appendix~A of \cite{Guarino:2019jef} (with indices $a$ there replaced with $\hat{\imath}$ here), and the quantities $M_{ij}$, etc., depend on the $D=4$ scalars and the $S^6$ embedding coordinates $\bm{\mu}$, $\bm{\nu}$. Specifically, they are given by the lengthy expressions:
{\setlength\arraycolsep{1pt}
\begin{eqnarray}
M_{ij} &=& \big[  \bm{\mu}^\textrm{T} \big( \bm{m} + \tfrac12 \bm{b}^\textrm{T} \bm{b} \big)  \bm{\mu} + e^{\varphi - \frac{1}{\sqrt{2}}  (\phi_1+\phi_2+\phi_3 )} \, \bm{\nu}^\textrm{T} \bm{\nu}  \big] \, (m^{-1})_{ij} 
\nonumber \\[5pt]
&& + \tfrac12 \, \epsilon_i{}^{k\ell} \epsilon_j{}^{hm} \, (  \bm{b}^\textrm{T} \bm{b} )_{kh}  \big( \Delta_1 m_{\ell m } - e^{\sqrt{2}  (\phi_1+\phi_2+\phi_3 )} \, m_{\ell r} \,  m_{ms} \, \mu^r \mu^s \big) 
\nonumber \\[5pt]
&& + \tfrac14 \, e^{\sqrt{2}  (\phi_1+\phi_2+\phi_3 )} \,  \Delta_1^{-1}  \epsilon_i{}^{k\ell} \epsilon_j{}^{hm} \, (  \bm{b}^\textrm{T} \bm{b} )_{kp} (  \bm{b}^\textrm{T} \bm{b} )_{hq} \, \mu^p \mu^q  \big( \Delta_1 m_{\ell m } - e^{\sqrt{2}  (\phi_1+\phi_2+\phi_3 )} \, m_{\ell r} \,  m_{ms} \, \mu^r \mu^s \big) 
\nonumber \\[5pt]
&& +\tfrac14 \, e^{\varphi+ \frac{1}{\sqrt{2}}  (\phi_1+\phi_2+\phi_3 )} \Big[ (  \bm{b}^\textrm{T} \bm{b}   \, \bm{b}^\textrm{T} \bm{b} )_{ij} -  (  \bm{b}^\textrm{T} \bm{b})_{ij} \, \textrm{tr}  (  \bm{b}^\textrm{T} \bm{b} ) +\tfrac12 \big[ \textrm{tr}  (  \bm{b}^\textrm{T} \bm{b} ) \big]^2 \delta_{ij} -\tfrac12 \textrm{tr}  (  \bm{b}^\textrm{T} \bm{b}   \, \bm{b}^\textrm{T} \bm{b} ) \, \delta_{ij}  \Big] \, \bm{\nu}^\textrm{T} \bm{\nu} %
\nonumber \\[5pt]
&& -\tfrac18 \big[ 1-2 e^{2\varphi} \chi^2 + 2 \, (\bm{b}^\textrm{T} \bm{b} )_{kh} \, (m^{-1})^{kh}  \big] \, \mu_i \mu_j \; , 
\nonumber \\[15pt]
M^{i \hat{\jmath}} &=& -(m^{-1})^{ih} \, \epsilon^{\hat{\jmath}}{}_{\hat k \hat \ell} \, b^{\hat{k}}{}_h \, b^{\hat{\ell}}{}_n \, \mu^n 
+ \sqrt{2} \, \chi \, e^{\varphi + \frac{1}{\sqrt{2}} (\phi_1+\phi_2+\phi_3) }  \epsilon^{ih\ell} \, b^{\hat{\jmath}}{}_h \, \big( m_{\ell n} + \tfrac12 ( \bm{b}^\textrm{T} \bm{b} )_{\ell n} \big) \, \mu^n %
\nonumber \\[5pt]
&& -\tfrac{1}{\sqrt{2}} \, \Delta_1^{-1} \, \chi \, e^{\varphi + \frac{3}{\sqrt{2}} (\phi_1+\phi_2+\phi_3) } b^{\hat{\jmath}}{}_q \, \epsilon^{ih\ell} \, m_{h n } \, ( \bm{b}^\textrm{T} \bm{b} )_{ \ell p}  \, \mu^q \mu^n \mu^p
\nonumber \\[5pt]
&& -\tfrac14 \, e^{\sqrt{2} (\phi_1+\phi_2+\phi_3) } \epsilon^{ikm} \, \epsilon^{\hat{\jmath}}{}_{\hat p \hat q} \, \epsilon^{pqn} \, b^{\hat p}{}_p \, b^{\hat q}{}_q \,  (  \bm{b}^\textrm{T} \bm{b})_{k\ell} \, \mu^\ell \, \big(  m_{mn} - \Delta_1^{-1} \,  e^{\sqrt{2}  (\phi_1+\phi_2+\phi_3 )} \, m_{mh} \,  m_{ns} \, \mu^h \mu^s \big) \; , 
\nonumber \\[15pt]
M^{\hat{\imath} \hat{\jmath}} &=& -\tfrac12 \, b^{\hat{\imath}}{}_k \, b^{\hat{\jmath}}{}_h \, \big[ (m^{-1})^{kh} +\Delta_1^{-1} \chi^2 \, e^{2\varphi +\sqrt{2} ( \phi_1 + \phi_2 + \phi_3)} \, \mu^k \mu^h \big]  \nonumber \\[5pt]
&& +\tfrac{1}{16} \, e^{\sqrt{2} (\phi_1+\phi_2+\phi_3 ) }   \, \Delta_1^{-1} \, \epsilon^{(\hat{\imath}}{}_{\hat k \hat h}  \, \epsilon^{\hat{\jmath})}{}_{\hat m \hat n}  \, \epsilon^{ijk} \, \epsilon^{pqr} \, b^{\hat{k}}{}_p \, b^{\hat{h}}{}_q b^{\hat{m}}{}_i b^{\hat{n}}{}_j   \big( \Delta_1 m_{rk } - e^{\sqrt{2}  (\phi_1+\phi_2+\phi_3 )} \, m_{rs} \,  m_{kt} \, \mu^s \mu^t \big) \nonumber \\[5pt]
&& -\tfrac{1}{2\sqrt{2}} \, \Delta_1^{-1} \, \chi \, e^{\varphi +\frac{3}{\sqrt{2}}  ( \phi_1 + \phi_2 + \phi_3)} \, b^{\hat{\imath}}{}_h \,  \mu^h \, \epsilon^{\hat{\jmath}}{}_{\hat k \hat h} \, \epsilon^{pqr} \,  b^{\hat k}{}_p \,  b^{\hat h}{}_q \, m_{rs} \, \mu^s  \; ,
\nonumber \\[15pt]
M_i & = & \nonumber  -\tfrac14 \,  \big[ 1 +2 \, e^{2\varphi}\chi^2  - 2\sqrt{2}  \, e^{\varphi+ \frac{1}{\sqrt{2}} (\phi_1 +\phi_2 +\phi_3 ) } \, \chi  \,  \textrm{det} \, \bm{b}  \big] \, \mu_i  \; , \\[15pt]
M & = & \nonumber  -\tfrac18 \,  \big[ 1 -2 \, e^{2\varphi}\chi^2 + 2\, \textrm{tr}  \big( \bm{b}^\textrm{T} \bm{b} \bm{m}^{-1} \big) \big]  \; , \\[15pt]
P &=& e^{-\varphi+\frac{1}{\sqrt{2}} (\phi_1+\phi_2+\phi_3 ) } \, \big(1+ e^{2\varphi} \chi^2 \big) \big[  \bm{\mu}^\textrm{T} \big( \bm{m} + \tfrac12 \bm{b}^\textrm{T} \bm{b} \big)  \bm{\mu} \big] + \big[ 1 + \tfrac12 \, \textrm{tr} \big( \bm{b}^\textrm{T} \bm{b} \bm{m}^{-1} \big) \big] \, \bm{\nu}^\textrm{T} \bm{\nu} \; .
\end{eqnarray}
}Indices $i$ and $\hat{\imath}$ here are raised and lowered with $\delta_{ij}$ and $\delta_{\hat{\imath}\hat{\jmath}}$, respectively. We have verified that these expressions reproduce the known uplifts of AdS fixed points, see Appendix~\ref{sec:GYsymmetriesGeom}.

We have also computed the embedding of the model of Section~\ref{section.minimal} (and, in fact, of the full $\textrm{SO}(3)_\textrm{R}$-invariant scalar sector \cite{Guarino:2019jef} of $\textrm{ISO}(7)$ supergravity) into the type IIA dilaton and Ramond-Ramond one-form. The consistent embedding into the dilaton reads
\begin{equation} \label{eq:dilaton}
e^{\hat \phi} = \Delta_1^{3/4} \Delta_2^{-1/2} \; ,
\end{equation}
in terms of the quantities introduced in (\ref{Delta1}), (\ref{Delta2}). The uplift of the scalars into the Ramond-Ramond one-form is given by
{\setlength\arraycolsep{2pt}
\begin{eqnarray} \label{eq:A1RR}
\hat{A}_\1 &=& - \tfrac12 \, g^{-1} \,  \epsilon^{ijk} \, a_{ij} \, d\mu_k  +\tfrac12 \, g^{-1} \,  e^{\sqrt{2} (\phi_1 +\phi_2 +\phi_3 ) } \, \Delta_1^{-1} \, \epsilon^{ijk}  \, m_{ih} \, (\bm{b}^\textrm{T} \bm{b} )_{j \ell} \, \mu^h \mu^\ell  \, d\mu_k \nonumber \\[4pt]
&&  +\tfrac{1}{\sqrt{2}} \, g^{-1} \, \chi \, e^{\varphi+ \frac{1}{\sqrt{2}} (\phi_1 +\phi_2 +\phi_3 ) } \, \Delta_{1}^{-1} \, b^{\hat{\imath} }{}_j \, \mu^j \big( \bm{\nu}^\textrm{T} J_{\hat{\imath} - } \bm{\nu}  \big)  \nonumber \\[4pt]
&&  +\tfrac{1}{4} \, g^{-1} \, e^{\sqrt{2} (\phi_1 +\phi_2 +\phi_3 ) } \, \Delta_{1}^{-1} \, \epsilon_{\hat{\imath} \hat{\jmath} \hat{k} } \, \epsilon^{ijk} \, b^{\hat{\imath} }{}_i \,  b^{\hat{\jmath}}{}_j \, m_{k \ell} \, \mu^\ell \,  \big( \bm{\nu}^\textrm{T} J^{\hat{k}} _- \, \bm{\nu}  \big)  \; ,
\end{eqnarray}
}where $a_{ij} = - a_{ji}$ are the $\textrm{SO}(3)_\textrm{R}$-invariant $D=4$ St\"uckelberg scalars introduced in \cite{Guarino:2019jef}. Note that these enter neither the IIA metric (\ref{eq:10Dmetric}) nor the dilaton (\ref{eq:dilaton}). We have verified that, when evaluated at the $\cN=2$ and $\cN=3$ critical points, equations (\ref{eq:dilaton}) and (\ref{eq:A1RR}) reproduce the corresponding expressions for the dilaton and Ramond-Ramond one-form given in \cite{Guarino:2015jca,Varela:2015uca,DeLuca:2018buk}, by making use of the $S^6$ embedding coordinates (\ref{eq:musN=2}), (\ref{eq:musN=3}) below.

The expressions for the B-field and the Ramond-Ramond three-form are left as an exercise. 

\subsection{Geometric realisation of the GY flow symmetries} \label{sec:GYsymmetriesGeom}

It is useful to introduce local coordinates on the deformed $S^6$ geometries to track how the different symmetries worked out in Appendix~\ref{eq:SymGYFT} act geometrically along the entire GY flow, including the endpoint geometries.

When evaluated at the $\cN=2$ $\textrm{SU}(3) \times \textrm{U}(1)$-invariant UV fixed point, the internal metric $ds^2_6 \equiv \Delta_2^{-1} \, d\bar{s}_6^2$ in (\ref{eq:10Dmetric}) reduces to equation (\ref{eq:N=2UVGeom}) of the main text, upon choosing the $S^6$ embedding coordinates $\mu^I = (\mu^i , \nu^a)$ as 
\begin{eqnarray} \label{eq:musN=2}
& \mu^1 +i \mu^2 = \sin \alpha \, \cos \xi  \, e^{i \psi}  \; , \qquad 
\mu^3 = \cos \alpha \; ,  \\[4pt]
& \nu^1 +i \nu^2 = -\sin \alpha \, \sin \xi \, \cos \tfrac{\theta}{2} \, e^{\frac{i}{2} ( 2\psi + \tau +\phi)}  \; , \qquad
\nu^3 +i \nu^4 =  -\sin \alpha \, \sin \xi \, \sin \tfrac{\theta}{2} \, e^{\frac{i}{2} ( 2\psi + \tau -\phi)}   \; , \nonumber
\end{eqnarray}
with
\begin{equation} 
0 \leq \alpha \leq \pi \; , \quad
0 \leq \psi < 2\pi \; , \quad 
0 \leq \tau < 2\pi \; , \quad
0 \leq \xi \leq \tfrac{\pi}{2} \; , \quad
0 \leq \theta \leq \pi \; , \quad
0 \leq \phi < 2\pi \; .
\end{equation}
The first four of these angles appear explicitly in the metric (\ref{eq:N=2UVGeom}), while the last two parametrise $\mathbb{CP}^1$, with
\begin{equation} \label{eq:musN=3}
ds^2 ( \mathbb{CP}^1 ) = \tfrac14 \big( d\theta^2 + \sin^2 \theta \, d\phi^2 \big) \; , \qquad
\sigma = \cos \theta \, d\phi \; .
\end{equation}
As discussed in the text, the metric inside of the brackets on the second line of (\ref{eq:N=2UVGeom}) is simply the Fubini-Study metric on $\mathbb{CP}^2$. Its isometry is thus SU(3). With the parametrisation (\ref{eq:musN=2}), only an $\textrm{SU}(2) \sim \textrm{SO}(3)_\textrm{R}$ symmetry, acting on the $\mathbb{CP}^1$ subspace, along with the $\textrm{U}(1)_\tau$ generated by $\partial_\tau$, is manifest. In these coordinates, $\textrm{SO}(3)_\textrm{R}$ is generated by the Killing vectors
\begin{equation} \label{eq:GenR}
R_1 = -\sin\phi \, \partial_\theta - \cos\phi \, \cot \theta \, \partial_\phi \; , \qquad
R_2 = \cos \phi \, \partial_\theta - \sin\phi \, \cot \theta \, \partial_\phi \; , \qquad
R_3 = \partial_\phi \; .
\end{equation}
These close into the Lie algebra
\begin{eqnarray} \label{eq:GenRComm}
[ R_i , R_j ] = -\epsilon_{ij}{}^k \, R_k \; ,
\end{eqnarray}
with the normalisation of (A.6) of \cite{Guarino:2019jef}. In addition to these flavour symmetries, the geometry (\ref{eq:N=2UVGeom}) is also invariant under the $\textrm{U}(1)_\psi$ generated by $\partial_\psi$. This $\textrm{U}(1)_\psi$ corresponds the R-symmetry of the dual UV $\cN=2$ field theory.

At the $\cN=3$ fixed point, it is convenient to choose a parametrisation that is better adapted to the symmetries of the solution. We have verified that the internal metric $ds^2_6 \equiv \Delta_2^{-1} \, d\bar{s}_6^2$ for the $\cN=3$ $\textrm{SO}(4)$ fixed point, (\ref{eq:N=3IRGeom}), can be recovered from (\ref{eq:10Dmetric}) by evaluating the latter at the corresponding $D=4$ scalars vevs and selecting the $S^6$ embedding coordinates as 
\begin{eqnarray} \label{eq:musN=3bis}
\mu^i = \cos \beta \, \tilde{\mu}^i \; , \quad
\nu^1 +i \nu^2 = -\sin \beta \, \cos \tfrac{\theta}{2} \, e^{\frac{i}{2} ( \psi^\prime -\phi)} \; , \quad
\nu^3 +i \nu^4 = -\sin \beta \, \sin \tfrac{\theta}{2} \, e^{(\psi^\prime +\phi)} \; .
\end{eqnarray}
Here, $\tilde{\mu}^i$, $i=1,2,3$ define a round $S^2$ through $\delta_{ij} \tilde{\mu}^i \tilde{\mu}^j =1$, and $0 \leq \beta \leq \tfrac{\pi}{2}$, $0 \leq \psi^\prime < 2\pi$. The angles $\theta$, $\phi$ in (\ref{eq:musN=3bis}) are the same that appear in (\ref{eq:musN=2}). Together with $\psi^\prime$, these now parametrise an $S^3$, with $\psi^\prime$ the coordinate along the Hopf fibre. The right-invariant one-forms on this $S^3$ are
\begin{equation}
\rho^1 = \cos \psi^\prime \, d\theta - \sin \psi^\prime \, \sin \theta \, d\phi \; , \;
\rho^2 =  \sin \psi^\prime \, d\theta + \cos \psi^\prime \, \sin \theta \, d\phi  \; , \;
\rho^3 = - \big( d\psi^\prime - \cos \theta \, d\phi \big) \; , \quad 
\end{equation}
and obey the Maurer-Cartan equations
\begin{equation} \label{eq:MC}
d\rho^i = -\tfrac12 \, \epsilon^i{}_{jk} \rho^j \wedge \rho^k \; .
\end{equation}

Equations (\ref{eq:MC}) are the dual, differential form version of the commutation relations (\ref{eq:GenRComm}). Indeed, the group $\textrm{SO}(3)_\textrm{R}$ generated by (\ref{eq:GenR}) is also a symmetry of the geometry (\ref{eq:N=3IRGeom}). In the IR $\cN=3$ field theory, this $\textrm{SO}(3)_\textrm{R}$ corresponds to the flavour symmetry. In addition, the solution  (\ref{eq:N=3IRGeom}) possesses an $\textrm{SO}(3)_\textrm{d}$ R-symmetry. To see how this acts, it is convenient to introduce further coordinates $\tilde{\theta}$, $\tilde{\phi}$ on the $S^2$ fibres by letting
\begin{equation} \label{eq:mutildes}
\tilde{\mu}^1 = \sin \tilde{\theta} \, \cos \tilde{\phi} \; , \qquad 
\tilde{\mu}^2 = \sin \tilde{\theta} \, \sin \tilde{\phi} \; , \qquad 
\tilde{\mu}^3 = \cos \tilde{\theta} \; . \qquad 
\end{equation}
The R-symmetry group $\textrm{SO}(3)_\textrm{d}$ is diagonally embedded into the $\textrm{SO}(3)^\prime$ that rotates $\tilde{\mu}^i$, and the left-invariant $\textrm{SO}(3)_\textrm{L}$ that acts on the $S^3$ parametrised by $\theta, \phi, \psi^\prime$. After some calculation, we find that $\textrm{SO}(3)_\textrm{d}$ is generated by the vectors
\begin{eqnarray} \label{eq:GenDiag}
D_1 & =&  \sin \tilde{\phi} \, \partial_{\tilde{\theta} } +  \cos \tilde{\phi} \, \cot \tilde{\theta}  \, \partial_{\tilde{\phi} } + \cos\psi^\prime \, \partial_\theta -\csc \theta \, \sin \psi^\prime \, \partial_\phi -\cot \theta \, \sin \psi^\prime \, \partial_{\psi^\prime} \; , \nonumber \\[4pt]
D_2 & =&  -\cos \tilde{\phi} \, \partial_{\tilde{\theta} } +  \sin \tilde{\phi} \, \cot \tilde{\theta}  \, \partial_{\tilde{\phi} } + \sin \psi^\prime \, \partial_\theta +\csc \theta \, \sin \psi^\prime \, \partial_\phi +\cot \theta \, \cos \psi^\prime \, \partial_{\psi^\prime} \; , \nonumber \\[4pt]
D_3 & =& -\partial_{\tilde{\phi} } - \partial_{\psi^\prime}  \; .
\end{eqnarray}
These can indeed be checked to be Killing vectors of the $\cN=3$ metric (\ref{eq:N=3IRGeom}), and to close on the $\textrm{SO}(3)_\textrm{d}$ commutation relations,
\begin{eqnarray} \label{eq:GenDiagComm}
[ D_i , D_j ] = \epsilon_{ij}{}^k \, D_k \; ,
\end{eqnarray}
normalised as in Appendix~A of \cite{Guarino:2019jef}. 

At intermediate energies along the GY flow, only an $\textrm{SO}(3)_\textrm{R} \times \textrm{U}(1)_\textrm{d}$ symmetry is preserved. This $\textrm{SO}(3)_\textrm{R}$ is generated by the Killing vectors (\ref{eq:GenR}) that act on the $\mathbb{CP}^1$ factor of both the UV, (\ref{eq:N=2UVGeom}), and IR, (\ref{eq:N=3IRGeom}), geometries. The $\textrm{U}(1)_\textrm{d}$ is the Cartan subgroup of the $\textrm{SO}(3)_\textrm{d}$ IR R-symmetry: it is, thus, generated by $D_3$ in (\ref{eq:GenDiag}). From this equation, it is apparent that $\textrm{U}(1)_\textrm{d}$ is the diagonal combination of the azimuthal $\textrm{U}(1)$ that acts on the $S^2$ parametrised by $\tilde{\mu}^i$, and the $\textrm{U}(1)_{\psi^\prime}$ that acts on the Hopf fibre of the $S^3$ base of the IR geometry. It is also interesting to determine how $\textrm{U}(1)_\textrm{d}$ is embedded into the $\textrm{U}(1)_\psi \times \textrm{U}(1)_\tau$ symmetry the UV geometry (\ref{eq:N=2UVGeom}).
 In order to elucidate this, we simply keep track of the diffeomorphism that relates the different coordinates in which the UV and IR geometries are expressed. Comparing the expressions for the $S^6$ embedding coordinates (\ref{eq:musN=2}) to (\ref{eq:musN=3bis}), (\ref{eq:mutildes}), and recalling that $\theta$, $\phi$ in both expressions are the same, we deduce
\begin{equation} \label{S6CoordsChange}
\cos \alpha = \cos \beta \, \cos \tilde{\theta} \; , \quad 
\sin \alpha \, \sin \xi = \sin \beta \; , \quad
\psi = \tilde{\phi} \; , \quad
\tau = \psi^\prime -2\psi \; .
\end{equation}
Therefore
\begin{equation} 
\label{eq:U1dinUVCoords}
D_3 = -\tfrac12 \,  (3 \, \partial_\psi + 2 \, \partial_\tau ) \; .
\end{equation}
This matches (\ref{eq:U1charges}) up to a rescaling of the $\tau$ coordinate.

\bibliography{references}
\end{document}